\RequirePackage{fix-cm}

\documentclass[twocolumn,epjc3]{svjour3}  

\smartqed  

\RequirePackage{graphicx}
\RequirePackage{cite}
\usepackage{atlasphysics}
\usepackage{subfig}
\usepackage{mathrsfs}
\usepackage{authblk}
\usepackage{multirow}
\usepackage{url}
\usepackage{xspace}
\usepackage[normalem]{ulem}
\usepackage{hyperref} 

\DeclareGraphicsExtensions{.eps}

\newcommand{\pom}{\ensuremath{{\rm I\!P}}\xspace}
\newcommand{\reg}{\ensuremath{{\rm I\!R}}\xspace}
\newcommand{\alphapom}{\ensuremath{\alpha_{\pom}(0)}\xspace}
\newcommand{\alphapomt}{\ensuremath{\alpha_{\pom}(t)}\xspace}
\newcommand{\alphapomprime}{\ensuremath{\alpha_{\pom}\prime}\xspace}
\newcommand{\alphareg}{\ensuremath{\alpha_{\reg}(0)}\xspace}
\newcommand{\xix}{\ensuremath{\xi_{X}}\xspace}
\newcommand{\xicut}{\ensuremath{\xi_{\rm Cut}}\xspace}
\newcommand{\xiy}{\ensuremath{\xi_{Y}}\xspace}
\newcommand{\mxovers}{\ensuremath{\frac{M_{X}^{2}}{s}}\xspace}
\newcommand{\myovers}{\ensuremath{\frac{M_{Y}^{2}}{s}}\xspace}
\newcommand{\logxix}{\ensuremath{\log_{10}\left(\xi_X\right)}\xspace}

\newcommand{\lappr}{\stackrel{<}{_{\sim}}\xspace}
\newcommand{\gappr}{\stackrel{>}{_{\sim}}\xspace}

\newcommand{\MX}{\ensuremath{M_{X}}\xspace}
\newcommand{\MY}{\ensuremath{M_{Y}}\xspace}
\newcommand{\Mp}{\ensuremath{M_{p}}\xspace}
\newcommand{\deta}{\ensuremath{\Delta\eta}\xspace}
\newcommand{\detaf}{\ensuremath{\Delta\eta^{F}}\xspace}
\newcommand{\detafcut}{\ensuremath{\Delta\eta^{F}_{\rm Cut}}\xspace}
\newcommand{\ptcut}{\ensuremath{p_\mathrm{T}^{\rm cut}}\xspace}
\newcommand{\pttrack}{\ensuremath{p_\mathrm{T}^{\rm Track}}\xspace}

\newcommand{\refeq}[1]{(\ref{#1})}
\newcommand{\reftab}[1]{Table~\ref{#1}}

\newcommand{\reffig}[1]{Figure~\ref{#1}}
\newcommand{\reffigs}[1]{Figures~\ref{#1}}
\newcommand{\refsec}[1]{Section~\ref{#1}}

\newcommand{\py}{{\sc pythia}}
\newcommand{\pysix}{{\sc pythia{\footnotesize{6}}}\xspace}
\newcommand{\pyeight}{{\sc pythia{\footnotesize{8}}}\xspace}
\newcommand{\pho}{{\sc phojet}\xspace}
\newcommand{\geant}{{\sc geant{\footnotesize{4}}}\xspace}
\newcommand{\hpp}{{\sc herwig{\scriptsize{++}}}\xspace}

\newcommand{\fnd}{\ensuremath{f\!_{N\!D}}\xspace}
\newcommand{\fsd}{\ensuremath{f\!_{S\!D}}\xspace}
\newcommand{\fdd}{\ensuremath{f\!_{D\!D}}\xspace}
\newcommand{\fcd}{\ensuremath{f\!_{C\!D}}\xspace}
\newcommand{\fd}{\ensuremath{f\!_{D}}\xspace}
\newcommand{\snd}{\ensuremath{\sigma\!_{N\!D}}\xspace}
\newcommand{\ssd}{\ensuremath{\sigma\!_{S\!D}}\xspace}
\newcommand{\sdd}{\ensuremath{\sigma\!_{D\!D}}\xspace}
\newcommand{\scd}{\ensuremath{\sigma\!_{C\!D}}\xspace}

\newcommand{\sqs}{\ensuremath{\sqrt{s} = 7}~\TeV\xspace}

\newcommand{\sig}{\ensuremath{S}\xspace}
\newcommand{\sigth}{\ensuremath{S_{\rm th}}\xspace}

\newcommand{\sigman}{\ensuremath{\sigma_\mathrm{noise}}\xspace}

\newcommand{\pc}{\,pC\xspace}

\journalname{Eur.\ Phys.\ J.\ C}

\begin{document}

\title{\vspace{-2.0cm}\flushleft{\normalsize\normalfont{CERN-PH-EP-2011-220}} \flushright{\vspace{-0.86cm}\normalsize\normalfont{Submitted to Eur. Phys. J. C}} \\[2.5cm] \flushleft{Rapidity Gap Cross Sections measured with the ATLAS Detector in {\boldmath $pp$} Collisions at {\boldmath $\sqrt{s} = 7 \TeV$}}}


\titlerunning{Rapidity Gap Cross Sections measured with the ATLAS Detector in $pp$ Collisions at $\sqrt{s} = 7 \TeV$}

\author{The ATLAS Collaboration}
\institute{}
\date{January 13, 2012}

\maketitle

\begin{abstract}
Pseudorapidity gap
distributions 
in proton-proton collisions at \sqs{}
are studied using a minimum bias data sample with an 
integrated luminosity of $7.1 \ {\rm \mu b^{-1}}$.
Cross sections are measured differentially in 
terms of \detaf,
the larger of the pseudorapidity regions 
extending to the limits of the ATLAS sensitivity,
at $\eta = \pm 4.9$, in which no final state
particles are produced above a transverse momentum 
threshold \ptcut. 
The measurements span the region $0 < \detaf < 8$ 
for $200 < \ptcut < 800 \MeV$.  
At small \detaf, the data  
test the reliability of hadronisation models in describing 
rapidity and transverse momentum fluctuations
in final state particle production.
The measurements at larger gap sizes
are dominated by contributions from
the single diffractive dissociation process ($pp \rightarrow Xp$), 
enhanced by double dissociation ($pp \rightarrow XY$)
where the invariant mass of the lighter of the
two dissociation systems satisfies $\MY \lappr 7 \GeV$. The
resulting cross section is
${\rm d} \sigma / {\rm d} \detaf \approx 1 \ {\rm mb}$ 
for $\detaf \gappr 3$.
The large rapidity gap data are used to constrain the value of the 
Pomeron intercept appropriate to triple Regge models of
soft diffraction. The cross section integrated over all gap sizes is
compared with other LHC inelastic cross section measurements.

\end{abstract}

\section{Introduction}

When two protons collide inelastically at 
a centre-of-mass energy of $7 \TeV$
in the Large Hadron Collider (LHC), 
typically around six charged particles 
are produced with transverse momentum\footnote{In the 
ATLAS coordinate system,
the $z$-axis points in the direction of 
the anti-clockwise beam viewed from above. Polar   
angles $\theta$ and transverse momenta \pt\ are measured with respect to 
this axis. 
The pseudorapidity $\eta = - \ln \, \tan( \theta / 2)$ 
is a good approximation to the rapidity of a particle 
whose mass is negligible compared with its energy and is used here, relative to the 
nominal $z = 0$ point at the centre of the apparatus, to describe regions of the detector.}
\pt\ $>$ 100 \MeV\ per unit of 
pseudorapidity in the central 
region \cite{:2010ir,arXiv:1011.5531,arXiv:1004.3514}.
On average, the rapidity difference between neighbouring particles 
is therefore around 0.15 units of rapidity, with 
larger gaps occurring due to statistical fluctuations 
in the hadronisation process. 
Such random
processes lead to an exponential suppression with gap 
size \cite{Bjorken:1992mj}, 
but very large gaps are produced
where a $t$-channel colour singlet
exchange takes place. 
This may be due to an electroweak
exchange, 
but occurs much more frequently through the exchange of 
strongly interacting states. 
At high energies
such processes are termed `diffractive'
and are associated with
`Pomeron' exchange \cite{pomeron,Chew:1961ev}.

\begin{figure*}
 \centering
 \subfloat[]{\includegraphics[width=0.3\textwidth]{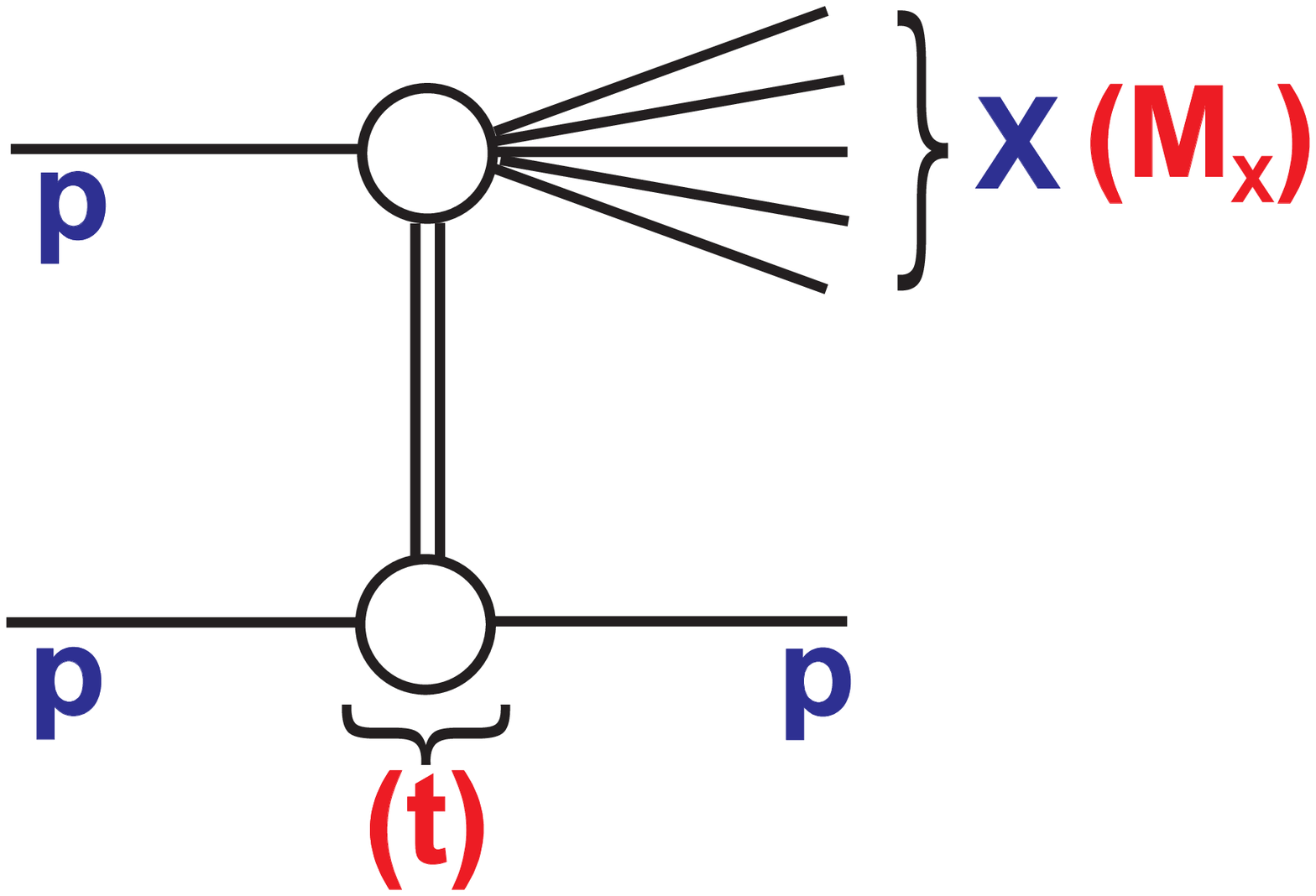}}   
 \hspace*{0.05\textwidth} 
 \subfloat[]{\includegraphics[width=0.3\textwidth]{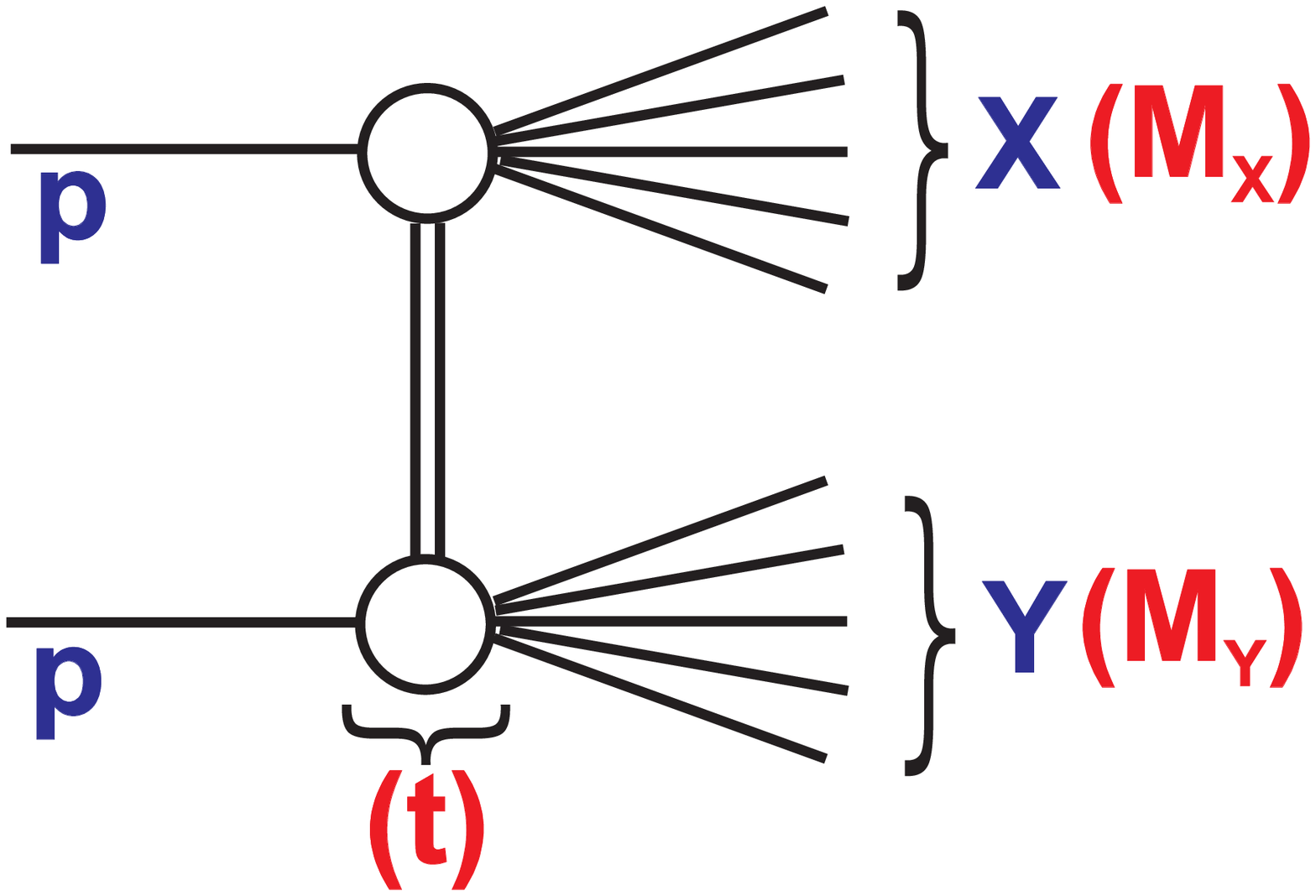}} 
 \hspace*{0.05\textwidth} 
 \subfloat[]{\includegraphics[width=0.3\textwidth]{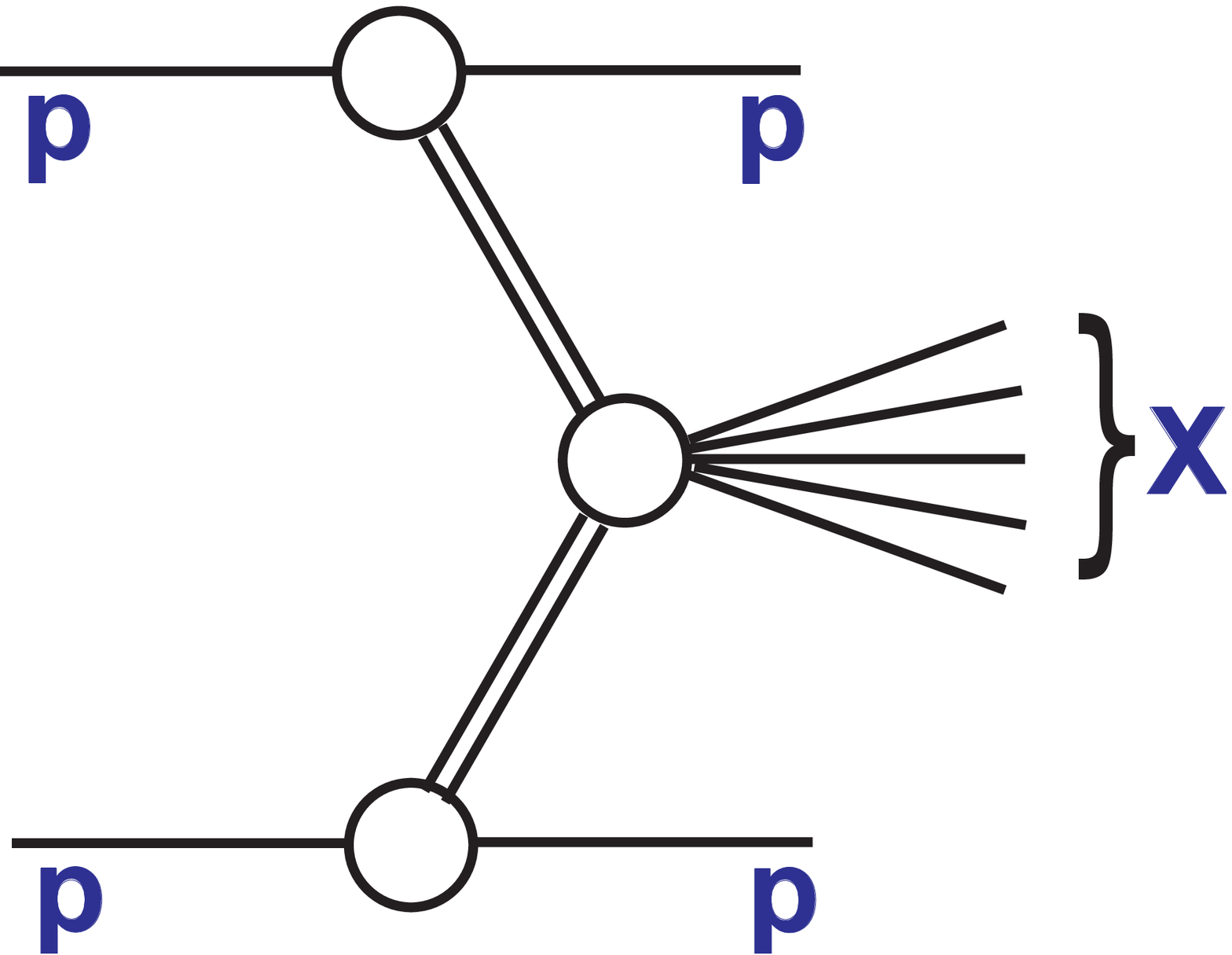}} 
 \caption[]{Schematic illustrations of the single-diffractive dissociation (a), 
double-diffractive dissociation
(b) and central diffractive (c) 
processes and the kinematic variables used
to describe them. By convention, the mass \MY is always smaller
than \MX in the double dissociation case 
and $\MY = \Mp$ in the single dissociation case, \Mp 
being the proton mass.}
 \label{Feynmans}
\end{figure*}

The total cross section in 
hadronic scattering experiments is commonly decomposed into 
four main components: elastic 
($pp \rightarrow pp$  in the LHC context), 
single-diffractive dissociation (SD, $pp \rightarrow Xp$ or $pp \rightarrow pX$,
\reffig{Feynmans}a),
double-diffractive dissociation (DD, $pp \rightarrow XY$,
\reffig{Feynmans}b) and
non-diffrac\-tive (ND) 
contributions. The
more complex 
central diff\-ractive configuration 
(CD,\linebreak $pp \rightarrow pXp$,
\reffig{Feynmans}c),  
in which final state particles are produced 
in the central region with intact protons on both sides, 
is suppressed relative to the SD process by a factor of around
10 at high energies \cite{Acosta:2003xi}.
Together, the diffractive channels contribute 
approximately 25--30\% of the total inelastic cross section 
at LHC energies \cite{lauren}.
Following measurements at the LHC of  
the elastic \cite{Antchev:2011zz}, total \cite{TOTEM:sigmatot} and total 
inelastic \cite{lauren,TOTEM:sigmatot} 
cross sections, this article contains the 
first detailed exploration of diffractive 
dissociation processes.

Understanding 
diffractive processes is important in its own right, 
as they are the dominant
contribution to high energy quasi-elastic
scattering between hadrons and,
via ideas derived from 
the optical theorem \cite{Mueller:1970fa}, 
are also related to the total cross section. 
They are often interpreted at the parton level 
in terms of the exchange of pairs of gluons \cite{Low:1975sv,Nussinov:1975mw}
and are thus sensitive to 
possible parton saturation effects in the 
low Bjorken-$x$ regime 
of proton structure \cite{Gribov:1984tu,GolecBiernat:1998js,Iancu:2000hn}.
Diffractive cross sections also
have relevance to cosmic ray physics \cite{Jung:2009eq} and   
may be related to the string 
theory of gravity \cite{Brower:2006ea}.
At the LHC, diffractive dissociation must be
well understood for a good 
description of the additional inelastic
proton-proton interactions
(pile-up) which accompany most events. It also
produces a significant uncertainty in approaches to luminosity 
monitoring which rely on measurements of the total, or total inelastic, cross
section \cite{Aad:2011dr}. 

Diffractive dissociation cross sections have been measured
previously over a wide range of centre-of-mass energies. 
Early measurements are reviewed 
in \cite{Goulianos:1982vk,Alberi:1981af,Zotov,Kaidalov:1979jz,Albrow:2010yb}.
SD measurements have been made in $p\bar{p}$ scattering at the
SPS \cite{Bernard:1986yh,Ansorge:1986xq} and the 
Tevatron \cite{Amos:1992jw,Abe:1993wu}, and also 
in photoproduction \cite{Adloff:1997mi,Breitweg:1997za} and deep inelastic 
scattering \cite{Aktas:2006hy,Chekanov:2008fh,H1:2010kz} at HERA.
Limited high energy 
DD \cite{Ansorge:1986xq,Adloff:1997mi,Affolder:2001vx} 
and CD \cite{Joyce:1993ru,Brandt:2002qr,Acosta:2003xi}
data are also available.
In most cases, the momentum transfer is too small
to permit an interpretation in terms of partonic degrees of 
freedom \cite{Ingelman:1984ns}.
Instead, phenomenological models such as those based on Regge theory
have been developed 
\cite{Kaidalov:1973tc,Field:1974fg,Zotov}, which underlie the 
Monte Carlo generators typically used to predict the properties of 
soft inelastic collisions \cite{pythia6,py8,phojet}. Mixed approaches have also 
been developed which employ perturbative QCD where 
possible \cite{Ryskin:2011qe,Gotsman:2010nw}.
Large theoretical uncertainties remain in the detailed dynamics expected
at the LHC.

Direct measurements of the masses 
\MX and \MY of the dissociated systems are 
difficult at ATLAS, since many of the 
final state particles are 
produced beyond the acceptance of the detector. 
However, the dissociation masses are closely correlated with the 
size of the rapidity region in which
particle production is completely suppressed
due to the net colour-singlet Pomeron exchange.
This correlation is exploited in this paper,
with cross sections reported as a function of 
the size of a pseudorapidity region which is
devoid of final state particle 
production. These 
unpopulated pseudorapidity regions are referred to in the following
as `rapidity gaps', or simply `gaps'.

To maximise the pseudorapidity coverage and sensitivity to charged and
neutral particle production, 
rapidity gaps are identified using both the 
ATLAS calorimeters and tracking detectors. 
The specific observable studied is \detaf, 
the larger of the two 
`forward' pseudorapidity
regions extending to at least $\eta = \pm 4.9$
in which no particles are produced with 
$\pt\ > \ptcut$, where \ptcut is varied between $200 \MeV$ and 
$800 \MeV$.
ND contributions appear at 
small gap sizes, with \ptcut and \detaf
dependences which are sensitive
to fluctuations in the hadronisation process. 
For small \ptcut choices, the large gap size region is
dominated by
SD events and DD events in which one of the dissociation
masses is small.

\section{Experimental Method}

\subsection{The ATLAS detector}
\label{sec:detector}

The ATLAS detector is described in detail 
elsewhere \cite{ATLAS_TDR}. The beam-line is
surrounded by the `inner detector' 
tracking system, which covers the pseudorapidity
range $|\eta| < 2.5$.
This detector consists of silicon pixel, 
silicon strip and straw tube detectors and is enclosed within a 
uniform $2 \, {\rm T}$ solenoidal magnetic field. 

The calorimeters lie outside the tracking system. A highly segmented 
electromagnetic (EM) liquid argon sampling calorimeter covers the range
$|\eta|<3.2$. The EM calorimeter also includes a
pre-sampler covering $|\eta|<1.8$.
The hadronic end-cap (HEC, $1.5<|\eta|<3.2$) 
and forward (FCal, $3.1<|\eta|<4.9$) calorimeters
also use liquid argon technology, with 
granularity decreasing with increasing $| \eta|$. Hadronic energy in
the central region ($|\eta|<1.7$)
is reconstructed in a steel/scintillator-tile 
calorimeter.

Minimum bias trigger scintillator (MBTS) detectors are mounted in front
of the end-cap calorimeters on both sides of the interaction point
and cover the pseudorapidity range $2.1 < |\eta| < 3.8$.
The MBTS is divided into inner and outer
rings, both of which have eight-fold segmentation. The MBTS is used
to trigger the events analysed here. 

In 2010, the luminosity was measured using a \v{C}eren\-kov light detector
which is located $17 \, {\rm m}$ 
from the interaction point. The luminosity
calibration is determined through van der Meer beam 
scans \cite{Aad:2011dr,LuminosityWorkingGroup:1330358}.

\subsection{Event selection and backgrounds}
\label{sec:evt_sel}

The data used in this analysis were collected during the first LHC run
at $\surd s = 7 \TeV$
in March 2010, when the LHC was filled with two bunches per beam, one pair colliding
at the ATLAS interaction point.
The peak instantaneous luminosity 
was $1.1 \times 10^{27} \ {\rm cm^{-2} \, s^{-1}}$.
Events were collected from colliding proton bunch
crossings in which the MBTS trigger recorded one or more 
inner or outer segments above threshold
on at least one side of ATLAS. 
After reconstruction, events are required to have  
hits in at least two of the 
MBTS segments
above a threshold of 0.15\pc. 
This threshold cut suppresses contributions from noise, 
which are well modelled by a Gaussian with 0.02\pc width. 
No further event selection requirements are
applied. 

The data 
sample corresponds to an integrated luminosity of 
$7.1 \pm 0.2 \ {\rm \mu b^{-1}}$ and the number of recorded events is 
422776. 
The mean number of interactions per bunch crossing is below 0.005,
which is consistent with the  approximately 400 events which have multiple 
reconstructed vertices. Pile-up contamination is thus negligible. 

The data sample contains a contribution from 
beam-induced background, mainly due to scattering 
of beam protons from
residual gas particles inside 
the detector region. 
This contamination is estimated using events collected in unpaired bunches
and is subtracted statistically in each measurement interval. 
Averaged over the full measurement region, it amounts to $0.2 \%$
of the sample.
More complex 
backgrounds in which beam-induced background is overlaid on a physics
event are negligible.
            
\subsection{Reconstruction of rapidity gaps}
\label{sec:gap-reco}

The analysis of final state activity in the central 
region ($|\eta| < 2.5$)
is based on combined information from inner detector
tracks and calorimeter modules. In the region
$2.5 < |\eta| < 4.9$, beyond the acceptance
of the inner detector, calorimeter information alone
is used. 
The track selection is as detailed in \cite{:2010ir}.
Energy deposits from final state particles in the calorimeters 
are identified using a topological 
clustering algorithm \cite{:2010wv,Lampl:1099735},
with a further requirement to improve the control over 
noise contributions, as described below. 

The identification of rapidity gap signatures relies crucially on the 
suppression of calorimeter noise contributions.  
The root-mean-squared cell energies due to noise vary
from around $20 \MeV$ in the most central region to
around $200 \MeV$ for the most forward region \cite{:2009zp}.
The shapes of the cell noise distributions in 
each calorimeter are well described by 
Gaussian distributions of standard deviation \sigman,
with the exception of the tile calorimeter, which has 
extended tails.   
The default 
clustering algorithm \cite{Lampl:1099735} is seeded by cells for which 
the significance of the measured energy, $E$, is
$\sig = E / \sigman > 4$. 
However, with this threshold
there are on average six clusters reconstructed per 
empty event due to fluctuations in the noise distributions.  
To suppress noise contributions to acceptable levels for gap finding, 
clusters of calorimeter energy deposits
are thus considered only if they contain at least one cell outside the tile calorimeter with an
energy significance above an $\eta$-dependent threshold, $\sigth$.
This threshold is determined separately in 
pseudorapidity slices of size $0.1$
such that the probability of finding 
at least one noisy cell in each $\eta$-slice has a common value, 
$1.4\times 10^{-4}$. 
This choice optimises
the resolution of the reconstructed gap
sizes with respect to the gaps in the generated final state particle
distributions according to MC studies.
Since the number of cells in an $\eta$-slice varies from 
about 4000 in
the central region to 10 in the
outer part of the FCal, the cell thresholds
vary between $\sigth = 5.8$ in the central region and 
$\sigth = 4.8$ at the highest $|\eta|$ values in the FCal. 

\begin{figure}
\centering
\subfloat[]{\includegraphics[width=0.46\textwidth]{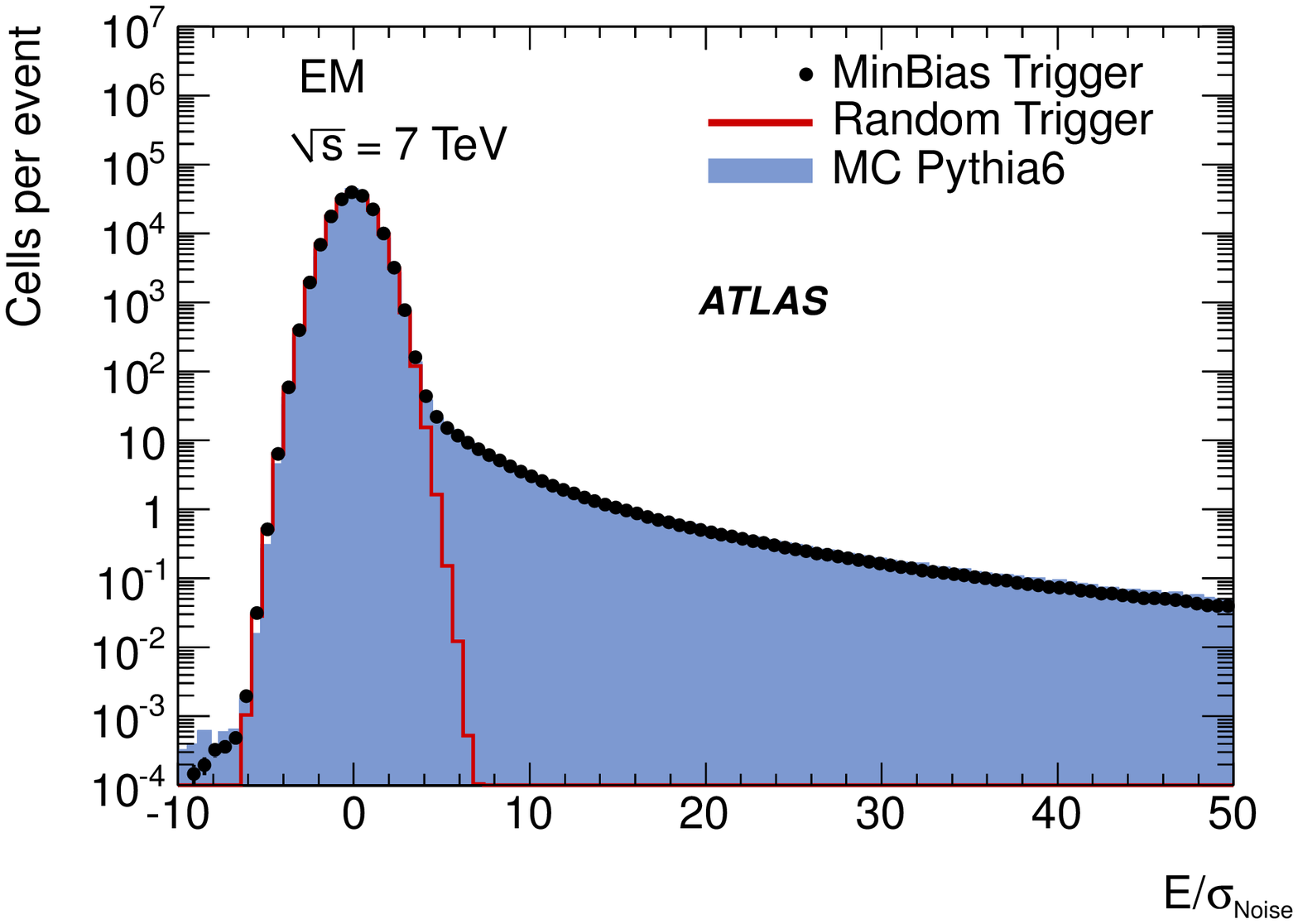}}\newline
\subfloat[]{\includegraphics[width=0.46\textwidth]{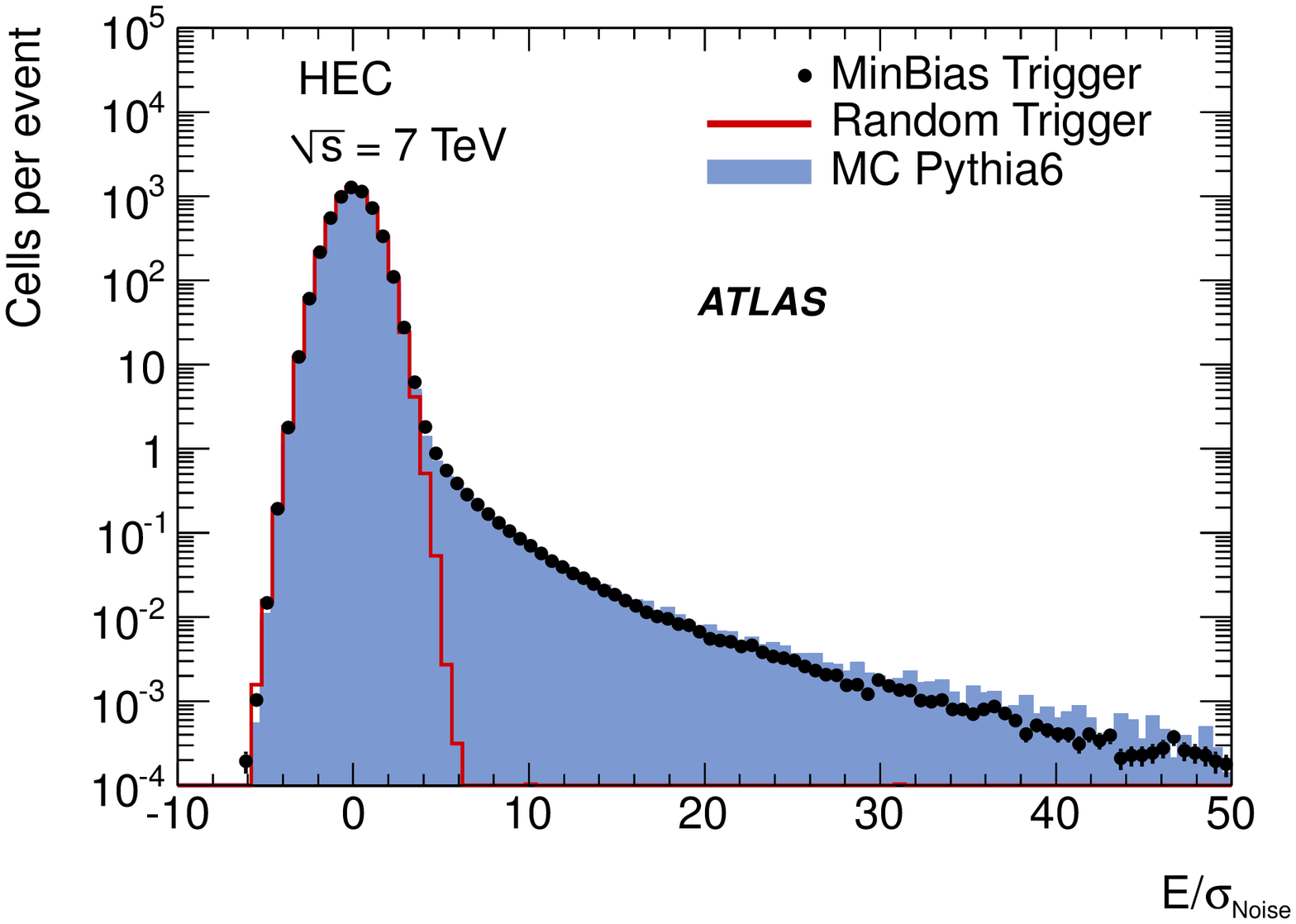}}\newline
\subfloat[]{\includegraphics[width=0.46\textwidth]{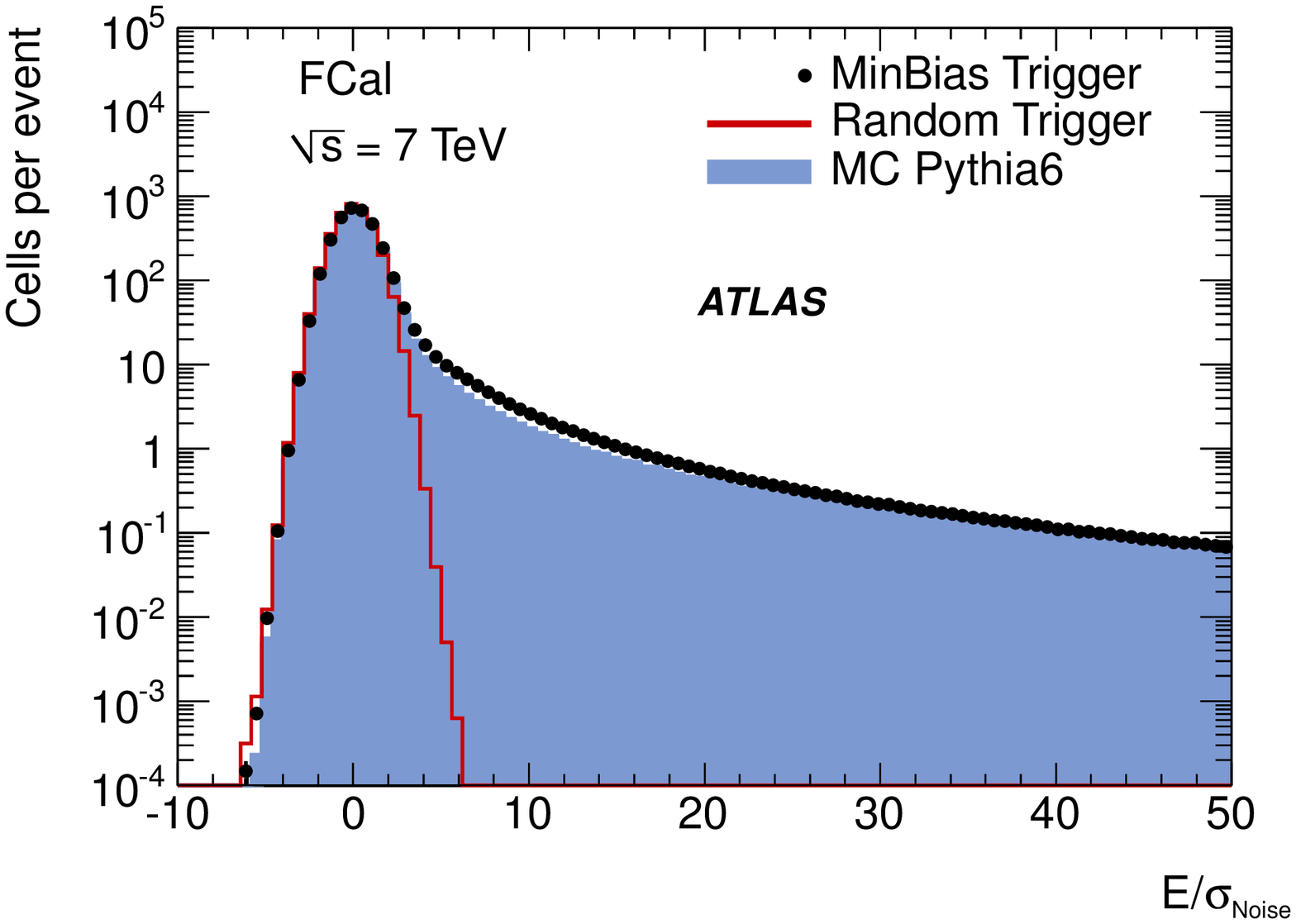}}\newline
\caption[Cell energy distribution]{Cell energy significance, 
$\sig = E/\sigman$, distributions 
for the EM (a), HEC (b) and FCal (c)  
calorimeters. Each cell used in the analysis is included
for every event, with the normalisation set to a single event.
MBTS-triggered minimum bias data (points) are compared
with events randomly triggered on empty 
bunch crossings (histograms) and with a Monte Carlo 
simulation (shaded areas).}
\label{Fig:Cell-Signif}
\end{figure}

The level of understanding of the calorimeter
noise 
is illustrated in \reffig{Fig:Cell-Signif}, which
shows the distributions of the 
cell significance 
$\sig$ for each of the liquid argon modules.
MBTS-triggered data from colliding bunch crossings 
are compared with 
a Monte Carlo simulation and with
events 
which are required to
exhibit no activity in the non-calorimeter components of the detector, 
triggered randomly on empty bunch crossings.
The signal from $pp$ collisions 
is clearly visible in the long positive tails, which are well described by the simulation.  
The data from the empty bunch crossings show the shape of 
the noise distribution with no influence from  
physics signals. The empty bunch crossing 
noise distributions are symmetric around zero and their negative sides closely  
match the negative parts of the MBTS-triggered data distributions.
The noise distribution is well described over seven orders of magnitude by the  
MC simulation, the small residual differences at positive significances
being attributable to deficiencies in the modelling of 
$pp$ collision processes. 

The measured energies of calorimeter clusters which pass the noise requirements
are discriminated using a given value of \ptcut, neglecting particle
masses. 
\linebreak
The 
calorimeter energy scale for electromagnetic showers
is determined from 
electron test-beam studies and 
$Z \rightarrow e^+ e^-$ data \cite{Aad:2010yt}, confirmed
at the relatively small
energies relevant to the gap finding algorithm through 
a dedicated study of $\pizero \rightarrow \gamma \gamma$ 
decays.
The calorimeter response to hadronic showers is substantially lower
than that to electromagnetic showers. 
In the central region,
the scale of the hadronic energy measurements 
is determined relative to the electromagnetic scale
through comparisons between the 
calorimeter and inner detector measurements of single isolated 
hadrons \cite{ATLAS-CONF-2010-017,ATLAS-CONF-2010-052,ATLAS-CONF-2011-028}.
Beyond the acceptance region of the tracking detectors, the difference
between the electromagnetic and the hadronic response 
is determined from test-beam 
results \cite{Pinfold:2008zzb,Archambault:2008zz,Dowler:2002zp}.
For the purposes of discriminating against thresholds in the gap finding
algorithm, all cluster energy measurements are taken at this hadronic 
scale. 
An interval in $\eta$ is deemed to contain final state particles 
if at least one cluster in that interval
passes the noise suppression requirements and 
has a transverse momentum above \ptcut, or 
if there is at least one good inner detector track 
with transverse momentum above \ptcut. 

\subsection{Definition of forward rapidity gap observable}
\label{sec:sigdef}

The reconstructed forward gap size $\detaf$ is defined by the 
larger of the two 
empty pseudorapidity regions extending between 
the edges 
of the detector acceptance at $\eta=4.9$ or $\eta=-4.9$
and the nearest track or calorimeter cluster passing the selection
requirements at smaller $|\eta|$.
No requirements are placed on particle production at $|\eta| > 4.9$
and no attempt is made to identify gaps in the central
region of the detector. 
The rapidity gap size relative to $\eta = \pm 4.9$
lies in the range $0 < \detaf < 8$, such that for example
$\detaf = 8$ implies that there is no 
reconstructed particle with 
$\pt > \ptcut$ in one of the regions $-4.9 < \eta < 3.1$ or
$-3.1 < \eta < 4.9$. The upper
limit on the gap size is constrained 
via the requirement of a high trigger efficiency by the acceptance of the 
MBTS detector.

The measurement is performed in \detaf intervals of 0.2,
except at the smallest values $\detaf < 2.0$, where the 
differential cross section varies fastest
with \detaf and the 
gap end-point determination is most strongly dependent on the 
relatively coarse cell granularity of the FCal. The
bin sizes in this region are increased to 0.4 pseudorapidity units,
commensurate with the resolution. 

The default value of the transverse momentum thre\-shold is chosen to be 
$\ptcut = 200 \MeV$. This value lies
within the acceptance of the track reconstruction for the inner detector
and ensures that the efficiency of the 
calorimeter cluster selection is
greater than $50 \%$ throughout the $\eta$ region which lies
beyond the tracking acceptance. 

As described in \refsec{sec:Unfolding}, the data 
are fully corrected for experimental effects using the Monte 
Carlo simulations introduced in \refsec{sec:mctune}.
The rapidity gap observable defining the 
measured differential cross sections are thus specified in terms of 
stable (proper lifetime $> 10 \ {\rm ps}$)
final state particles (hereafter referred to as the `hadron level'),
with transverse momentum larger than the 
threshold, 
\ptcut, used in the gap reconstruction algorithm.

\section{Theoretical Models and Simulations}\label{sec:Event-Gen}

\subsection{Kinematic Variables and Theory}
\label{theory}

As illustrated in \reffig{Feynmans}a and b,
diffractive dissociation kinematics
can be described in terms of the invariant
masses \MX and \MY of the dissociation systems $X$ and $Y$, respectively
(with $\MY = \Mp$ in the SD case),   
and the 
squared four-momentum 
transfer $t$. In the following, the convention $\MY<\MX$ is adopted. 
The cross section
is vastly dominated by small values of 
$|t| \lappr 1 \GeV^{2}$, such that
the intact proton in SD events is scattered through only a small
angle, gaining transverse momentum $\pt \simeq \surd |t|$. 
Further commonly used kinematic variables are defined as
\begin{equation}
\hspace*{1cm}
\xix = \mxovers \ , 
\hspace*{1.5cm} 
\xiy = \myovers \ ,
\end{equation}
where $s$ is the
square of the centre-of-mass energy.

Diffractive dissociation cross sections
can be modelled using Regge 
phenomenology \cite{Kaidalov:1973tc,Field:1974fg,Roy:1974wq}, 
with Pom\-eron
exchange being the dominant process at small $\xi_X$ values.
For the SD case, the amplitude is factorised into a Pomeron 
flux associated with the proton which remains intact,
and a total probability for the interaction of the Pomeron with the 
dissociating proton. The latter can be described in terms of a further
Pomeron exchange using Muller's generalisation of the optical 
theorem \cite{Mueller:1970fa}, which is applicable
for $s \gg M_{X}^2 \gg m_p^2$. The SD cross section can 
then be expressed as
a triple Pomeron ($\rm \pom \pom \pom$) amplitude,
\begin{eqnarray}
  \frac{ {\rm d} \sigma}{{\rm d}t \, {\rm d} M_X^2} = G_{3 \pom}(0) 
  s^{2 \alpha_{\pom}(t) - 2} \left( M_X^2 \right)^{\alpha_{\pom}(0) - 2 \alpha_{\pom}(t)} f(t) 
  \label{pomonly}
\end{eqnarray}
where $G_{3 \pom}(0)$ is a product of couplings and
$\alpha_{\pom}(t) = \alpha_{\pom}(0) + \alpha_{\pom}^\prime t$ is the
Pomeron trajectory. The term $f(t)$ is usually taken to be 
exponential 
such that 
${\rm d} \sigma / {\rm d} t \propto e^{B (s, M_X^2) \: t}$ at fixed $s$ and 
$M_X$, $B$ being the slope parameter. 
With $\alpha_{\pom}(0)$ close to unity and $|t|$ small, 
equation~\refeq{pomonly}\ 
leads to an approximately constant
${\rm d} \sigma / {\rm d} \ln \xi_X$
at fixed $s$. The DD cross section 
follows a similar dependence at fixed \xiy.
The deviations from this behaviour are sensitive to 
the intercept \alphapom of the Pomeron 
trajectory \cite{Donnachie:1992ny,Cudell:1996sh}
and to absorptive corrections associated with unitarity
constraints \cite{Ryskin:2011qe,Gotsman:2010nw}.

The rapidity gap size and its location are closely correlated with the 
variables \xix and \xiy. For the SD process,
the size $\Delta \eta$ of the rapidity gap between the final state proton and
the $X$ system satisfies
 \begin{equation}
  \hspace*{3cm}
  \deta \simeq - \ln \xix \ .
  \label{gap:eqn}
\end{equation}
The $\detaf$ observable studied here 
differs from $\Delta \eta$  
in that $\detaf$ takes no account of
particle production at $| \eta | > 4.9$. For the SD process, where
the intact proton has 
$\eta \simeq \pm \frac{1}{2} \ln (s/m_p^2) \simeq \pm 8.9$, the gap
variables are related by $\detaf \simeq \Delta \eta - 4$.
Equations~\refeq{pomonly} and~\refeq{gap:eqn} thus lead to 
approximately constant predicted cross sections 
${\rm d} \sigma / {\rm d} \Delta \eta^F$
for SD and low \MY DD events.
With the high centre-of-mass energy
of the LHC and 
the extensive acceptance of the ATLAS detector, events with 
\xix between around $10^{-6}$ and $10^{-2}$ can be selected on the
basis of their rapidity gap signatures, corresponding
approximately to $7 < \MX < 700 \GeV$.

Previous proton-proton scattering \cite{Bernard:1986yh} 
and photoproduction \cite{Adloff:1997mi,Breitweg:1997za} 
experiments have observed enhancements
relative to triple-Pomeron behaviour at the smallest \MX values
in the triple Regge region. This effect has been interpreted in terms
of a further triple Regge term ($\rm \pom \pom \reg$) in which the
reaction still proceeds via Pomeron exchange, but where the
total Pomeron-proton cross section is described by a sub-leading
Reggeon (\reg) with intercept 
$\alphareg \simeq 0.5$ \cite{Donnachie:1992ny}. 
This leads by analogy with equation~\refeq{pomonly}\ 
to a contribution to the cross section which falls as 
${\rm d} \sigma / {\rm d} M_X^2 \propto 1 / M_X^3$.
In the recent model of Ryskin, Martin and Khoze (RMK) \cite{Ryskin:2011qe},
a modified triple-Pomeron approach to the large \xix region is 
combined with a 
dedicated treatment of low mass diffractive dissociation, motivated
by the original $s$-channel picture of Good and 
Walker \cite{Good:1960ba}, in which
proton and excited proton eigenstates scatter elastically
from the target with different absorption coefficients.
This leads
to a considerable enhancement in the low \xix cross section which
is compatible with that observed in the pre-LHC data \cite{Bernard:1986yh}. 

\subsection{Monte Carlo Simulations}
\label{sec:mctune}

Triple Pomeron-based parameterisations 
are implemen\-ted in the commonly used Monte Carlo (MC) event generators, 
\py~\cite{pythia6,py8} and \pho~\cite{phojet,Bopp:1998rc}. These generators are 
used to correct the data for experimental effects and as a 
means of comparing the corrected data with theoretical models.

By default, the \py\ model of diffractive dissociation processes 
uses the Schuler and Sj\"{o}strand 
parameterisation \cite{py6_diff}
of the Pomeron flux, which 
assumes a Pomeron intercept of
unity and an exponential $t$ dependence $e^{B(\xix, \xiy) t}$. 
Three alternative flux models are also implemented. The Bruni and 
Ingelman version \cite{Bruni:1993ym} is similar to Schuler and Sj\"{o}strand,
except that its $t$ dependence is given by the sum of two exponentials. 
In the Berger and Streng \cite{Berger:1986iu,Streng:1988hd} 
and Donnachie and Landshoff \cite{Donnachie:1984xq}
models, 
the Pomeron trajectory is linear, with variable parameters, 
the default being $\alpha_{\pom}(t) = 1.085 + 0.25 t$ \cite{Jung:1993gf}, 
consistent with results from 
fits to total \cite{Donnachie:1992ny,Cudell:1996sh} and 
elastic \cite{Abe:1993xx} hadronic cross
section data. Whilst the model attributed to 
Berger and Streng has an
exponential $t$ dependence, the Donnachie and Landshoff version 
is based on a dipole model of the proton elastic form factor.
For all flux parameterisations in \py, 
additional factors are applied
to modify the distributions in kinematic regions in which a
triple-Pomeron approach is known to be inappropriate. Their main effects are to 
enhance the low mass components of the dissociation spectra, 
to suppress the
production of very large masses and, in the DD case, to
reduce the probability of the systems $X$ and $Y$ overlapping in rapidity
space~\cite{py8,py6_diff}.

Above the very low mass resonance region, 
dissociation systems are treated in the 
\pysix\ generator using the Lund string 
model \cite{Andersson:1983ia}, with final state hadrons 
distributed in a longitudinal phase space with limited 
transverse momentum.
In \pyeight,
diffractive parton distribution functions from HERA \cite{Aktas:2006hy} are 
used to include diffractive final states which are characteristic
of hard partonic collisions, whilst preserving the 
\xix, \xiy, $s$ and $t$ dependences of the diffractive
cross sections from the \pysix model \cite{Navin:2010kk}.
This approach yields a significantly harder 
final state particle transverse momentum spectrum in SD 
and DD processes in \pyeight compared with \pysix, in better agreement with the present data.
The default \py\ multiple parton interaction model is applied to ND events and, in the case of \pyeight, 
also within the dissociated systems in SD and DD events.

The specific versions 
used to correct the data  are
\py{\footnotesize{6.4.21}} (with the AMBT1 tune performed by 
\linebreak
ATLAS \cite{MB15T}) 
and
\py{\footnotesize{8.145}} (with the 4C tune~\cite{ATL-PHYS-PUB-2011-009}).
Updated versions, \py{\footnotesize{8.150}} 
and \py{\footnotesize{6.4.25}} 
(using the 4C and AMBT2B tunes, 
respectively),
are used for comparisons with 
the corrected data (see \reftab{tab:diff_fraction}).
The 4C tune of \pyeight takes account of the 
measurement of the 
diffractive fraction $f_D$
of the inelastic cross section in \cite{lauren}, whilst keeping the total
cross section fixed,
resulting in a somewhat smaller diffractive cross section than 
in \pysix.

The \pho model uses the two component dual parton 
model \cite{Capella:1992yb} to combine features of Regge phenomenology
with AGK cutting rules \cite{Abramovsky:1973fm} and leading 
order QCD.
Diffractive dissociation is described in a two-channel eikonal model,
combining a triple Regge approach to soft processes 
with lowest order QCD for processes
with parton scattering transverse momenta above $3 \GeV$.
The Pomeron intercept is taken to be $\alphapom = 1.08$ and
for hard diffraction, the diffractive parton densities are taken 
from \cite{Capella:1994uy,Capella:1995pn}. 
Hadronisation follows the Lund string model, as for \py.
The CD process is 
included at the level of $1.7\%$ of the total inelastic cross section.
The specific version used is
\pho{\footnotesize{1.12.1.35}}, 
with fragmentation and hadronisation as in 
\py{\footnotesize{6.1.15}}.

\begin{table}
\centering
   \caption{
    Predicted ND, SD, DD and CD cross sections, 
together with the fractions of the total inelastic
cross section 
$\fnd$, $\fsd$, $\fdd$ and $\fcd$ 
attributed to each process according to the default versions of the MC models
(\py{\footnotesize{8.150}}, 
\py{\footnotesize{6.4.25}} 
and \pho{\footnotesize{1.12.1.35}}), 
used 
for comparisons with the measured cross sections. 
The modified
fractions used in the trigger efficiency and migration 
unfolding procedure, tuned as explained
in the text, are also given. 
   \label{tab:diff_fraction} 
}
\begin{tabular}{lcccccc}
\hline\noalign{\smallskip}
\multicolumn{4}{c}{\bf Cross section at {\boldmath \sqs}} \\ 
\noalign{\smallskip}\hline\noalign{\smallskip}
Process & \py{\scriptsize{6}} & \py{\scriptsize{8}} & {\bf \pho}  \\
\noalign{\smallskip}\hline\noalign{\smallskip}
$\snd$ ($\rm mb$)  &  48.5 & 50.9 & 61.6 \\
$\ssd$ ($\rm mb$)  &  13.7 & 12.4 & 10.7 \\
$\sdd$ ($\rm mb$) & \phantom{0}9.2 & \phantom{0}8.1 &  \phantom{0}3.9 \\
$\scd$ ($\rm mb$) & \phantom{0}0.0 & \phantom{0}0.0 &  \phantom{0}1.3 \\
\noalign{\smallskip}\hline
Default $\fnd$ (\%) & 67.9 & 71.3 & 79.4  \\
Default $\fsd$ (\%) & 19.2 & 17.3 & 13.8  \\
Default $\fdd$ (\%) & 12.9 & 11.4 & \phantom{0}5.1  \\
Default $\fcd$ (\%) & \phantom{0}0.0 & \phantom{0}0.0 & \phantom{0}1.7  \\
\noalign{\smallskip}\hline
Tuned $\fnd$ (\%) & 70.0 & 70.2 & 70.2  \\
Tuned $\fsd$ (\%) & 20.7 & 20.6 & 16.1  \\
Tuned $\fdd$ (\%) & \phantom{0}9.3 & \phantom{0}9.2 & 11.2  \\
Tuned $\fcd$ (\%) & \phantom{0}0.0 & \phantom{0}0.0 & \phantom{0}2.5  \\
\noalign{\smallskip}\hline
   \end{tabular}
\end{table}

After integration over $t$, \xix and \xiy, the 
cross sections for the diffractive processes vary considerably
between the default
MC models, as shown in \reftab{tab:diff_fraction}. 
The DD variation is particularly
large, due to the lack of experimental constraints. 
For use in the data correction procedure, the overall 
fractional 
non-diffractive (\fnd) and
diffractive ($\fd = \fsd + \fdd + \fcd = 1 - \fnd$)
contributions to the total inelastic cross section
are modified to match the results
obtained 
in the context of each model in a previous ATLAS analysis \cite{lauren}.
Despite the close agreement
between the diffractive fractions $\fd \sim 30 \%$ determined for the three 
default models 
(see the `Tuned' fractions in \reftab{tab:diff_fraction}),
the \fd parameter 
is rather sensitive to the choice of Pomeron flux model and to 
the value of $\alphapom$, for example reaching $\fd \sim 25 \%$ 
for the Bruni and Ingelman flux in \pyeight \cite{lauren}.

The default \pho and \py\ models do not take into account Tevatron data
which are relevant to the 
decomposition of the diffractive
cross section into SD, DD and CD components, so 
these fractions are also adjusted for the present analysis. Based on CDF
SD \cite{Abe:1993wu} and DD \cite{Affolder:2001vx} cross section data, extrapolated to the 
full diffractive kinematic ranges in each of the models, 
constraints of $0.29 < \sdd / \ssd < 0.68$ and
$0.44 < \sdd / \ssd < 0.94$ are derived for the \py\ and \pho
models of diffraction, respectively. 
The tuned ratios used in the correction procedure are 
taken at the centres of these bounds. 
The CD contribution in \pho is compatible with the measured Tevatron value of 9.3\% of the SD cross
Section \cite{Acosta:2003xi} and $\scd/\ssd$ is therefore kept fixed, with 
$\fcd$ increasing in proportion to $\fsd$.
\reftab{tab:diff_fraction}\ summarises
the tuned decomposition of the inelastic cross section for each MC model.

Despite the substantial differences between the approaches to diffraction
taken in \pho and \py, the two models both employ the 
Lund string model \cite{Andersson:1983ia} of hadronisation. In order to investigate the
sensitivity of the data at small gap sizes to the hadronisation model for ND
processes, comparisons 
of the measured cross sections are also made
with the \hpp generator \cite{Bahr:2008pv} 
(version 2.5.1 with the UE7-2 tune \cite{Gieseke:2011xy,HERWIG:more}), 
which uses an alternative 
cluster-based model.
The \hpp minimum bias generator takes the total inelastic cross section 
to be $81 \ {\rm mb}$, based on a Donnachie-Landshoff model \cite{Donnachie:2004pi}.
Perturbatively treated semi-hard processes are distinguished from 
soft processes 
according to whether they
produce objects with transverse momentum above a 
fixed threshold which is taken to be 
$3.36 \GeV$.
Partons produced from the 
parton shower are combined into colour singlet pairs called clusters, 
which can be interpreted as excited hadronic resonances. The clusters are then 
successively split into new clusters until they reach the required mass to form
hadrons. The most recent \hpp\ versions contain a mechanism to reconnect partons 
between cluster pairs via a colour reconnection (CR) algorithm, which
improves the modelling of charged 
particle multiplicities in $pp$ collisions \cite{Gieseke:2011na}.
Similarly to \py, \hpp contains an eikonalised underlying event model,
which assumes that separate scatterings in the same event
are independent. At fixed impact 
parameter, this leads to 
Poisson distributions for both the number of soft scatters and the number of semi-hard processes
per event.  
There is thus a small probability for `empty' events to occur with no scatterings of either type.
Under these circumstances, particle production occurs only in association with the 
dissociation of the beam protons, in a manner which is reminiscent of diffractive
dissociation processes.

\subsection{Comparisons between Monte Carlo simulations and uncorrected data}

For use in the correction procedure, MC events
are processed through the ATLAS detector simulation 
program \cite{:2010wqa}, 
which is based on \geant\ \cite{Agostinelli:2002hh}. They are then
subjected to the same reconstruction and analysis chain as is used
for the data.

\begin{figure*}
\centering
\subfloat[]{\includegraphics[width=0.475\textwidth]{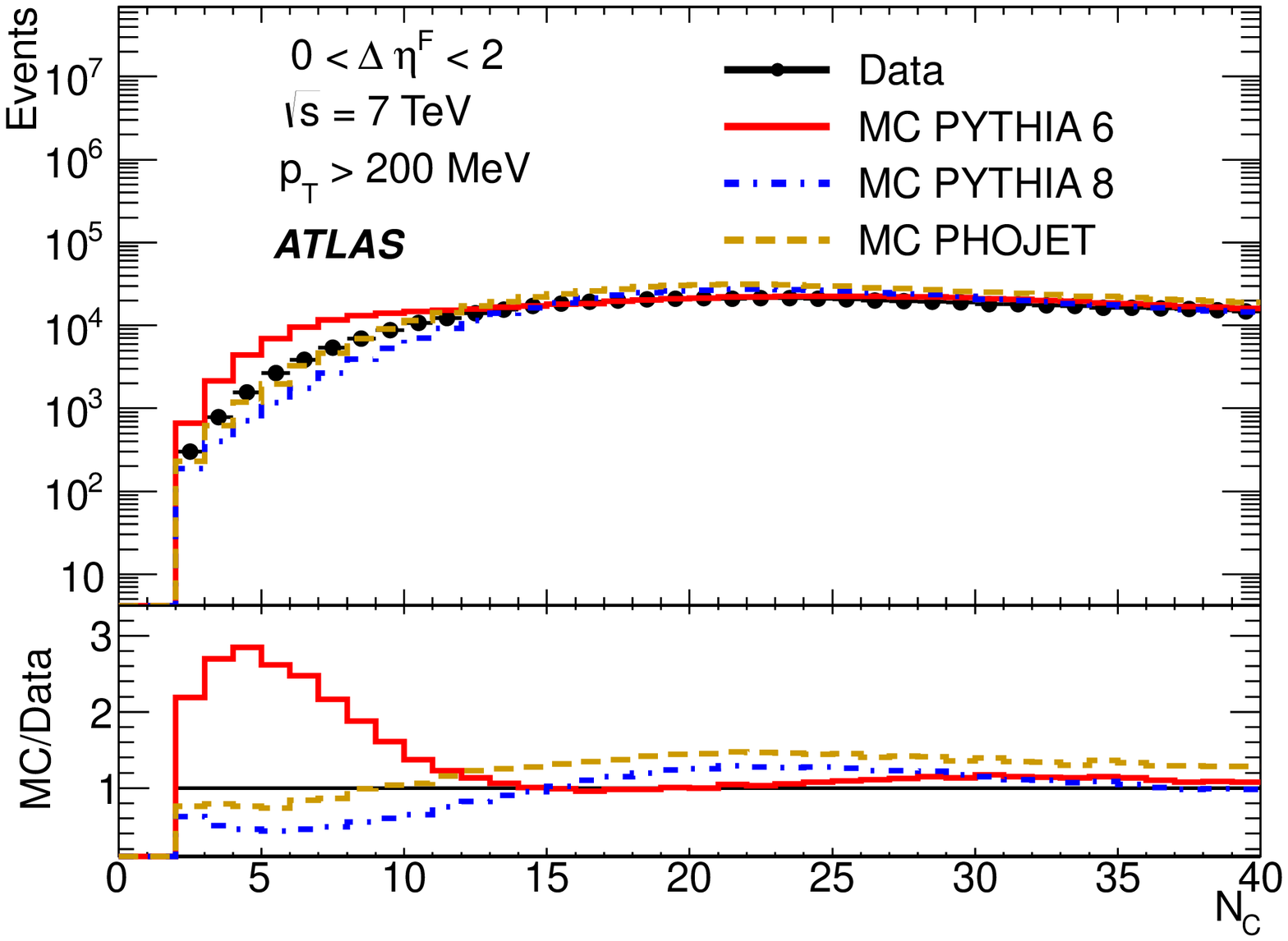}}
\subfloat[]{\includegraphics[width=0.475\textwidth]{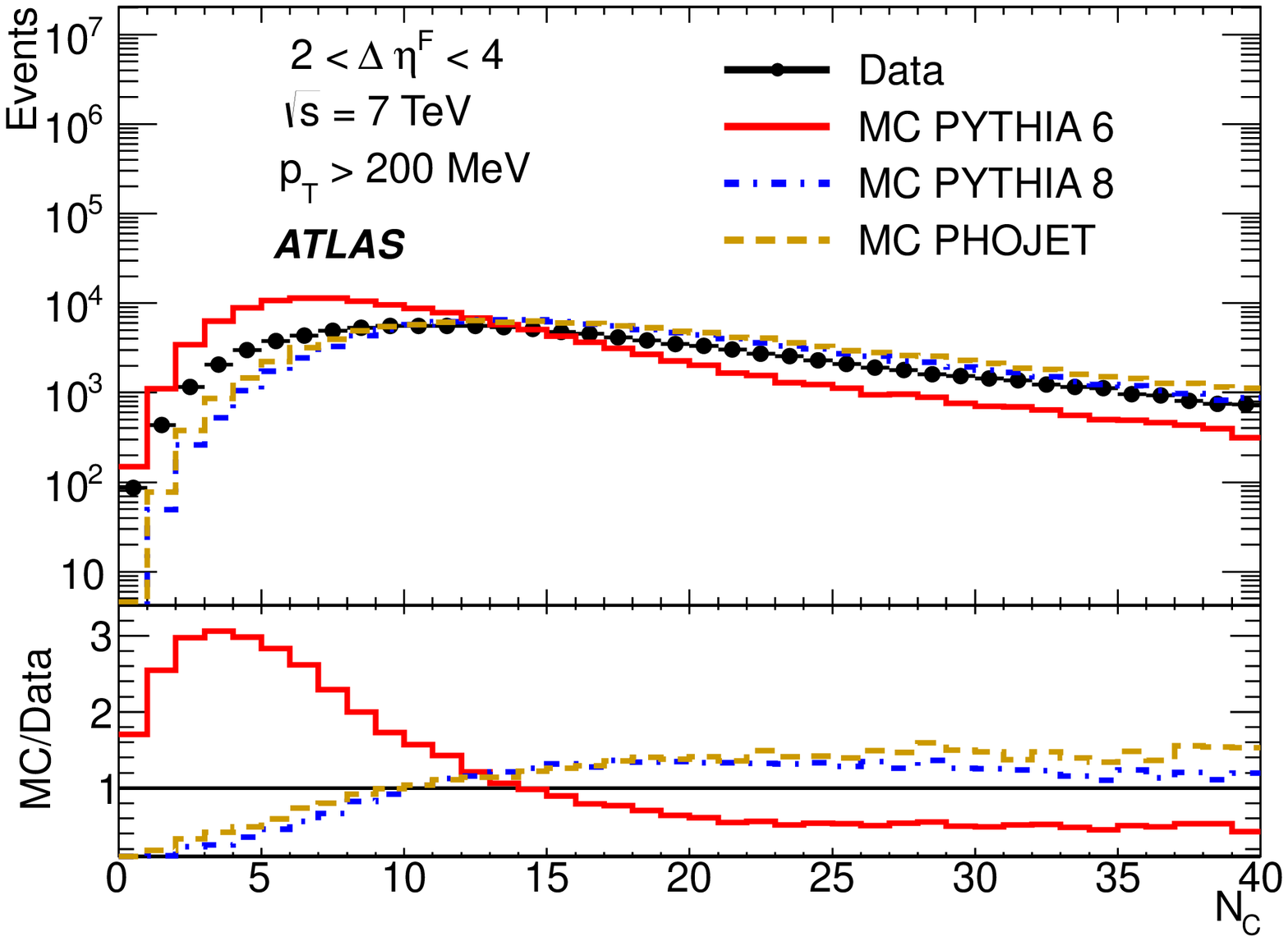}}
\newline
\subfloat[]{\includegraphics[width=0.475\textwidth]{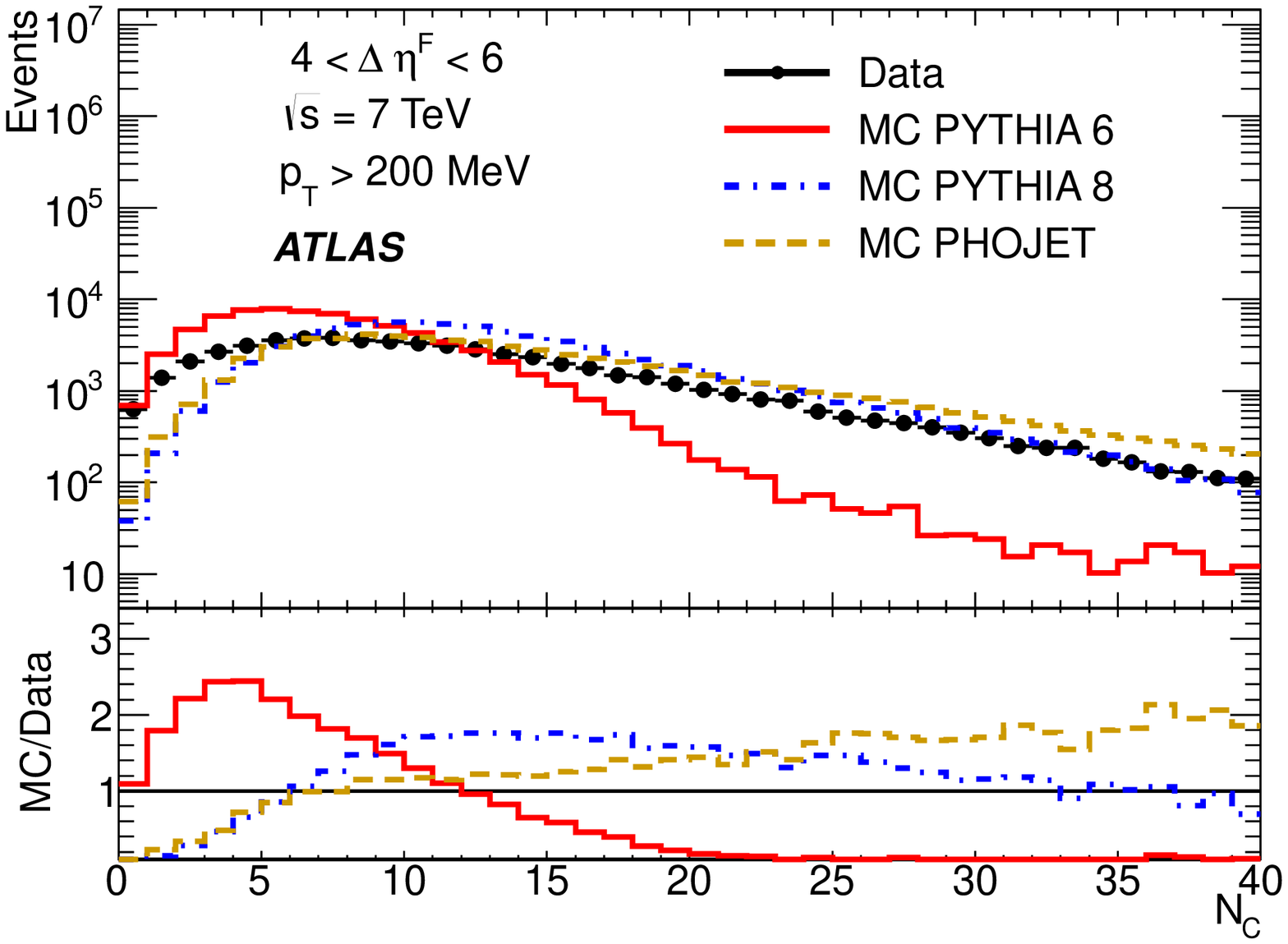}}
\subfloat[]{\includegraphics[width=0.475\textwidth]{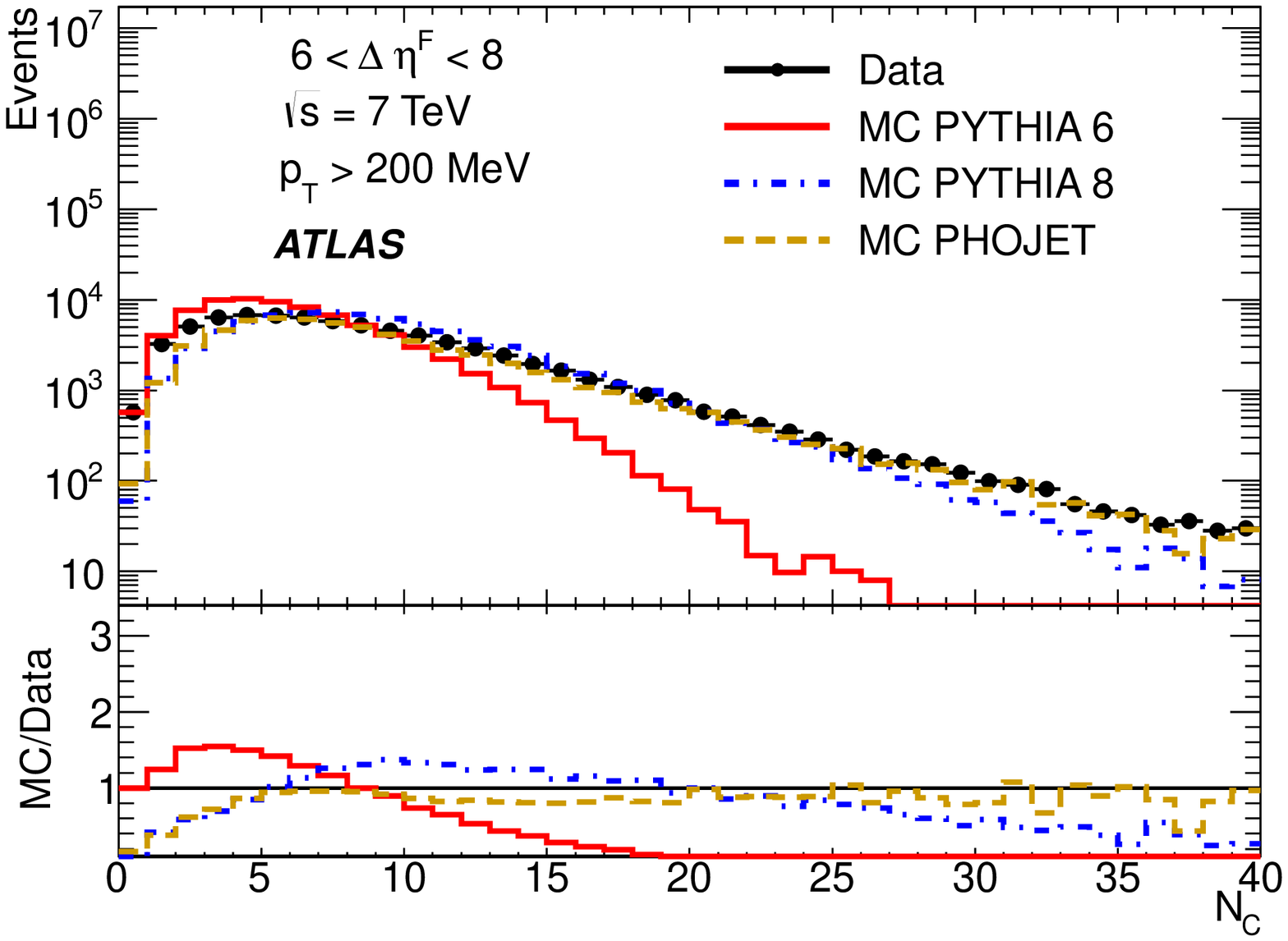}}
\newline
\subfloat[]{\includegraphics[width=0.475\textwidth]{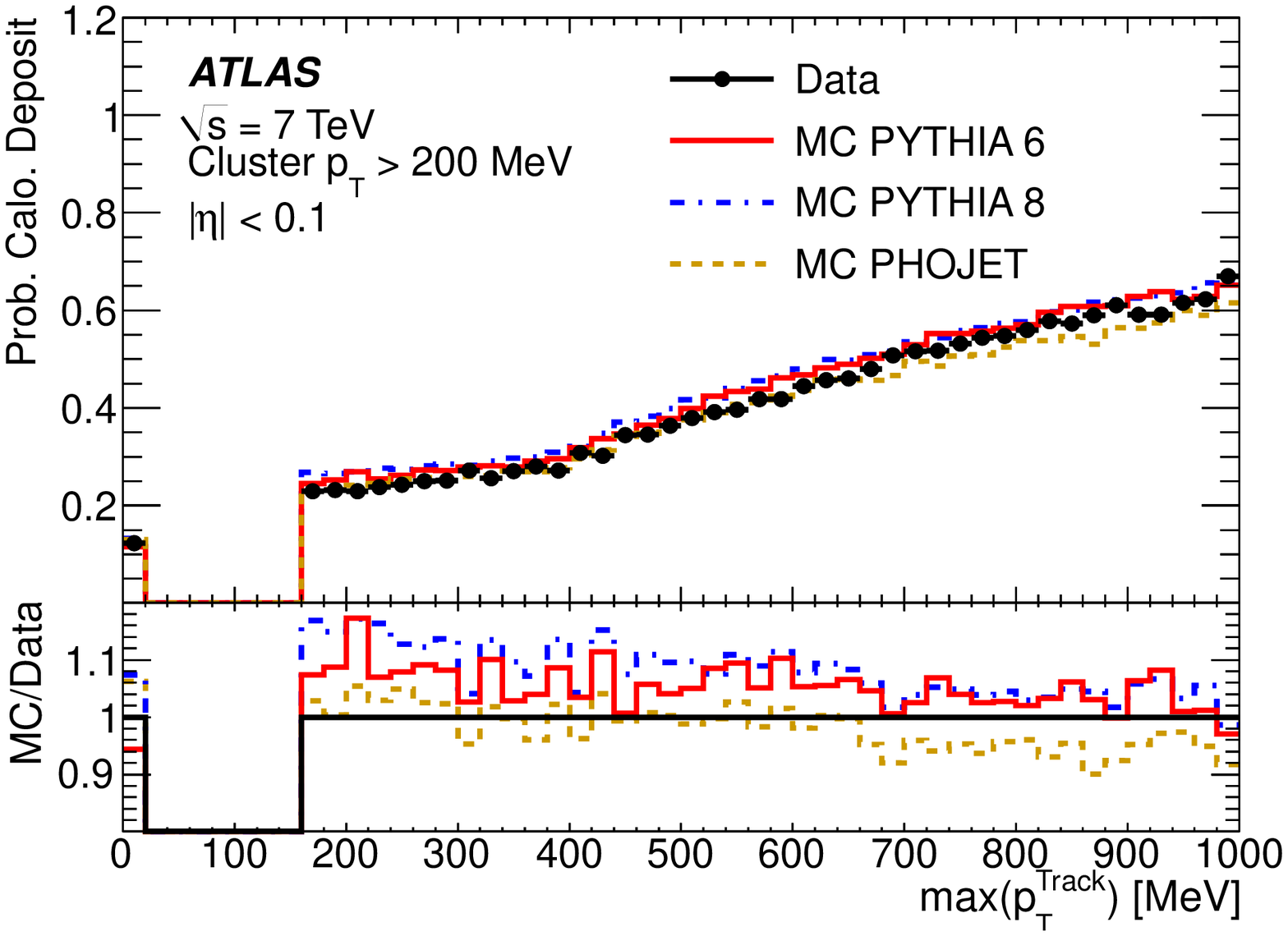}}
\subfloat[]{\includegraphics[width=0.475\textwidth]{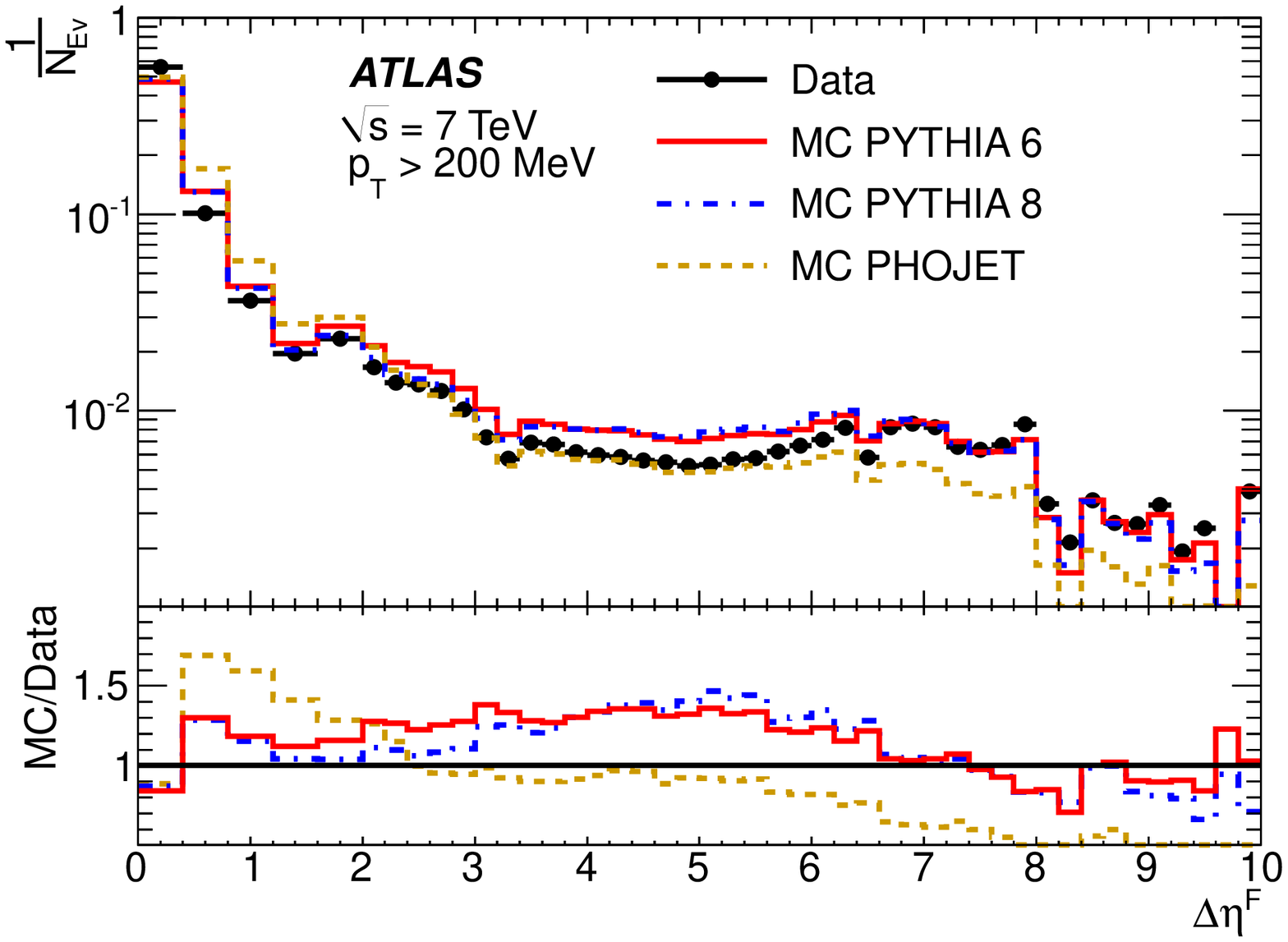}}
\caption{Comparisons of uncorrected distributions
between data and MC models. 
(a)-(d) Total calorimeter cluster multiplicities $N_C$ for events
reconstructed with (a) $0 < \detaf < 2$, (b) $2 < \detaf < 4$,
(c) $4 < \detaf < 6$ and (d) $6 < \detaf < 8$.
(e) Probability 
of detecting significant calorimeter energy in the most central region
$|\eta| < 0.1$ as a function of
the highest transverse momentum ${\rm max} (\pttrack)$
of the tracks reconstructed in the inner detector
in the same $|\eta|$ range. The bin at zero corresponds
to events where no charged track with $\pt > 160 \MeV$
is reconstructed. 
(f) Forward rapidity gap distribution for $\ptcut = 200 \MeV$.
The final bin at 
\detaf = 10 corresponds to cases where no reconstructed particles have 
$\pt > \ptcut$.}
\label{Fig:gapsDataMC}
\end{figure*}

The quality of the MC description of the 
most important
distributions 
for the correction procedure is tested through a set of control plots
which compare the uncorrected data and MC distributions. These include
energy flows, track and calorimeter cluster multiplicities and
transverse momentum distributions, as well as  
leading cell energy significances in different pseudorapidity regions.
All such distributions are reasonably well described. Examples
are shown in \reffigs{Fig:gapsDataMC}a-d, where the 
total multiplicities
of calorimeter clusters which pass the selection described in 
\refsec{sec:gap-reco}\ 
are shown for events in four different regions of reconstructed
forward
rapidity gap size. Whilst none of the MC models gives a
perfect description, particularly at small multiplicities, the three
models tend to bracket the data, with \pysix showing an excess
at low multiplicities and \pyeight and \pho showing a deficiency in the
same region. 

A further
example control distribution is shown in \reffig{Fig:gapsDataMC}e. The
probability of detecting at least one calorimeter cluster
passing the noise requirements 
with $\pt > \ptcut = 200 \MeV$ in 
the most central region ($|\eta| < 0.1$) 
is shown as a function of the \pt\ of the leading track
reconstructed in the same $\eta$ region. 
In cases where this track has \pt\
below around $400$ \MeV, it spirals in the
solenoidal field outside the acceptance of the EM calorimeter.
The plotted quantity then corresponds to the detection probability for neutral
particles in the vicinity of a track.
Good agreement is observed between MC and data.

The shape of the
uncorrected $\detaf$ distribution for $\ptcut = 200 \MeV$
is compared between the data
and the MC models in \reffig{Fig:gapsDataMC}f.
The binning reflects that used in the
final result (\refsec{sec:sigdef}) except that contributions with
$\detaf > 8$, where the trigger efficiency becomes small, are
also shown.  
None of the models considered are able to describe the data
over the full $\detaf$ range, with the largest deviations observed for small
non-zero gaps in \pho. 
All of the models give an acceptable description of the shape
of the distribution for large gaps up to the limit of the measurement at
$\detaf = 8$ and beyond.

Considering all control plots together,  
\pyeight provides the best description of the shapes of the distributions.
Hence this generator is chosen to correct the data.
The deviations from \pyeight of 
\pysix and \pho, which often lie in
opposite directions and tend to enclose the data, 
are used to evaluate the systematic uncertainties on the unfolding procedure. 

\subsection{Corrections for Experimental Effects}
\label{sec:Unfolding}

After the statistical subtraction of the beam-induced background 
in each interval of $\detaf$ (\refsec{sec:evt_sel}), the data are
corrected for the influence of the limited acceptance and 
small particle detection inefficiencies of the MBTS using the
MC simulation. For the ND, SD and DD processes, the trigger efficiency
is close to $100 \%$ for $\detaf < 7$, dropping to around 
$80 \%$ at $\detaf = 8$.
Since the topology of CD events sometimes involves hadronic activity in
the central region of the detector, with gaps on either side, a larger fraction 
fail the trigger requirement, with efficiencies of close to $100 \%$ for $\detaf < 3$ 
and between $85\%$ and $95 \%$ for
$3 < \detaf < 8$. 

\begin{figure}
  \includegraphics[width=0.5\textwidth]{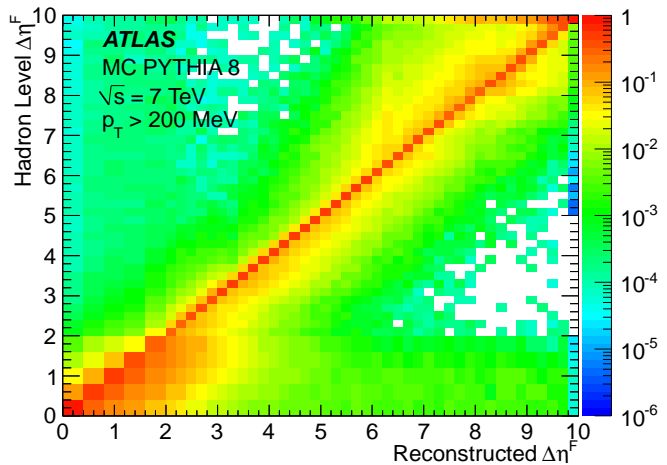}    
  \caption{Migration matrix between the reconstructed and hadron level
values of $\detaf$
for $\ptcut = 200 \MeV$, according to \pyeight. 
The distribution is normalised to unity in columns and is shown to
beyond the limit of the measurement at $\detaf = 8$.}
\label{Fig:Migration}
\end{figure}

The data are corrected for migrations between the reconstructed
and hadron level
\detaf values,  
due to 
\linebreak
missed or spurious activity
and cases where a final state particle is observed in a 
different $\eta$ interval from that in which it is produced. 
The migration corrections are obtained using
a Bayesian unfolding method \cite{D'Agostini:1994zf} 
with a single iteration. 
The priors for the unfolding procedure with each MC model 
are taken after
tuning the diffractive cross sections 
as described in \refsec{sec:mctune}.
The migration matrix between the 
reconstructed and hadron level forward gap distributions
according to the \pyeight MC is shown for $\ptcut = 200 \MeV$
in \reffig{Fig:Migration}. An approximately diagonal
matrix is obtained. 

\section{Systematic Uncertainties} \label{sec:systematics}

The sources of systematic uncertainty on the measurement are outlined
below. 

\paragraph{MC Model and Unfolding Method Dependence:} 
The trigger efficiency and migration correction procedure is carried out using 
each of the \pysix, \pyeight and \pho models. 
The deviation of the data unfolded with \pho from 
those obtained with \pyeight
is used to obtain a systematic
uncertainty due to the assumed $\xi_X$, $\xi_Y$ and $t$ dependences in the unfolding
procedure. The model dependence due to the details of the final state
particle production is obtained 
from the difference between the 
results obtained with \pysix and \pyeight. 
Both of these model dependences
are evaluated separately in each measurement interval and 
are applied symmetrically as upward and downward uncertainties. 
They
produce the largest 
uncertainty on the measurement over most of the measured range. 
For $\ptcut = 200 \MeV$, 
the contributions from the \pysix
and \pho variations are of similar size.
Their combined effect is typically at the 
$6 \%$ level for large $\detaf$, growing to $20 \%$ for gaps 
of around 1.5 pseudorapidity units.
At larger \ptcut values, the \pysix source becomes dominant.
The dependence on the unfolding technique has also been studied
by switching between the default
Bayesian method \cite{D'Agostini:1994zf}, 
a method using a Singular Value Decomposition of the unfolding 
matrix \cite{Hocker:1995kb} and a 
simple bin-to-bin method. The resulting variations in the measured
cross section are 
always within the systematic uncertainty defined by varying the MC
model. 

\paragraph{Modelling of Diffractive Contributions:}
\label{difffracsec}
In addition to the differences between the Monte Carlo generators, 
additional systematic uncertainties are applied on the modelling
of the fractional diffractive cross sections. The SD and DD cross sections in the 
\pyeight model are each 
varied to the limits of the constraints from Tevatron data described 
in \refsec{sec:mctune}. The fraction
\fdd is enhanced to 11.3\% of 
the total inelastic cross section, with \fsd reduced 
to 18.5\% to compensate.
At the opposite extreme, \fsd is enhanced to 23.2\% of the cross section, 
with \fdd reduced to 6.6\%.
These changes result in an uncertainty at the $1 \%$ level for
$\detaf > 3$. 
A systematic uncertainty on the CD cross section is obtained by 
varying the CD and SD cross sections in 
\pho between the tuned values and $\scd / \ssd = 0.093$, 
corresponding to
the CDF measurement in \cite{Acosta:2003xi}. 
This variation also results in a $1 \%$ uncertainty in the large gap
region.

\paragraph{Calorimeter Energy Scale:}
The uncertainty on the calo\-rimeter energy scale is constrained 
to be below the $5 \%$ level down to energies of a few hundred 
\MeV\ in
the central region, $|\eta| < 2.3$, through comparisons between 
isolated calorimeter cluster energy measurements and momentum determinations
of matched tracks in the inner 
detector \cite{ATLAS-CONF-2010-017,ATLAS-CONF-2010-052,ATLAS-CONF-2011-028}.
This method is not available for larger $|\eta|$ values beyond the tracking acceptance.
However,
as $|\eta|$ grows, the default $\ptcut = 200 \MeV$ threshold corresponds to 
increasingly large energies, reaching beyond $10 \GeV$ 
at the 
outer limits of the FCal. 
The uncertainty on the response to electromagnetic showers in
this energy range is
determined as a function of $|\eta|$ from the maximum observed
deviations between the data and the MC simulation in the peaks 
of $\pizero \rightarrow \gamma \gamma$ signals, 
under 
a variety of assumptions on background shapes and cluster
energy resolutions.  
The relative response to charged pions compared with the 
electromagnetic scale has been studied in the relevant energy range for the 
FCal \cite{Archambault:2008zz,Kiryunic} and 
HEC \cite{Pinfold:2008zzb,Kiryunic} 
test-beam data, with systematic uncertainties of
$8\%$ and $4\%$, respectively, determined from the difference between data and MC.
Adding the uncertainties in the electromagnetic scale and in the relative response to 
hadrons in quadrature, energy scale uncertainties
of $5 \%$ for $|\eta| < 1.37$, 
$21 \%$ for $1.37 < |\eta| < 1.52$ (transition region between barrel and end-cap),
$5 \%$ for $1.52 < |\eta| < 2.3$,
$13 \%$ for $2.3 < |\eta| < 3.2$ and 
$12 \%$ for $3.2 < |\eta| < 4.9$ are ascribed. 
In addition to the absolute calorimeter response,
these values account for systematic effects arising from dead material uncertainties and
from the final state decomposition into different particle species.
In the unfolding procedure, the 
corresponding systematic variation is applied to energy depositions from 
simulated 
final state particles in MC, 
with noise contributions left unchanged. The clustering algorithm is 
then re-run over the modified calorimeter cells.
The scale uncertainty variation is thus considered both in the 
application of the \ptcut threshold to the clusters and in the 
discrimination of
cells within selected clusters against the significance cut used to veto noise.
The resulting fractional uncertainties on the differential cross 
sections at the default $\ptcut = 200 \MeV$ are 
largest (reaching $\sim$ 12\%) in the region $\detaf \lappr 3$, 
where the gap identification relies most strongly on the
calorimeter information. For larger gaps, the well measured 
tracks play an increasingly important role in defining the 
gap size and the cross section is dominated by low \xix 
diffractive events for which particle production in the gap region
is completely suppressed. 
The sensitivity to the calorimeter 
scale is correspondingly reduced to a few percent. 

\paragraph{MBTS Efficiency:}
The description of the MBTS efficiency in the MC models leads to 
a potential systematic effect on the trigger efficiency and on the off-line MBTS
requirement. Following \cite{lauren}, the
associated uncertainty is evaluated by increasing the 
thresholds of all MBTS counters in the simulation to match the 
maximum variation in the measured response in data according to studies with
particles extrapolated from the tracker or FCal.
This systematic error amounts to typically 
$0.5 - 1\%$ for
$\detaf > 2$ and is negligible at the smallest $\detaf$.

\paragraph{Tracking Efficiency:}
The dominant uncertainty in the charged particle track reconstruction 
efficiency arises due to possible inadequacies in the modelling of the
material through which the charged particles pass \cite{:2010ir}. 
This uncertainty is quantified by studying the influence on the 
data correction procedure of using an MC sample produced with a 
10\% enhancement in the support material in the inner detector. 
The resulting uncertainty is smaller than 3.5\% throughout the 
measured distribution.

\paragraph{Luminosity:}
Following the van der Meer scan results 
in \cite{LuminosityWorkingGroup:1330358},
the normalisation uncertainty due to the uncertainty on the integrated luminosity is $3.4 \%$. \\ 

Each of the systematic uncertainties is determined with correlations 
between bins
taken into account in the unfolding by repeating the full analysis using data or MC 
distributions after application of the relevant systematic shift. 
The final systematic error on the differential cross section is 
taken to be the sum in quadrature of all sources. 
Compared with the systematic uncertainties, the statistical errors are 
negligible at the smallest gap sizes, where the differential cross section is largest. For gap
sizes $\detaf \gappr 3$, 
the statistical errors are at the $1\%$ level and are typically
smaller than the systematic errors by factors between five and ten.

\section{Results}
\label{sec:Gap-size-dist}

\subsection{Differential cross section for forward rapidity gaps}

\begin{figure*}
\centering
\subfloat[]{\includegraphics[width=0.5\textwidth]{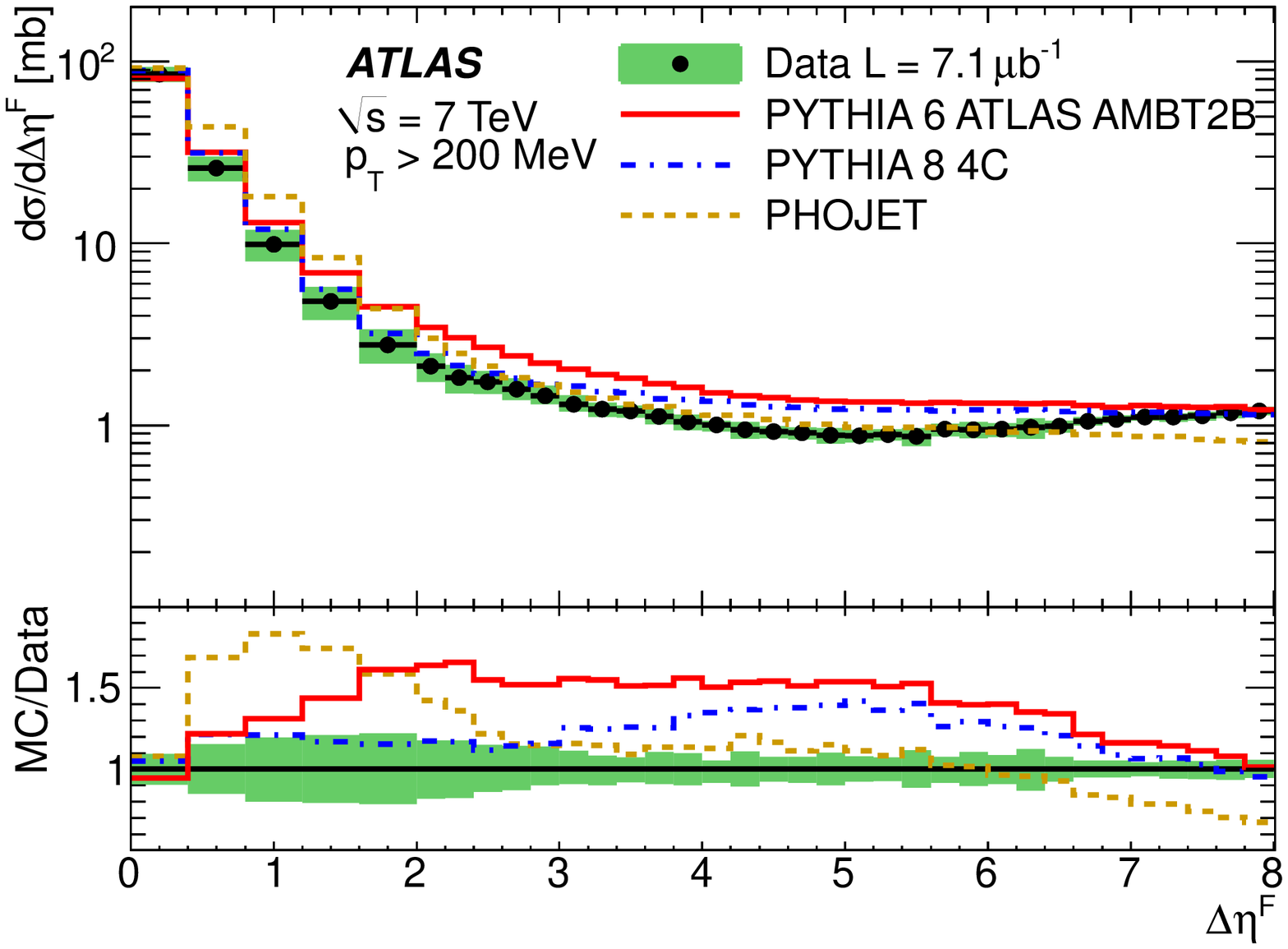}}
\subfloat[]{\includegraphics[width=0.5\textwidth]{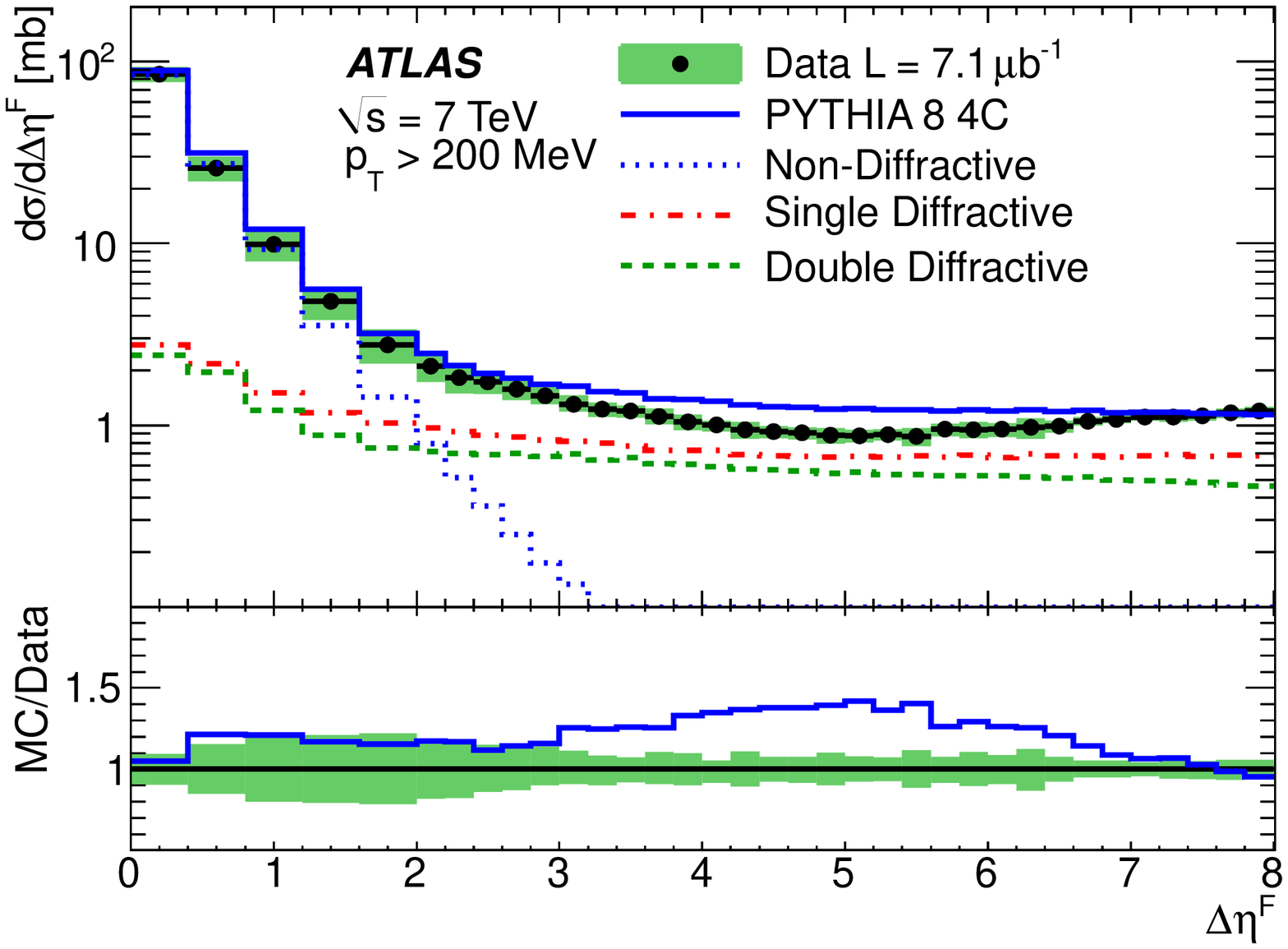}}
\newline
\subfloat[]{\includegraphics[width=0.5\textwidth]{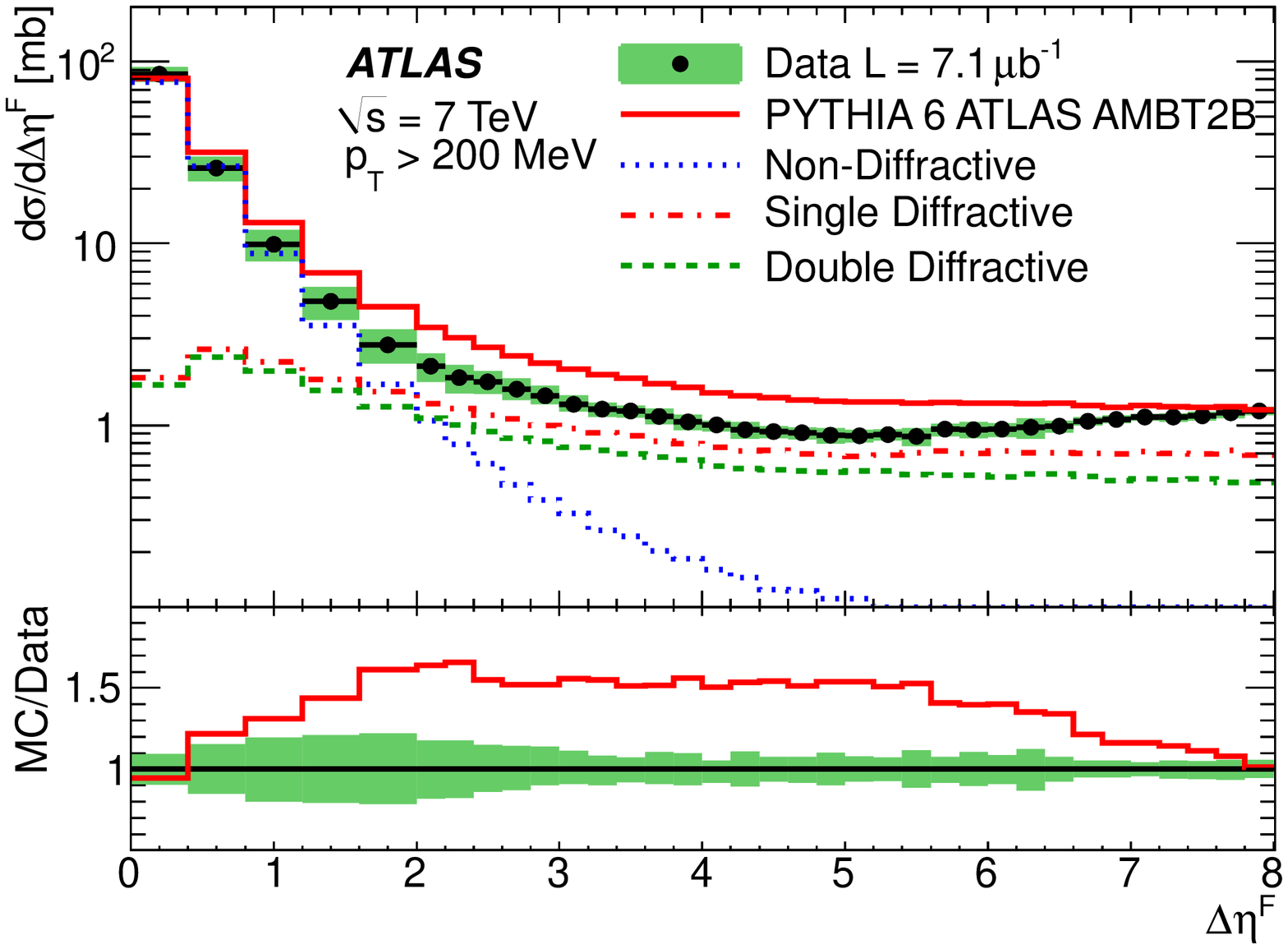}}
\subfloat[]{\includegraphics[width=0.5\textwidth]{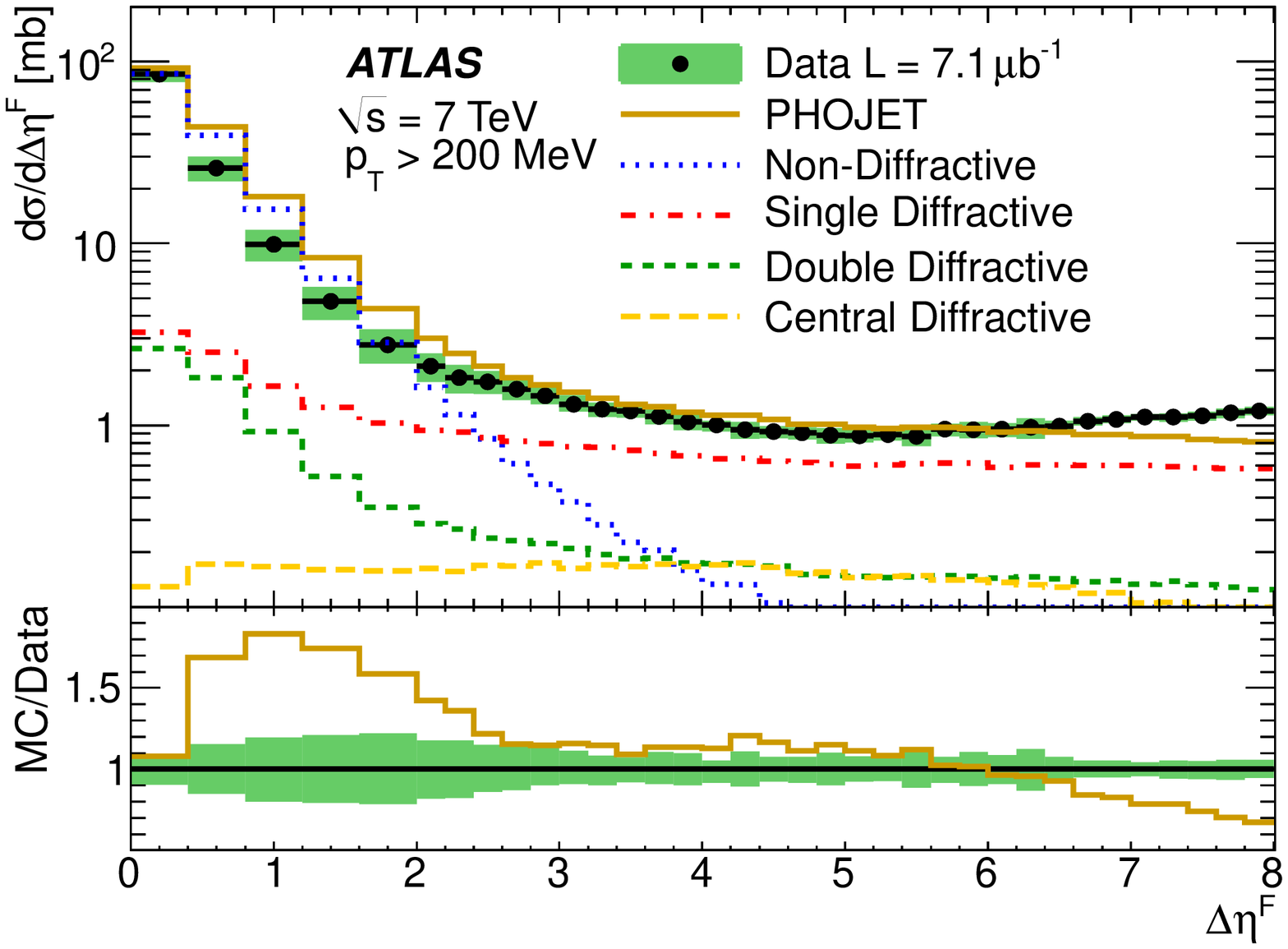}}
\caption{Inelastic cross section differential in forward gap size 
$\detaf$ for particles with 
$\pt > 200 \MeV$. The shaded bands represent 
the total uncertainties. The full lines show the
predictions of \pho
and the default versions of \pysix and \pyeight\.  
The dashed lines in (b-d) represent the
contributions of the ND, SD and DD components according to the models. The
CD contribution according to \pho is also shown in (d).}
\label{Fig:forwardgap_det}
\end{figure*}

In this section, 
measurements are presented 
of the inelastic cross section 
differential in forward 
rapidity gap size, $\detaf$, as defined in \refsec{sec:sigdef}.
The data cover the range $0 < \detaf < 8$.
In the large gap region which is populated by diffractive processes,
the cross section corresponds
to a $t$-integrated sum of SD events in which 
either of the colliding protons dissociates and DD events with
$\xiy \lappr 10^{-6}$ ($\MY \lappr 7 \GeV$). The data span the range
$\xix \gappr 10^{-5}$. Diffractive events with 
smaller \xix values are subject to large MBTS trigger inefficiencies and thus lie
beyond the kinematic range of the measurement.

As discussed in \refsec{sec:sigdef}, the lowest transverse
momentum requirement for the gap definition which is directly 
accessible experimentally is $\ptcut = 200 \MeV$.
\reffig{Fig:forwardgap_det}a shows the 
differential gap cross section for this choice of \ptcut, which is
also given numerically in \reftab{Tab:DPs}.
The uncertainty on the measurement is typically less than $8 \%$ for
$\detaf > 3$, growing to around 
$20 \%$ at $\detaf = 1.5$ before improving to around $10\%$
for events with little or no forward gap. 
The data are compared with the 
predictions of the default settings of 
the \pysix (labelled 
`{\sc pythia{\footnotesize 6} atlas ambt{\footnotesize 2}b}')
\pyeight (`{\sc pythia{\footnotesize 8 4}c}') and
\pho models. 
In \reffigs{Fig:forwardgap_det}b-d, the results are
compared with each of the MC
models separately, with the default decomposition 
of the cross section into ND, SD, DD and CD contributions according to the 
models (\reftab{tab:diff_fraction}) also indicated. 

\subsection{Small gap sizes and constraints on hadronisation models}

\begin{figure*}
 \centering
\subfloat[]{\includegraphics[width=0.5\textwidth]{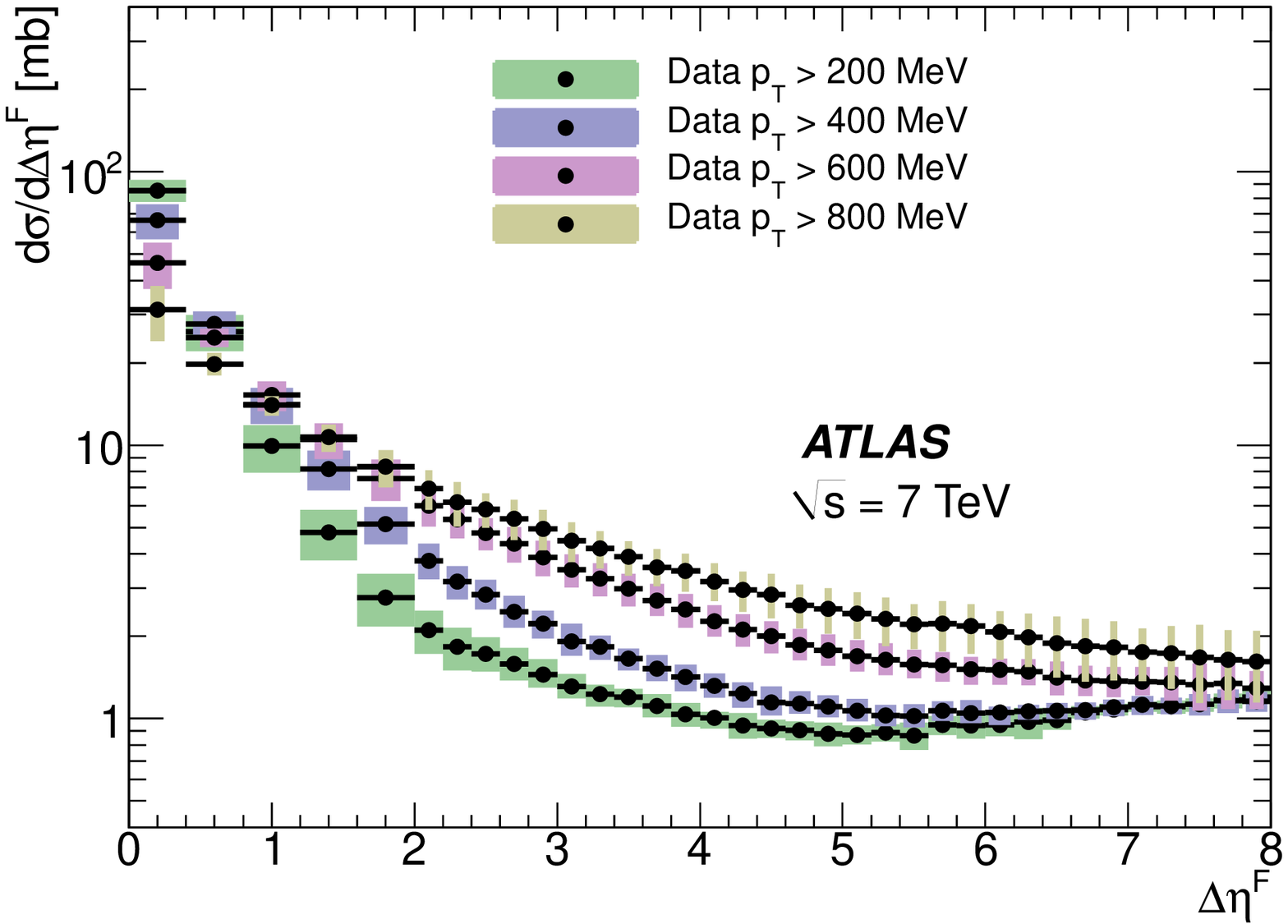}}
\subfloat[]{\includegraphics[width=0.5\textwidth]{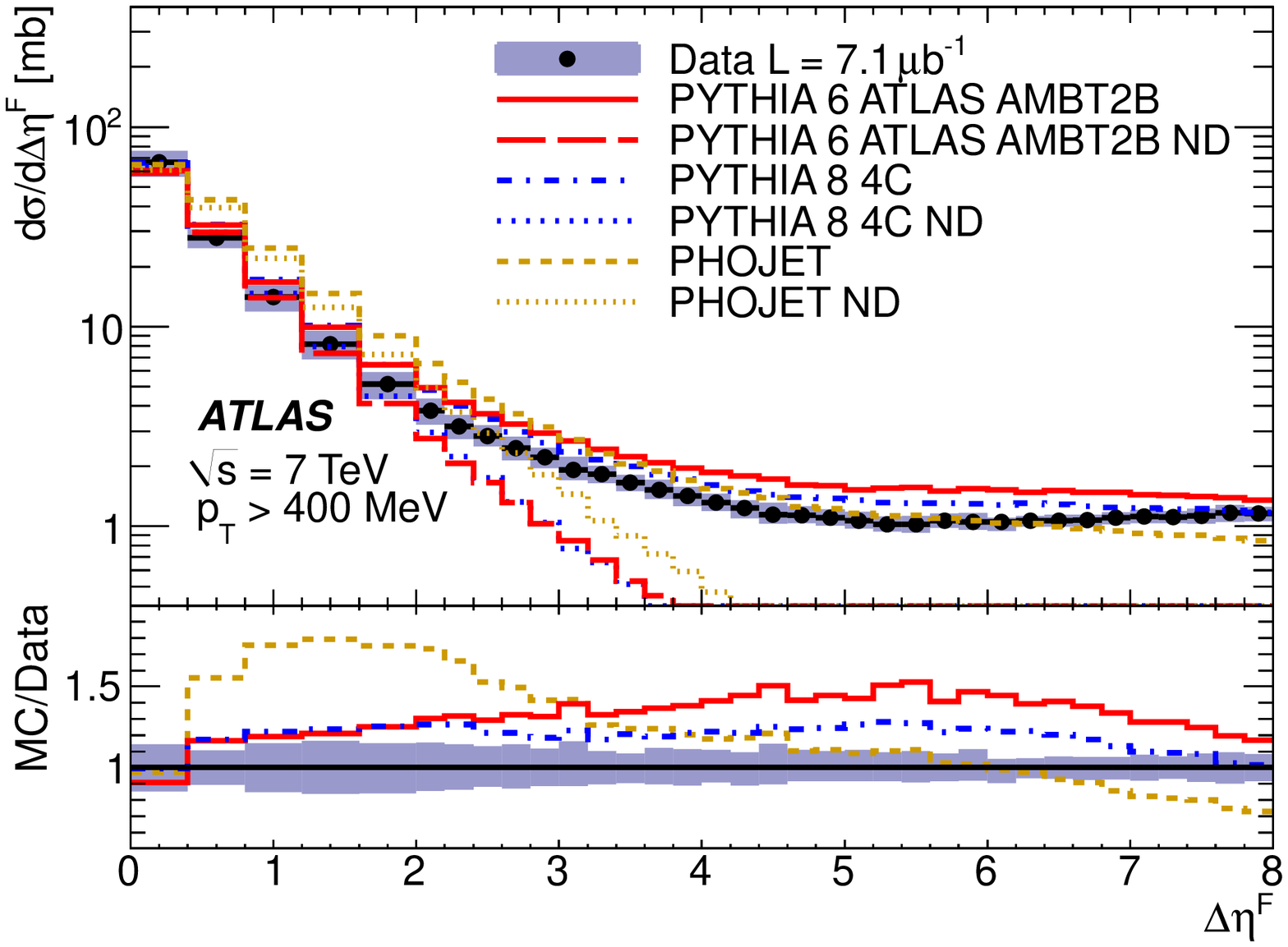}}
\newline
\subfloat[]{\includegraphics[width=0.5\textwidth]{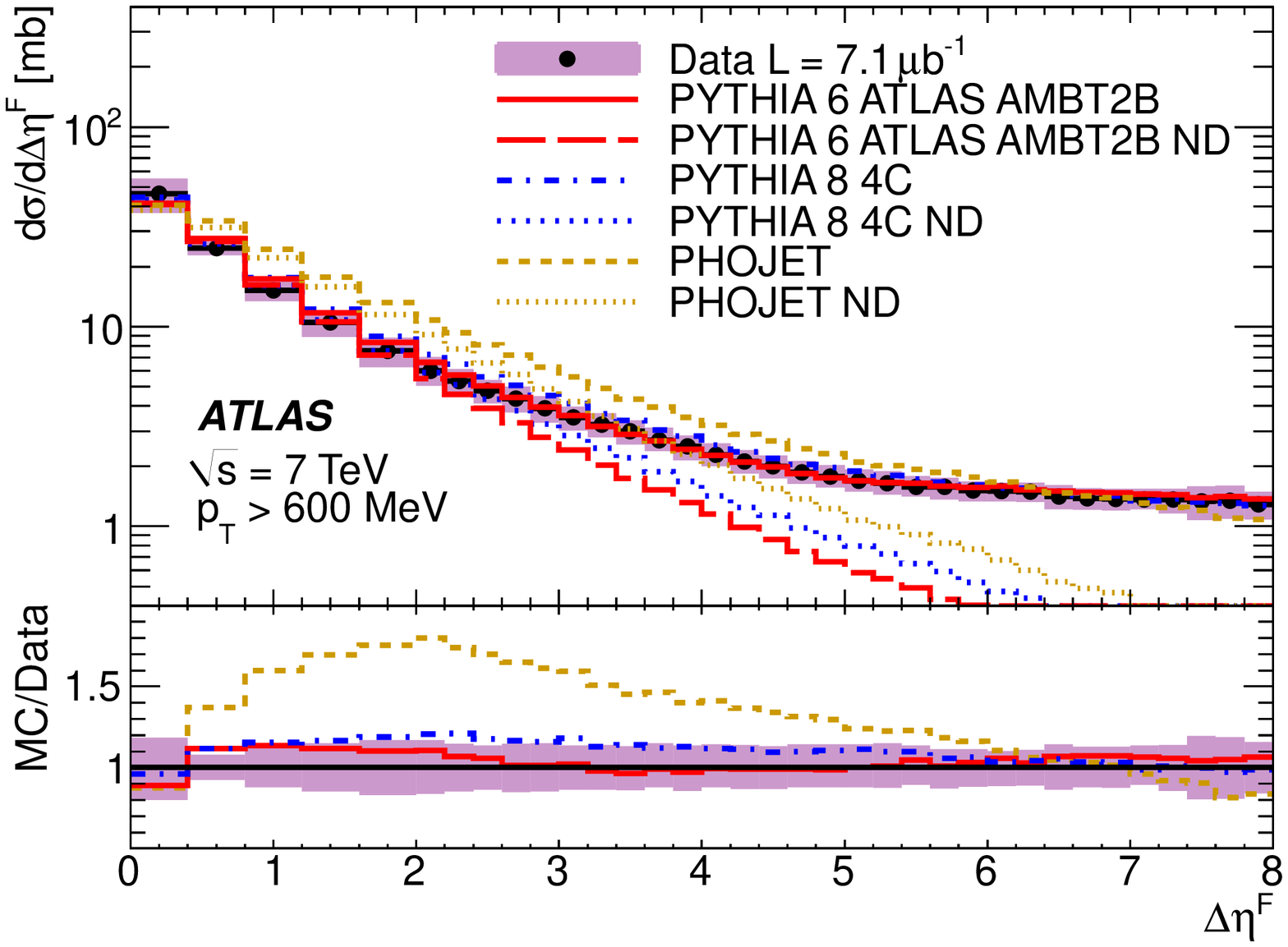}}
\subfloat[]{\includegraphics[width=0.5\textwidth]{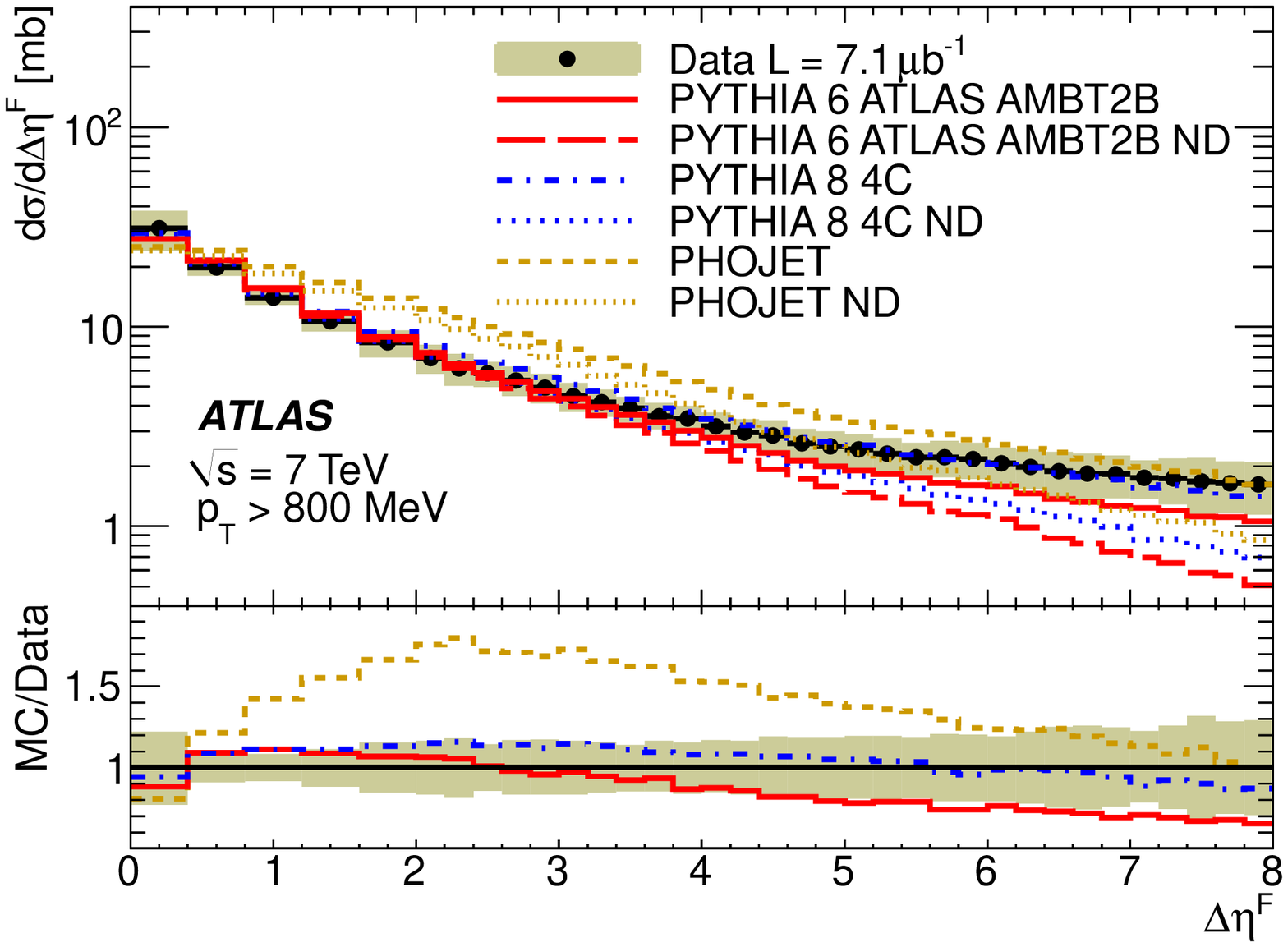}}
 \caption{Inelastic cross section differential in forward gap size 
\detaf for different \ptcut values. 
(a) Comparison between the measured cross sections. The full uncertainties are
shown. They are correlated between the different \ptcut choices.
(b-d) Comparison between the data and the MC models for 
$\ptcut = 400, 600$ and $800 \MeV$.
The non-diffractive component in each MC model is also shown.}
 \label{Fig:ptcut}
\end{figure*}

At $\detaf \lappr 2$, 
all models agree that the ND process is dominant and
the expected \cite{Bjorken:1992mj} exponential decrease of the cross section with increasing
gap size, characteristic of hadronisation fluctuations, 
is the dominant feature of the data. According to the models, 
this region also contains DD events which have $\xiy \gappr 10^{-6}$,
such that the $Y$ system extends into the ATLAS detector acceptance,
as well as both SD and DD events
with very large \xix, such that no 
large rapidity gap is 
present within the region $|\eta| < 4.9$.
The default MC models tend to lie above the data in this region, a result which is 
consistent with the overestimates of the 
total inelastic cross section observed for the 
same models in \cite{lauren}. 
The \pyeight model is closest in shape to the data, which is partly due to the 
modification of \fd in the 
most recent versions made in light of the previous ATLAS data \cite{lauren}. 
Both \py\ models are closer to the 
small $\detaf$ data than \pho, 
which exhibits an excess of almost a factor of two for $\detaf \sim 1$. 

As can be inferred from comparisons between the predicted
shapes of the ND contributions in the different MC models 
(\reffigs{Fig:forwardgap_det}b-d),
there are considerable uncertainties in the probability of obtaining
large hadronisation fluctuations 
among low transverse momentum
final state
particles \cite{springerlink:10.1140/epjc/s10052-010-1392-5}. 
Studying the dependence of the measured differential cross section
on \ptcut provides a detailed probe of 
fluctuations in the 
hadronisation process in soft 
scattering and of hadronisation models in 
general. 
The measurement is thus repeated with different choices of
\ptcut, applied both in the rapidity gap reconstruction 
and in the definition of the measured hadron level cross section. 
To avoid cases where the largest gap switches from one side of the
detector to the other when low \pt\ particles are excluded by
the increased \ptcut choice, 
the side of the detector on which the gap is located 
is fixed to that determined at
$\ptcut = 200 \MeV$ for all measured cross sections. 

A comparison between the results with 
$\ptcut = 200$, $400$, $600$ and $800 \MeV$ is shown in
\reffig{Fig:ptcut}a. 
\reffigs{Fig:ptcut}b-d show the results for 
$\ptcut = 400, 600$ and $800 \MeV$, respectively, 
compared with the \pyeight, \pysix and 
\pho MC models. The ND contributions according to each of the models are
also shown. 
As \ptcut increases, the
exponential fall becomes less steep, so 
larger \detaf values become more heavily populated
and the non-diffractive and diffractive contributions in the
models become similar. Also, the uncertainties
due to the MC model dependence of the unfolding procedure grow. 

\begin{figure*}
 \centering
\subfloat[]{\includegraphics[width=0.5\textwidth]{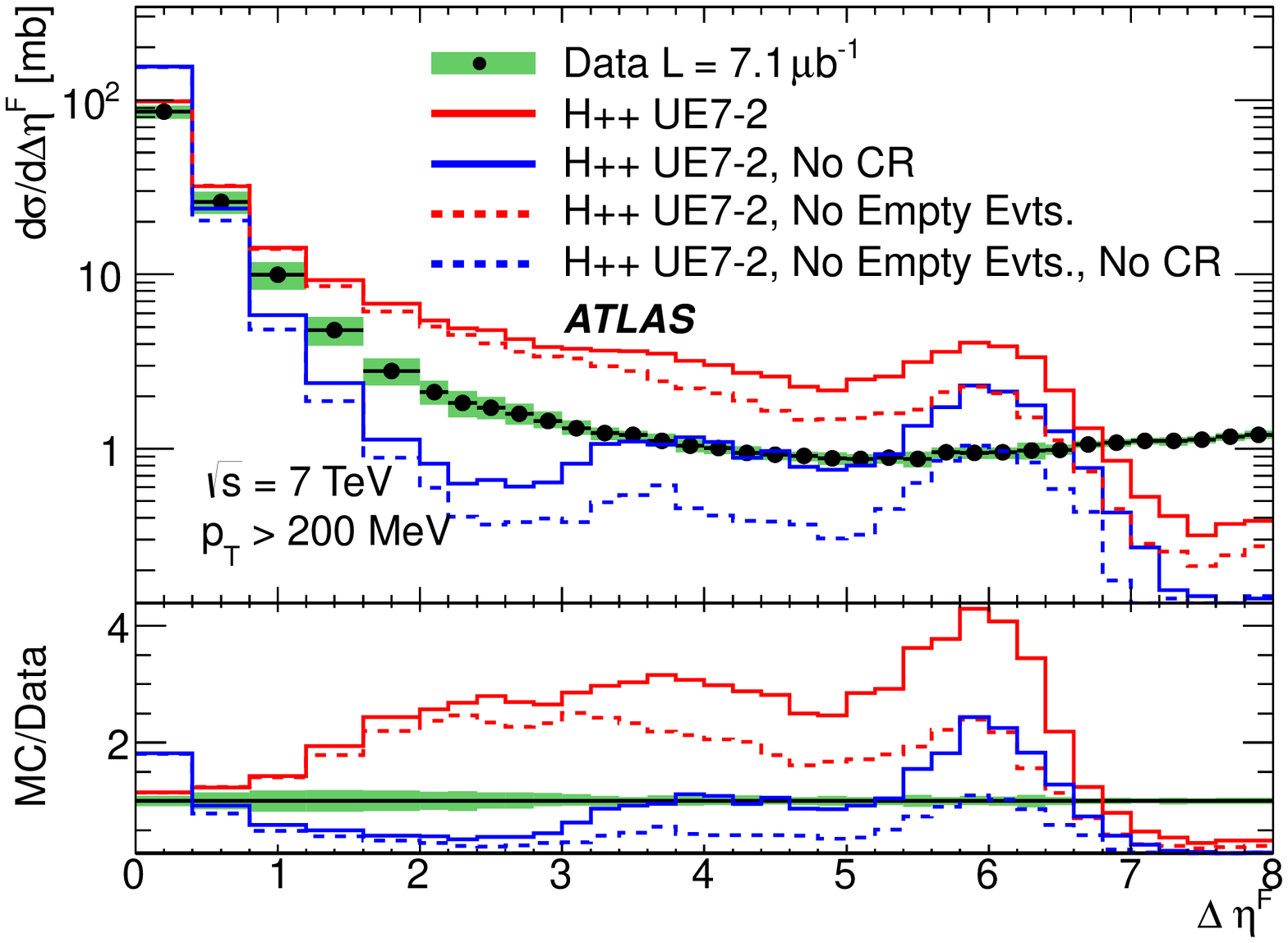}}
\subfloat[]{\includegraphics[width=0.5\textwidth]{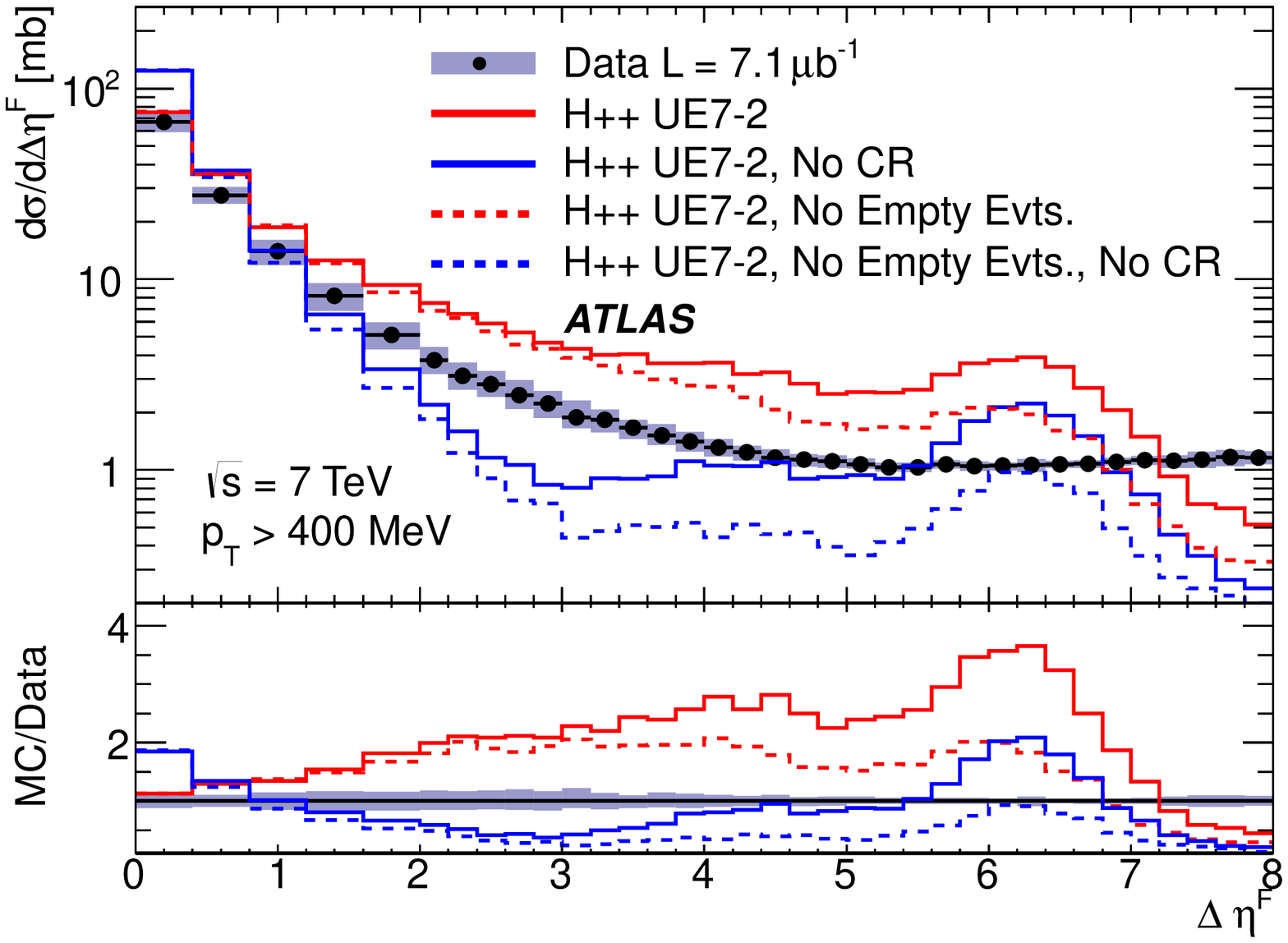}}
\newline
\subfloat[]{\includegraphics[width=0.5\textwidth]{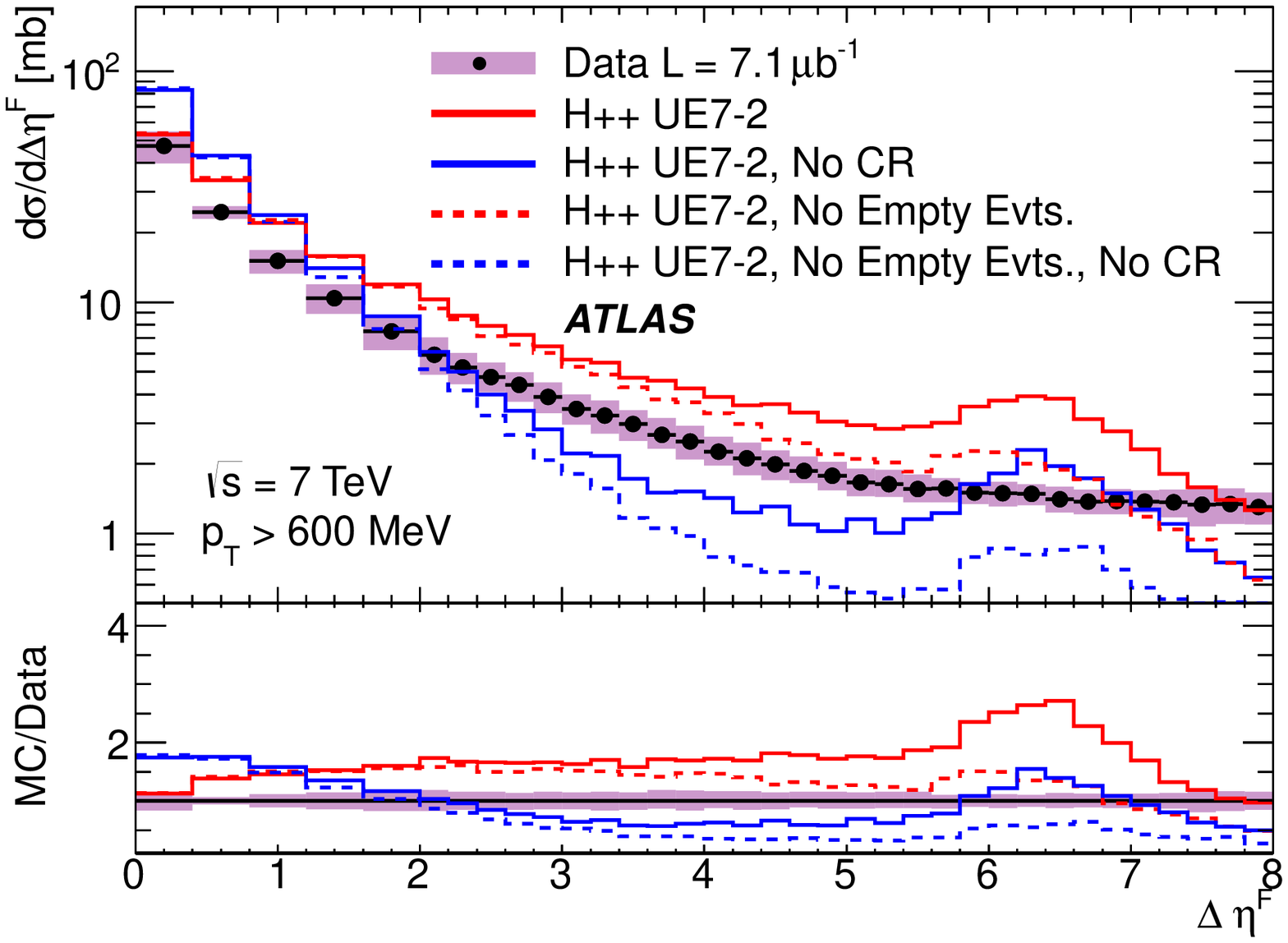}}
\subfloat[]{\includegraphics[width=0.5\textwidth]{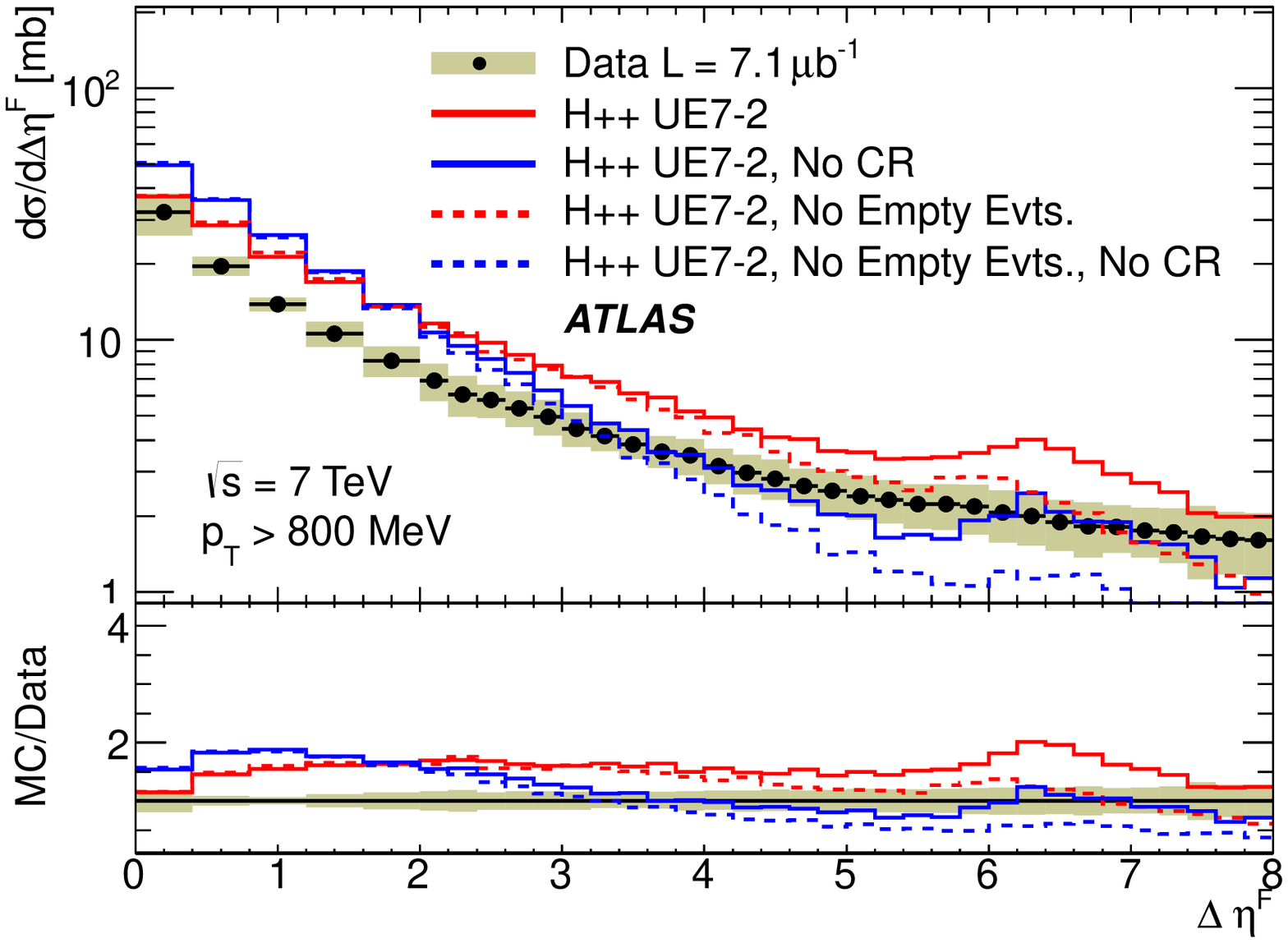}}
 \caption{Inelastic cross section differential in forward gap size 
$\detaf$ for $\ptcut =$ (a) $200 \MeV$,
(b) $400 \MeV$,
(c) $600 \MeV$ and
(d) $800 \MeV$. The data are 
compared with the UE7-2 tune of the \hpp model. 
In addition to the default tune, versions are 
shown in which the colour reconnection 
model is switched off and in which events with zero scatters are 
excluded (see text for further details).}
 \label{Fig:Hpp}
\end{figure*}

The influence of changing from 
$\ptcut = 200 \MeV$ to 
$\ptcut = 400 \MeV$
is small at large \detaf, where the cross section
is dominated by small \xix diffractive events and particle
production is kinematically forbidden over a wide range of 
pseudorapidity.
For $\detaf \lappr 4$, where ND contributions become important,  
a significant fraction of events are assessed as having larger 
gaps for $\ptcut = 400 \MeV$ than for
$\ptcut = 200 \MeV$.
As the value of \ptcut increases to $600$ and $800 \MeV$,
soft ND events migrate to larger \detaf values, giving 
significant contributions throughout most of the distribution 
and confirming \cite{:2010ir} that the production of final state particles with
more than a few hundred \MeV\ is rare in minimum bias
events, even at LHC energies.   
All MC models are able to reproduce the general trends of the data, though
none provides a full description.  

It is interesting to investigate the extent to which 
the alternative cluster-based approach to hadronisation in the 
non-diffractive \hpp model is able 
to describe the data at small gap sizes, 
where the contribution from ND processes is dominant.
A comparison of the data at each of the \ptcut values
with \hpp is shown in \reffig{Fig:Hpp}. 
Four versions 
of the UE7-2 tune are shown,
with variations in the details of the model which are 
expected to have the largest influence on 
rapidity gap distributions.
These are the 
default version (UE7-2), a version in which the colour reconnection
model is switched off (UE7-2, No CR) and similar versions which 
exclude events with no scatterings
of either the soft or semi-hard types (UE7-2, No Empty Evts and
UE7-2, No Empty Evts, No CR).  
At small gap sizes, all versions of the model produce an
exponential fall with increasing gap size, though the dependence on
\detaf is not steep enough in the default model and is too steep when
colour recombination effects are switched off.

Despite not containing an explicit diffractive component,  
the default \hpp\ minimum bias model 
produces a sizeable fraction of events with large gaps,
overshooting the measured cross section 
by up to factor of four in the interval $2 < \detaf < 7$ and
producing an enhancement centred 
around $\detaf = 6$.
When colour reconnection is switched off, this large gap contribution 
is reduced considerably, but remains at a similar level to that measured
in the range $3<\detaf<5$. The enhancement near
$\detaf \approx 6$ is still present.
The events with zero scatters
in the \hpp underlying event model provide a
partial explanation for the large gap contribution.
Removing this contribution reduces the predicted large
gap cross section, but the non-exponential tail and large \detaf
enhancement persist. For all scenarios considered, the alternative cluster based 
hadronisation model in \hpp shows structure which is incompatible with the data.

\subsection{Large gap sizes and sensitivity to diffractive dynamics}
\label{diff:results}

\begin{figure*}
 \centering
\subfloat[]{\includegraphics[width=0.5\textwidth]{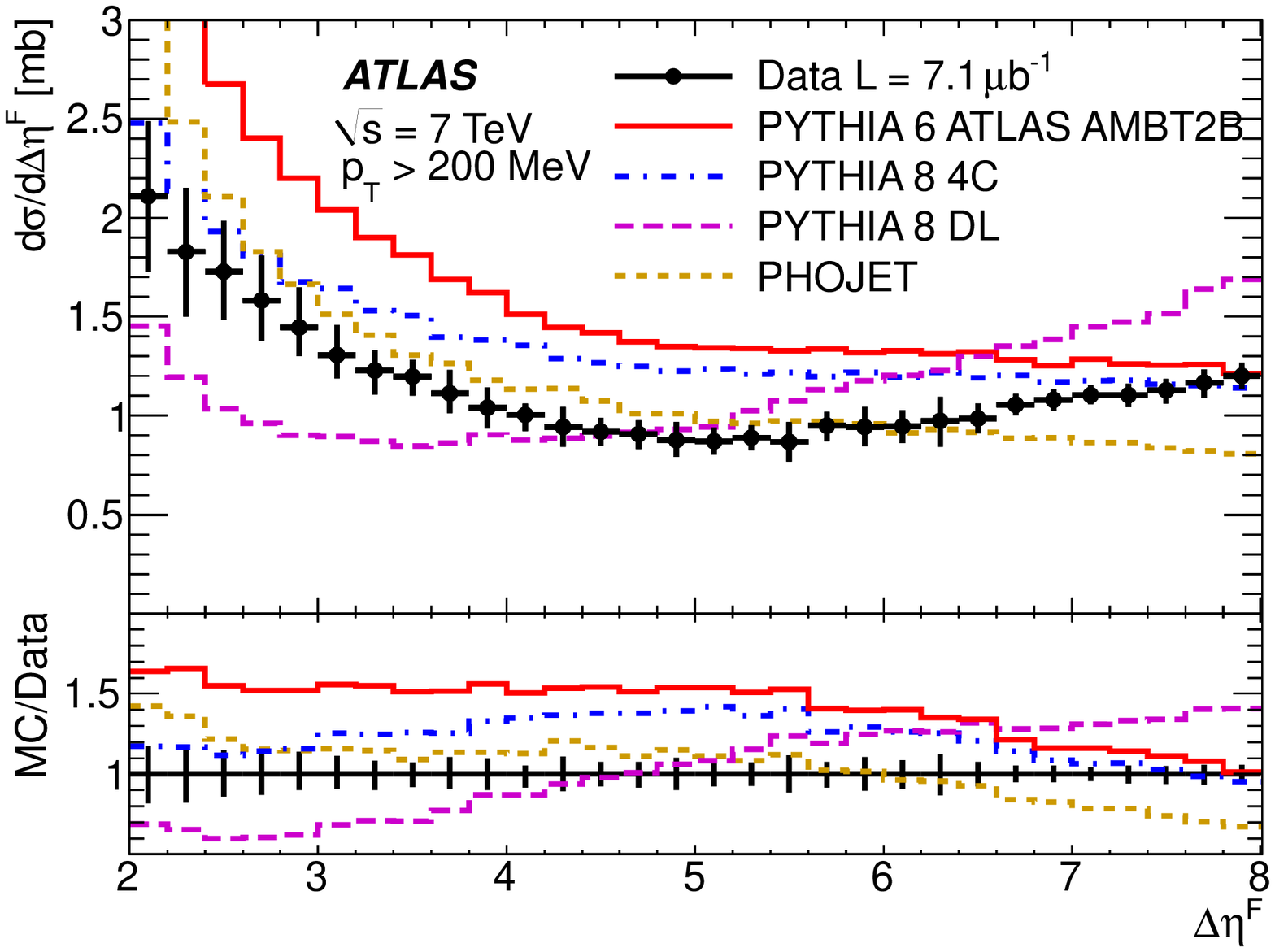}}
\subfloat[]{\includegraphics[width=0.5\textwidth]{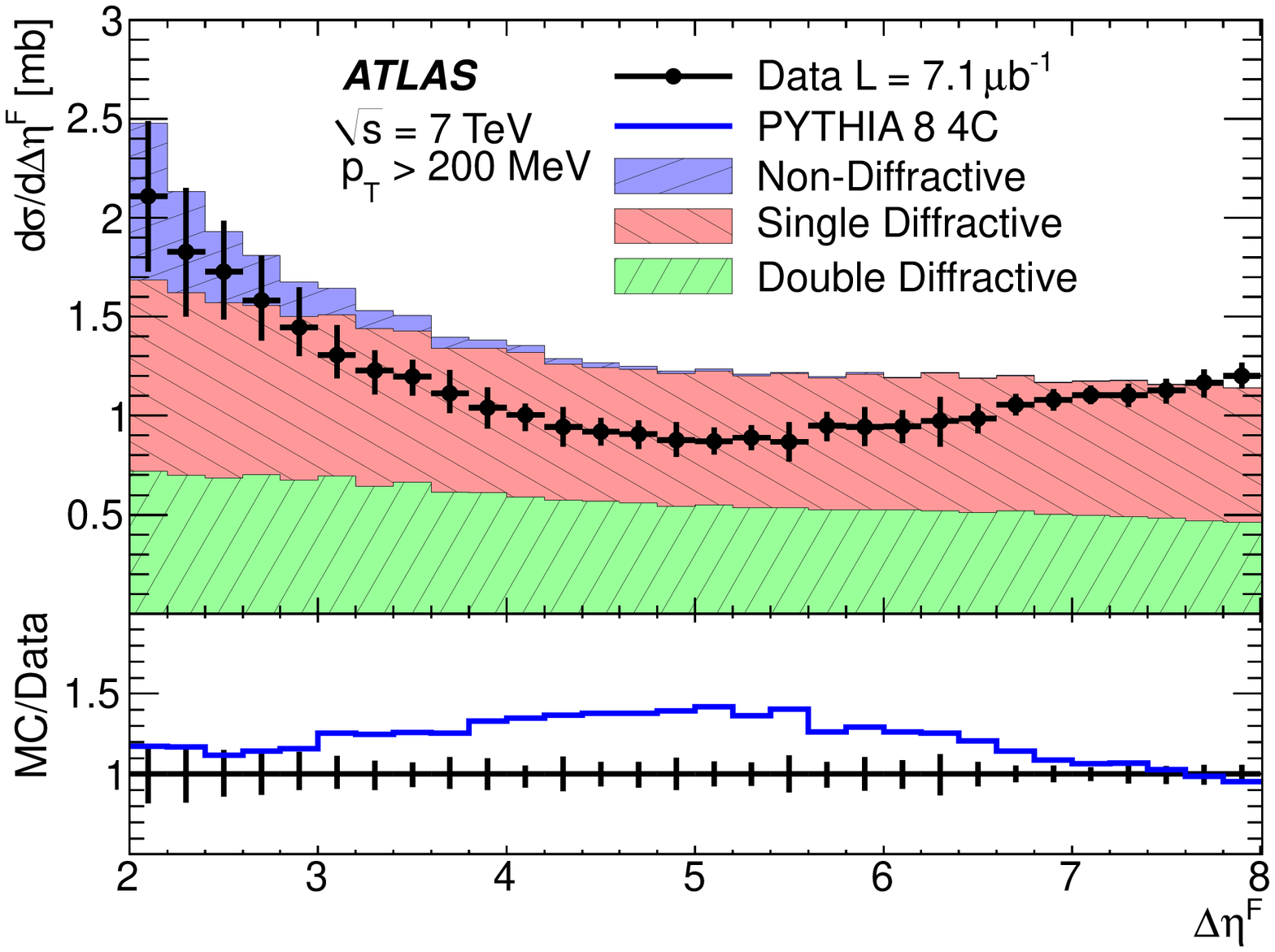}}
\newline
\subfloat[]{\includegraphics[width=0.5\textwidth]{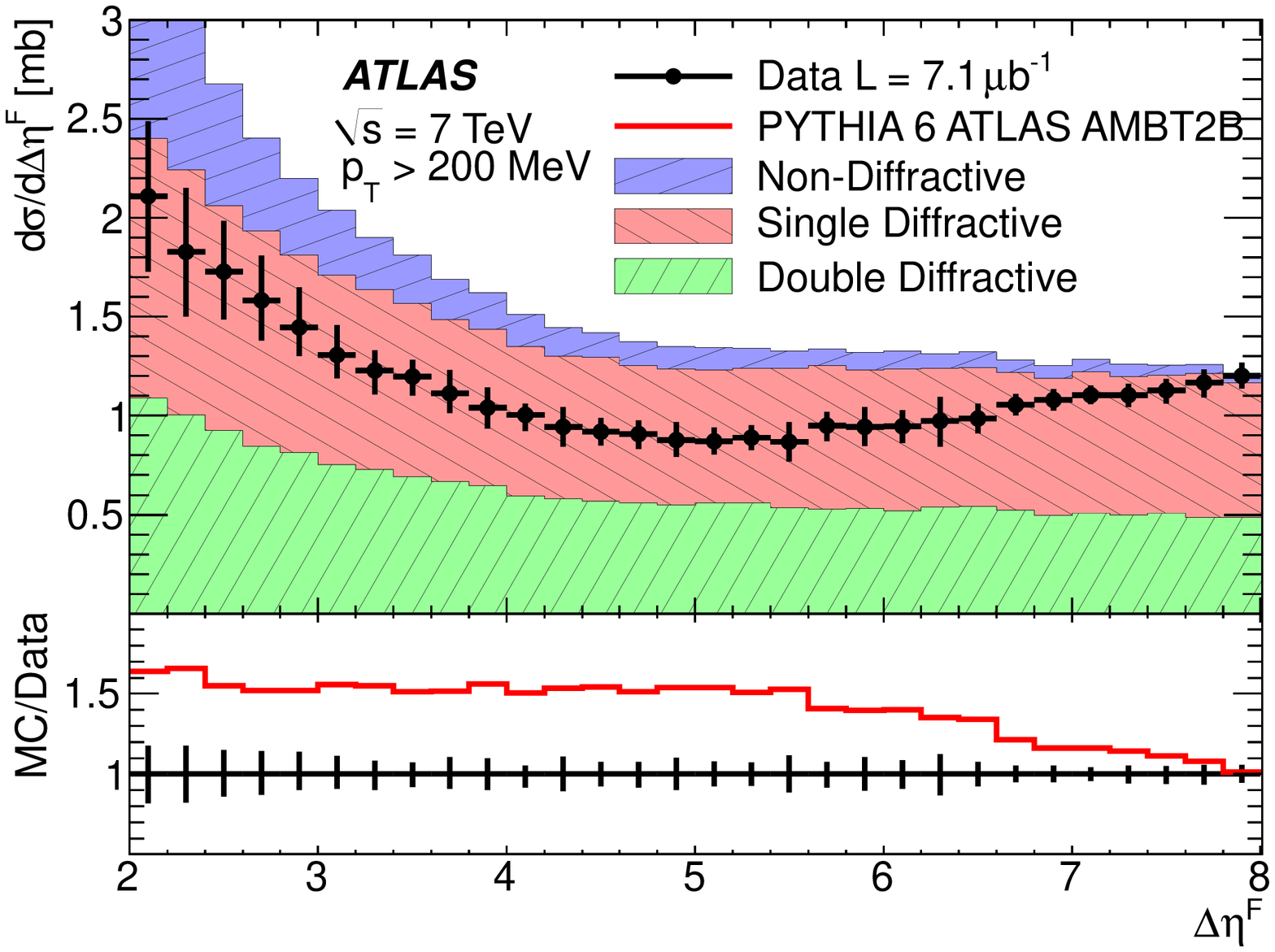}}
\subfloat[]{\includegraphics[width=0.5\textwidth]{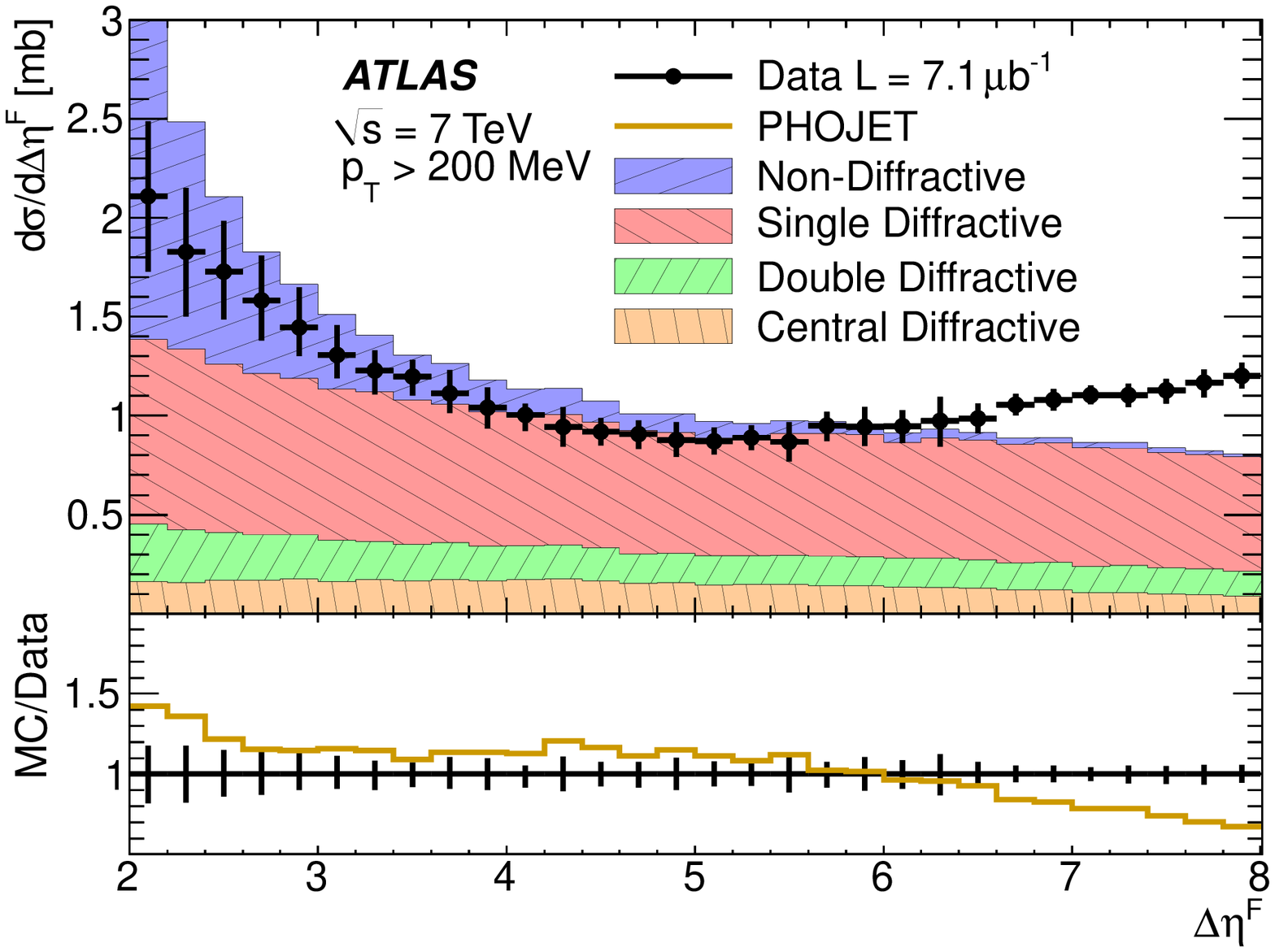}}
 \caption{Inelastic cross section differential in 
forward gap size \detaf for particles 
with $\pt > 200 \MeV$ and $\detaf > 2$. 
The error bars indicate the total uncertainties.
In (a), the full lines show the
predictions of \pho, the default versions of \pysix and \pyeight, and
\pyeight with the Donnachie-Landshoff Pomeron flux.  
The remaining plots show the contributions 
of the SD, DD and ND components according to each generator. 
The CD contribution according to \pho is also shown in (d).}
 \label{xsec:largedeta}
\end{figure*}

At large \detaf, the 
differential cross section exhibits a plateau, which is 
attributed mainly to diffractive 
processes (SD events, together with DD events at
$\xiy \lappr 10^{-6}$) and is shown in
detail in \reffig{xsec:largedeta}. 
According to the \pho MC model, the
CD contribution is also distributed fairly uniformly across this region. 
Over a wide range of gap sizes with $\detaf \gappr 3$,
the differential cross section is roughly constant at around   
$1 \ {\rm mb}$ per unit of rapidity gap size.  
Given the close correlation between $\detaf$ and $- \ln \xi$ 
(\refsec{theory}), 
this behaviour is expected as a consequence of the dominance
of soft diffractive processes.
All MC models roughly reproduce the diffractive plateau, though none gives a
detailed description of the shape as a function of 
\detaf.

When absolutely normalised, the \py\ predictions overshoot
the data throughout most of the diffractive region, despite the 
tuning of
\fd to previous ATLAS data \cite{lauren}
in these models. The excess 
here is partially a reflection
of the $10\%$ overestimate of the \py\ prediction in the total inelastic
cross section 
and may also be associated 
with the large DD cross section 
in the measured region, which exceeds that expected based on Tevatron
data \cite{Affolder:2001vx} and
gives rise to almost
equal SD and DD contributions at large $\detaf$. 
For \pho, the 
underestimate of the diffractive fraction \fd is
largely compensated by the excess in the total 
inelastic cross section, such that the large gap cross section  
is in fair agreement
with the measurement up to $\detaf \approx 6$. 
The DD contribution to the cross section in \pho is heavily suppressed
compared with that in the \py\ models. 

Integrated over
the diffractive-dominated region 
\linebreak
$5 < \detaf < 8$, 
corresponding approximately to
\linebreak
$-5.1 \lappr \logxix \lappr -3.8$ 
according to the 
MC models, the measured cross
section is $3.05 \pm 0.23 \ {\rm mb}$, approximately $4 \%$
of the total inelastic cross section. This can be compared with
$3.58 \ {\rm mb}$, $3.89 \ {\rm mb}$ and $2.71 \ {\rm mb}$ for 
the default versions of \pyeight, \pysix and \pho, respectively.

As can be seen in \reffig{xsec:largedeta}, the differential 
cross section rises slowly with increasing \detaf 
for $\detaf \gappr 5$. Non-diffractive contributions in this region are
small and fall with increasing \detaf 
according to all models, so this rise is attributable to
the dynamics
of the SD and DD processes. Specifically the rising cross section
is as expected from the $\pom \pom \pom$ term in triple Regge models with 
a Pomeron intercept in excess of unity (see equation \refeq{pomonly}). 
In \reffig{xsec:largedeta}a, a comparison is made with 
the \pyeight model, after replacing the 
default Schuler and Sj\"{o}strand Pomeron flux
with the Donnachie and Landshoff (DL) version using
the default Pomeron trajectory,
$\alphapomt = 1.085 + 0.25 t$ (`{\sc pythia{\footnotesize 8} dl}').
It is clear that the 
data at large \detaf are not perfectly described with this choice. 

Whilst the data are insensitive to the choice of 
$\alphapomprime$, there
is considerable sensitivity to the value of $\alphapom$.
The data in the cleanest diffractive region $\detaf > 6$
are used to obtain a best estimate of the 
appropriate choice of the Pomeron
intercept to describe the data. 
SD and DD \pyeight samples are generated with
the DL Pomeron flux for a range of \alphapom
values.
In each case, the default 
$\alphapomprime$ value of $0.25 \GeV^{-2}$ is taken
and the 
tuned ratios of the SD and DD 
contributions appropriate to \pyeight from
\reftab{tab:diff_fraction}\ are 
used.
The $\chi^{2}$ value for the best fit 
to the data in the region $6 < \detaf < 8$ 
is obtained for each of the 
samples with different \alphapom values, with 
the cross section 
integrated over the fitted region 
allowed to float as a free parameter.
The optimum \alphapom is
determined from the minimum of the 
resulting $\chi^2$ parabola. 

The full procedure is repeated for
data points shifted according to each of the systematic effects described in 
\refsec{sec:systematics}, such that correlations between the 
uncertainties on the data points
are taken into account in evaluating the uncertainties. 
The systematic uncertainty is dominated by the MC model dependence of the 
data correction procedure, in particular the
effect of unfolding using 
\pysix in place of \pyeight, which leads to a significantly flatter
dependence of the data on $\detaf$ at large gap sizes.

The result obtained in the context of the \pyeight
model with the DL flux
parameterisation is
\begin{equation}
\hspace*{0.8cm}
\alphapom = 1.058\pm0.003(\textrm{stat.})^{+0.034}_{-0.039}(\textrm{syst.}) \ . 
\end{equation}
The data are thus compatible with a value of 
$\alphapom$ which matches that appropriate to the description of 
total hadronic
cross sections \cite{Donnachie:1992ny,Cudell:1996sh}. 
When the Berger-Streng Pomeron flux,
which differs from the DL version in the modelling of the $t$ dependence,  
is used in the fit procedure, the result is modified to
$\alphapom = 1.056$. 
The effects of varying $\alphapomprime$ 
between $0.1 \GeV^{-2}$ and $0.4 \GeV^{-2}$  
and of varying the $\fsd$ and $\fdd$ 
fractions assumed in the fit in the ranges given in
\refsec{sec:systematics}\ 
are also smaller than the statistical uncertainty.
Compatible results are obtained by fitting the higher \ptcut data.

\begin{figure}
 \centering
\includegraphics[width=0.5\textwidth]{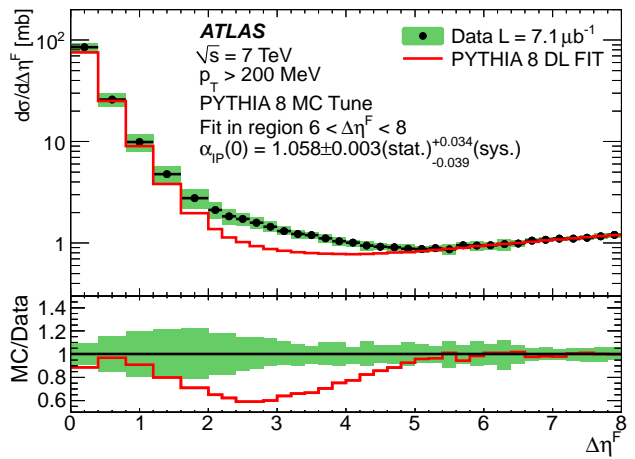}
\caption{Inelastic cross section differential in 
forward gap size \detaf for particles 
with $\pt > 200 \MeV$. The data are compared with 
a modified version of the \pyeight model with the DL flux, in which the 
Pomeron intercept \alphapom is determined from fits to the data in the region
$6 < \detaf < 8$. See text for further details.}
 \label{Fig:fit}
\end{figure}

A comparison between the data and
a modified version of \pyeight,
with \alphapom
as obtained from the fit,
is shown in \reffig{Fig:fit}. 
Here, the diffractive contribution to the inelastic cross section
$\fd = 25.6\%$
is matched\footnote{Since only data at large \detaf are included in the fit, 
the result for \alphapom
is insensitive to systematic variations in \fd.} 
to the 
fitted value of \alphapom using the results in \cite{lauren}.
Together with the cross section integrated over the region $6 < \detaf < 8$
as obtained from the fit and the tuned ratio $\fdd / \fsd$ 
from \reftab{tab:diff_fraction},
this fixes the normalisation of the full distribution. 
The description of
the data at large \detaf is excellent
and 
the exponential fall at small \detaf is also adequately
described. 
There is a discrepancy in the region $2 < \detaf < 4$, 
which may be a consequence of the uncertainty in 
modelling large hadronisation fluctuations
in ND events (compare the ND tails to large $\detaf$ in
\reffig{xsec:largedeta}b, c and d). It may also be attributable 
to sub-leading trajectory 
exchanges \cite{Adloff:1997mi,Aktas:2006hy} or to 
the lack of a CD component in the \py\ model.

\subsection{The integrated inelastic cross section}

By summing over the \detaf distribution from zero to a 
maximum gap size $\detafcut$, 
the integrated inelastic cross section can be obtained, 
excluding the contribution from
events with very large 
gaps $\detaf > \detafcut$. 
As discussed in \refsec{theory},  
there is a strong correlation between the 
size of the gap and the kinematics of diffraction
(see e.g. equation \refeq{gap:eqn}\ for the SD process).
The cross section integrated over a given range of gap size can thus 
be converted into an integral over the inelastic $pp$ cross  
section down to some minimum value \xicut of \xix.
The variation in the 
integrated inelastic cross section with 
$\detafcut$
can then be used to
compare inelastic cross section results 
with different lower limits, \xicut.  

The integral of the forward gap cross section 
\begin{eqnarray*}
\hspace*{2cm}
  \int_0^{\detafcut} \frac{{\rm d} \sigma}{{\rm d} \detaf} \
{\rm d} \detaf
\end{eqnarray*}
is obtained for $\detafcut$ values
varying between $3$ and $8$ by 
cumulatively adding the cross section contributions from 
successive bins of the measured gap distribution. 
The correspondence between maximum gap size and 
minimum \xix used 
here is determined from the
\pyeight model to be
$\log_{10} \xicut = -0.45 \, \detafcut - 1.52$. 
The uncertainty 
on this correlation is small; for example
the \pho model results in the same slope of $-0.45$
with an intercept of $-1.56$. This correlation is applied to
convert to an integral 
\begin{eqnarray*}
\hspace*{2.5cm}
  \int_{\xicut}^1 \frac{{\rm d} \sigma}{{\rm d} \xi_X} \
{\rm d} \xi_X \ .
\end{eqnarray*}
A small correction is applied to account for the fact
that the gap cross section neglects particles 
with\footnote{The finite
\ptcut value in the measured gap cross sections 
tends to increase gap sizes slightly relative to 
$\ptcut = 0$. However, MC
studies indicate that this effect has the biggest influence on the
exponentially falling distribution at small gap sizes, whereas the 
difference for the \detaf 
values which are relevant to the integrated cross section   
are relatively small. According to the MC models, 
the cross section integrated over $5 < \detaf < 8$
decreases by 2\% when changing from  
$\ptcut = 200 \MeV$ to $\ptcut = 0$.}
$\pt < \ptcut = 200 \MeV$ 
and includes a contribution from ND processes.
This correction factor is calculated using 
\linebreak
\pyeight with the 
DL flux, and 
the optimised \alphapom and \fd values, as determined in 
\refsec{diff:results}. 
The integration range is chosen such that the correction is always smaller
than $\pm 1.3\%$.
The systematic uncertainty on the correction factor, evaluated 
by comparison with results obtained using \pho or \pyeight  
with the default Schuler and Sj\"{o}strand flux, together with the 
systematic variations of the tuned fractions \fsd and \fdd as 
in \refsec{sec:systematics}, is also small.

\begin{figure}
 \centering
\includegraphics[width=0.5\textwidth]{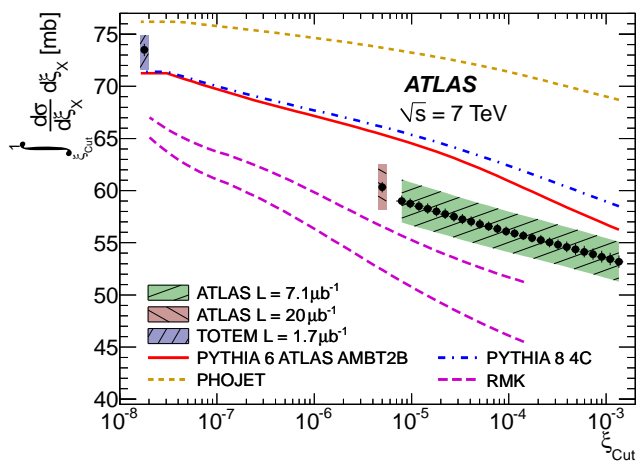}
\caption{Inelastic cross section excluding 
diffractive processes with $\xix < \xicut$, 
obtained by integration of the differential cross section 
from gap sizes of zero to a 
variable maximum. 
The results from the present 
analysis (`ATLAS ${\rm L} = 7.1 \ {\rm \mu b^{-1}}$') 
are compared with a previous
ATLAS result \cite{lauren} and a TOTEM 
measurement integrated over
all kinematically accessible \xix values \cite{TOTEM:sigmatot}. The predictions of the
default versions of the \pysix, \pyeight and \pho models are also
shown, along with two versions of the RMK
model \cite{Ryskin:2011qe} (see text).
The vertical error bars on the ATLAS measurements
denote the systematic uncertainties excluding that on the 
luminosity measurement, whilst the shaded area represents 
the full systematic uncertainty. 
For the TOTEM point, the error bar represents the statistical uncertainty 
whilst the shaded area represents the full uncertainty. 
The uncertainties on the data points obtained in the present analysis
are strongly correlated between neighbouring points.}
\label{Fig:integratexi}
\end{figure}

The integrated inelastic cross section is shown 
as a function of \xicut
in \reffig{Fig:integratexi}, where it is also 
compared with a previous ATLAS result \cite{lauren} and 
with the TOTEM extraction of the full inelastic cross 
section \cite{TOTEM:sigmatot}, derived from a measurement of the
elastic cross section via the optical theorem.
The errors on all of the experimental data points are dominated
by the luminosity uncertainties. 
The previous ATLAS result was also based on MBTS-triggered data, but 
is quoted at the \xix value corresponding to 
$50 \%$ trigger efficiency, which is 
slightly beyond the range accessed here.  
Extrapolating according to the measured 
dependence on \xicut, the new data are in 
good agreement with the previous result, 
the small apparent difference being well within the  
uncertainty due to run-to-run luminosity measurement
variations. 

It is instructive to compare the TOTEM result with the ATLAS measurements, 
since the latter 
omit the poorly understood lowest
$\xix$ region. By comparing the lowest \xicut 
data point from the present
analysis with the TOTEM measurement and neglecting any correlations 
between the ATLAS and TOTEM uncertainties, the inelastic cross section
integrated over 
\linebreak
$\xix < 8 \times 10^{-6}$ is inferred to be
$14.5 ^{+ 2.0}_{-1.5} \ {\rm mb}$.
Significantly smaller 
contributions are predicted by the default versions of
\py\ ($\sim 6 \ {\rm mb}$) and \pho 
\linebreak
($\sim 3 \ {\rm mb}$).
\reffig{Fig:integratexi}\ also shows two versions of the 
RMK model (see \refsec{theory}), corresponding to  
versions (i) (upper curve) and (ii) (lower curve) in \cite{Ryskin:2011qe}.
These versions differ in the radii attributed to the elastically
scattered eigenstates comprising the 
low \xix contribution which is added to the more standard triple Pomeron
calculation in the model, (ii) being the 
favoured version and (i) being 
indicative
of the flexibility in the model whilst preserving an
acceptable description of pre-LHC data. 
The additional low \xix processes enhance the inelastic cross section by
$5.5 \ {\rm mb}$ and $6.7 \ {\rm mb}$ in versions (i) and (ii),
respectively.  
Although the RMK model 
lies below the data in general, the low \xix enhancement
is compatible with that observed \cite{Martin:2011gi}.
The shape of the distribution at low \xix is not predicted in the
model, but is compatible with the data if, as shown 
here \cite{Ryskin:private},
it is assumed to have the steep \xix dependence associated
with the $\pom \pom \reg$, rather than the 
$\pom \pom \pom$ triple Regge term. Similar conclusions have been reached
previously from proton-proton \cite{Bernard:1986yh} 
and photoproduction \cite{Adloff:1997mi,Breitweg:1997za} data.

\section{Summary}
\label{sec:Conclusion}

A novel algorithm has been devised for identifying rapidity gaps in 
the final state of minimum bias ATLAS data, 
leading to measurements in which particle production is considered down to
transverse momentum thresholds \ptcut
between $200 \MeV$ and $800 \MeV$. 
The differential cross section ${\rm d} \sigma / {\rm d} \detaf$ 
is measured
for 
forward rapidity gaps of size $0 < \detaf < 8$, 
corresponding to the larger of the two gaps
extending to $\eta = \pm 4.9$, with no requirements on
activity at $| \eta| > 4.9$. An
exponentially falling non-diffractive contribution is observed at
small gap sizes, which is also a feature of the \py, \pho and 
\hpp Monte Carlo models. However, none of the models describes the 
\detaf or \ptcut dependence of 
this region in detail.  
At large gap sizes, 
the differential cross section exhibits a plateau, 
which corresponds to 
a mixture of the single-diffractive dissociation 
process and double dissociation
with $\xiy \lappr 10^{-6}$. 
This plateau amounts to a cross section close to $1 \ {\rm mb}$
per unit of gap size and its magnitude is 
roughly described by the \py\ and \pho 
Monte Carlo models.
None of the default models reproduce the rise of the
differential cross section as a function of gap size at the 
largest \detaf values. This rise is interpreted 
within the triple Pomeron-based approach of the \pyeight 
model with a Donnachie-Landshoff
Pomeron flux in terms of a Pomeron intercept of 
$\alphapom = 1.058\pm0.003(\textrm{stat.})
\linebreak 
^{+0.034}_{-0.039}(\textrm{syst.})$.
Since the bulk of the inelastic $pp$ cross section is contained within the 
measured 
range, integrated cross sections are also obtained and compared with
previous measurements.
The contribution to the total inelastic cross section from the 
region $\xix < 10^{-5}$ is determined to be around 
$20 \%$, which is considerably larger
than is predicted by most models. 

\section*{Acknowledgements}

We are grateful to the \hpp collaboration for their 
assistance in investigating the low \pt\ rapidity gap spectrum 
produced by their generator and to M.~Ryskin, A.~Martin and V.~Khoze 
for supplying their prediction for the 
behaviour of the 
inelastic cross section as a function of \xicut. 

We thank CERN for the very successful operation of the LHC, as well as the
support staff from our institutions, without whom ATLAS could not be
operated efficiently.

We acknowledge the support of ANPCyT, Argentina; YerPhI, Armenia; ARC,
Australia; BMWF, Austria; ANAS, Azerbaijan; SSTC, Belarus; CNPq and FAPESP,
Brazil; NSERC, NRC and CFI, Canada; CERN; CONICYT, Chile; CAS, MOST and NSFC,
China; COLCIENCIAS, Colombia; MSMT CR, MPO CR and VSC CR, Czech Republic;
DNRF, DNSRC and Lundbeck Foundation, Denmark; ARTEMIS and ERC, European Union;
IN2P3-CNRS, CEA-DSM/IRFU, France; GNAS, Georgia; BMBF, DFG, HGF, MPG and AvH
Foundation, Germany; GSRT, Greece; ISF, MINERVA, GIF, DIP and Benoziyo Center,
Israel; INFN, Italy; MEXT and JSPS, Japan; CNRST, Morocco; FOM and NWO,
Netherlands; RCN, Norway; MNiSW, Poland; GRICES and FCT, Portugal; MERYS
(MECTS), Romania; MES of Russia and ROSATOM, Russian Federation; JINR; 

MSTD, Serbia; MSSR, Slovakia; ARRS and MVZT, Slovenia; DST/NRF, South Africa;
MICINN, Spain; SRC and Wallenberg Foundation, Sweden; SER, SNSF and Cantons of
Bern and Geneva, Switzerland; NSC, Taiwan; TAEK, Turkey; STFC, the Royal
Society and Leverhulme Trust, United Kingdom; DOE and NSF, United States of
America.

The crucial computing support from all WLCG partners is acknowledged
gratefully, in particular from CERN and the ATLAS Tier-1 facilities at
TRIUMF (Canada), NDGF (Denmark, Norway, Sweden), CC-IN2P3 (France),
KIT/GridKA (Germany), INFN-CNAF (Italy), NL-T1 (Netherlands), PIC (Spain),
ASGC (Taiwan), RAL (UK) and BNL (USA) and in the Tier-2 facilities
worldwide.

\appendix

\begin{table*}
 \centering
 \caption{The measured differential cross section
data points for $\ptcut = 200 \MeV$, with 
each value corresponding to an average over the given $\detaf$ range.
Also quoted are the percentage
statistical ($\delta_{\textrm{stat}}$) uncertainty
and the upward ($+\delta_{\textrm{tot}}$) and downward ($-\delta_{\textrm{tot}}$) 
total uncertainties, obtained by adding all uncertainties in
quadrature.
The remaining columns contain the percentage shifts 
due to each of the contributing systematic sources, which are correlated
between data points. Those due to the 
modelling of final state particle production ($\delta_{\textrm{py6}}$),
the modelling of
the $\xi_X$, $\xi_Y$ and $t$ dependences ($\delta_{\textrm{pho}}$) and
variation of the
CD ($\delta_{\textrm{cd}}$) cross section
in the unfolding procedure are applied symmetrically
as upward and downward uncertainties, as are
those due to the dead material budget in the 
tracking region ($\delta_{\textrm{mat}}$) and the MBTS 
response ($\delta_{\textrm{mbts}}$).
The uncertainties due to variations in the relative energy scale in
data and MC 
are evaluated separately for upward 
($\delta_{e+}$) and downward ($\delta_{e-}$) shifts, as are the
modelling uncertainties
due to enhancing ($\delta_{\textrm{sd}}$) or 
reducing ($\delta_{\textrm{dd}}$)
the \ssd/\sdd cross section ratio.
Minus signs appear where the shift in a variable is anti-correlated
rather than correlated with the shift in the differential cross section. 
The $3.4\%$ normalisation uncertainty due to the luminosity measurement
is also included in the $\pm \delta_{\textrm{tot}}$ values. 
These data points can be obtained from the 
HEPDATA database \cite{hepdata}, 
along with their counterparts for
$\ptcut = 400$, $600$ and $800 \MeV$. A Rivet \cite{rivet} routine 
is also available.}
 \label{Tab:DPs}
 \begin{tabular}{c|cccc|ccccccccc}
\hline\noalign{\smallskip}
\multirow{2}{*}{\scriptsize{$\detaf$}} & \scriptsize{${\rm d}\sigma/{\rm d}\detaf$} & \scriptsize{$\delta_{\textrm{stat}}$} &\scriptsize{$+\delta_{\textrm{tot}}$} & \scriptsize{$-\delta_{\textrm{tot}}$} &\scriptsize{$\delta_{\textrm{py6}}$} & \scriptsize{$\delta_{\textrm{pho}}$} & \scriptsize{$\delta_{\textrm{e+}}$} & \scriptsize{$\delta_{\textrm{e-}}$} & \scriptsize{$\delta_{\textrm{sd}}$} & \scriptsize{$\delta_{\textrm{dd}}$} & \scriptsize{$\delta_{\textrm{cd}}$} & \scriptsize{$\delta_{\textrm{mat}}$} & \scriptsize{$\delta_{\textrm{mbts}}$} \\ 
 & \scriptsize{[mb]} & \scriptsize{[\%]} & \scriptsize{[\%]} &  \scriptsize{[\%]} & \scriptsize{[\%]} & \scriptsize{[\%]} & \scriptsize{[\%]} & \scriptsize{[\%]} & \scriptsize{[\%]} & \scriptsize{[\%]} & \scriptsize{[\%]} & \scriptsize{[\%]} & \scriptsize{[\%]} \\ 
\noalign{\smallskip}\hline\noalign{\smallskip}
\scriptsize{0.0--0.4} & \scriptsize{85} & \scriptsize{0.2} & \scriptsize{9.4} & \scriptsize{-9.5} & \scriptsize{-2.9} & \scriptsize{-7.6} & \scriptsize{3.0} & \scriptsize{-3.2} & \scriptsize{0.1} & \scriptsize{-0.1} & \scriptsize{0.0} & \scriptsize{-1.1} & \scriptsize{-0.1} \\ 
\scriptsize{0.4--0.8} & \scriptsize{26} & \scriptsize{0} & \scriptsize{15} & \scriptsize{-15} & \scriptsize{4} & \scriptsize{14} & \scriptsize{-3} & \scriptsize{3} & \scriptsize{-0} & \scriptsize{0} & \scriptsize{0} & \scriptsize{2} & \scriptsize{0} \\ 
\scriptsize{0.8--1.2} & \scriptsize{10} & \scriptsize{0} & \scriptsize{20} & \scriptsize{-20} & \scriptsize{5} & \scriptsize{18} & \scriptsize{-6} & \scriptsize{5} & \scriptsize{-0} & \scriptsize{0} & \scriptsize{0} & \scriptsize{3} & \scriptsize{0} \\ 
\scriptsize{1.2--1.6} & \scriptsize{5} & \scriptsize{0} & \scriptsize{21} & \scriptsize{-21} & \scriptsize{10} & \scriptsize{17} & \scriptsize{-6} & \scriptsize{7} & \scriptsize{0} & \scriptsize{-0} & \scriptsize{0} & \scriptsize{2} & \scriptsize{-0} \\ 
\scriptsize{1.6--2.0} & \scriptsize{2.8} & \scriptsize{0} & \scriptsize{22} & \scriptsize{-22} & \scriptsize{15} & \scriptsize{13} & \scriptsize{-7} & \scriptsize{9} & \scriptsize{0} & \scriptsize{-0} & \scriptsize{-0} & \scriptsize{3} & \scriptsize{0} \\ 
\scriptsize{2.0--2.2} & \scriptsize{2.1} & \scriptsize{1} & \scriptsize{18} & \scriptsize{-18} & \scriptsize{15} & \scriptsize{5} & \scriptsize{-9} & \scriptsize{8} & \scriptsize{-0} & \scriptsize{0} & \scriptsize{-0} & \scriptsize{1} & \scriptsize{0} \\ 
\scriptsize{2.2--2.4} & \scriptsize{1.8} & \scriptsize{1} & \scriptsize{18} & \scriptsize{-18} & \scriptsize{14} & \scriptsize{7} & \scriptsize{-8} & \scriptsize{8} & \scriptsize{-0} & \scriptsize{0} & \scriptsize{-0} & \scriptsize{1} & \scriptsize{0} \\ 
\scriptsize{2.4--2.6} & \scriptsize{1.7} & \scriptsize{1} & \scriptsize{15} & \scriptsize{-14} & \scriptsize{9} & \scriptsize{2} & \scriptsize{-9} & \scriptsize{10} & \scriptsize{-0} & \scriptsize{0} & \scriptsize{-0} & \scriptsize{-3} & \scriptsize{0} \\ 
\scriptsize{2.6--2.8} & \scriptsize{1.6} & \scriptsize{1} & \scriptsize{14} & \scriptsize{-13} & \scriptsize{5} & \scriptsize{1} & \scriptsize{-11} & \scriptsize{13} & \scriptsize{-0} & \scriptsize{0} & \scriptsize{-0} & \scriptsize{-0} & \scriptsize{0} \\ 
\scriptsize{2.8--3.0} & \scriptsize{1.4} & \scriptsize{1} & \scriptsize{14} & \scriptsize{-10} & \scriptsize{5} & \scriptsize{2} & \scriptsize{-8} & \scriptsize{12} & \scriptsize{-0} & \scriptsize{0} & \scriptsize{-1} & \scriptsize{-1} & \scriptsize{0} \\ 
\scriptsize{3.0--3.2} & \scriptsize{1.3} & \scriptsize{1} & \scriptsize{11} & \scriptsize{-9} & \scriptsize{4} & \scriptsize{2} & \scriptsize{-7} & \scriptsize{10} & \scriptsize{-1} & \scriptsize{0} & \scriptsize{-0} & \scriptsize{1} & \scriptsize{1} \\ 
\scriptsize{3.2--3.4} & \scriptsize{1.2} & \scriptsize{1.3} & \scriptsize{8.4} & \scriptsize{-9.9} & \scriptsize{3.5} & \scriptsize{3.7} & \scriptsize{-7.5} & \scriptsize{5.4} & \scriptsize{-0.6} & \scriptsize{0.4} & \scriptsize{-0.4} & \scriptsize{1.1} & \scriptsize{0.6} \\ 
\scriptsize{3.4--3.6} & \scriptsize{1.20} & \scriptsize{1.3} & \scriptsize{7.3} & \scriptsize{-8.2} & \scriptsize{4.4} & \scriptsize{0.4} & \scriptsize{-4.9} & \scriptsize{3.4} & \scriptsize{-0.7} & \scriptsize{0.6} & \scriptsize{-0.5} & \scriptsize{-2.6} & \scriptsize{0.9} \\ 
\scriptsize{3.6--3.8} & \scriptsize{1.1} & \scriptsize{1} & \scriptsize{11} & \scriptsize{-9} & \scriptsize{6} & \scriptsize{5} & \scriptsize{-4} & \scriptsize{7} & \scriptsize{-1} & \scriptsize{0} & \scriptsize{-1} & \scriptsize{-1} & \scriptsize{1} \\ 
\scriptsize{3.8--4.0} & \scriptsize{1.0} & \scriptsize{2} & \scriptsize{10} & \scriptsize{-10} & \scriptsize{7} & \scriptsize{4} & \scriptsize{-4} & \scriptsize{4} & \scriptsize{-0} & \scriptsize{0} & \scriptsize{-1} & \scriptsize{-2} & \scriptsize{1} \\ 
\scriptsize{4.0--4.2} & \scriptsize{1.01} & \scriptsize{1.6} & \scriptsize{5.7} & \scriptsize{-8.5} & \scriptsize{3.1} & \scriptsize{2.6} & \scriptsize{-6.2} & \scriptsize{0.2} & \scriptsize{-0.7} & \scriptsize{0.6} & \scriptsize{-1.0} & \scriptsize{-0.6} & \scriptsize{0.8} \\ 
\scriptsize{4.2--4.4} & \scriptsize{0.9} & \scriptsize{1} & \scriptsize{11} & \scriptsize{-11} & \scriptsize{5} & \scriptsize{8} & \scriptsize{-2} & \scriptsize{3} & \scriptsize{-1} & \scriptsize{1} & \scriptsize{-1} & \scriptsize{-3} & \scriptsize{1} \\ 
\scriptsize{4.4--4.6} & \scriptsize{0.92} & \scriptsize{1.8} & \scriptsize{7.8} & \scriptsize{-7.8} & \scriptsize{3.9} & \scriptsize{5.0} & \scriptsize{-2.1} & \scriptsize{2.1} & \scriptsize{-0.3} & \scriptsize{0.3} & \scriptsize{-1.0} & \scriptsize{-0.7} & \scriptsize{0.4} \\ 
\scriptsize{4.6--4.8} & \scriptsize{0.91} & \scriptsize{1.7} & \scriptsize{7.8} & \scriptsize{-8.4} & \scriptsize{4.5} & \scriptsize{4.8} & \scriptsize{-3.3} & \scriptsize{0.8} & \scriptsize{-0.9} & \scriptsize{0.7} & \scriptsize{-0.9} & \scriptsize{-0.3} & \scriptsize{1.1} \\ 
\scriptsize{4.8--5.0} & \scriptsize{0.88} & \scriptsize{2} & \scriptsize{10} & \scriptsize{-10} & \scriptsize{6} & \scriptsize{7} & \scriptsize{-0} & \scriptsize{3} & \scriptsize{-1} & \scriptsize{1} & \scriptsize{-1} & \scriptsize{-1} & \scriptsize{1} \\ 
\scriptsize{5.0--5.2} & \scriptsize{0.87} & \scriptsize{1.6} & \scriptsize{8.2} & \scriptsize{-7.8} & \scriptsize{5.3} & \scriptsize{4.0} & \scriptsize{0.5} & \scriptsize{2.5} & \scriptsize{-0.7} & \scriptsize{0.5} & \scriptsize{-0.9} & \scriptsize{-0.4} & \scriptsize{0.9} \\ 
\scriptsize{5.2--5.4} & \scriptsize{0.89} & \scriptsize{1.8} & \scriptsize{7.3} & \scriptsize{-7.5} & \scriptsize{5.5} & \scriptsize{2.5} & \scriptsize{-1.5} & \scriptsize{-1.6} & \scriptsize{-0.7} & \scriptsize{0.5} & \scriptsize{-0.8} & \scriptsize{-0.5} & \scriptsize{0.9} \\ 
\scriptsize{5.4--5.6} & \scriptsize{0.9} & \scriptsize{1} & \scriptsize{12} & \scriptsize{-12} & \scriptsize{8} & \scriptsize{7} & \scriptsize{1} & \scriptsize{1} & \scriptsize{-1} & \scriptsize{1} & \scriptsize{-1} & \scriptsize{2} & \scriptsize{1} \\ 
\scriptsize{5.6--5.8} & \scriptsize{0.95} & \scriptsize{1.2} & \scriptsize{7.5} & \scriptsize{-8.4} & \scriptsize{5.1} & \scriptsize{3.8} & \scriptsize{-2.2} & \scriptsize{-3.5} & \scriptsize{-1.0} & \scriptsize{0.8} & \scriptsize{-0.8} & \scriptsize{0.8} & \scriptsize{1.3} \\ 
\scriptsize{5.8--6.0} & \scriptsize{0.9} & \scriptsize{1} & \scriptsize{11} & \scriptsize{-10} & \scriptsize{7} & \scriptsize{6} & \scriptsize{3} & \scriptsize{0} & \scriptsize{-1} & \scriptsize{0} & \scriptsize{-1} & \scriptsize{1} & \scriptsize{1} \\ 
\scriptsize{6.0--6.2} & \scriptsize{0.95} & \scriptsize{1.4} & \scriptsize{8.6} & \scriptsize{-9.3} & \scriptsize{7.0} & \scriptsize{2.7} & \scriptsize{0.2} & \scriptsize{-3.5} & \scriptsize{-0.6} & \scriptsize{0.5} & \scriptsize{-1.1} & \scriptsize{1.5} & \scriptsize{1.0} \\ 
\scriptsize{6.2--6.4} & \scriptsize{1.0} & \scriptsize{1} & \scriptsize{12} & \scriptsize{-13} & \scriptsize{6} & \scriptsize{7} & \scriptsize{7} & \scriptsize{-8} & \scriptsize{-1} & \scriptsize{1} & \scriptsize{-1} & \scriptsize{4} & \scriptsize{1} \\ 
\scriptsize{6.4--6.6} & \scriptsize{0.99} & \scriptsize{1.3} & \scriptsize{7.8} & \scriptsize{-7.9} & \scriptsize{3.6} & \scriptsize{5.7} & \scriptsize{-0.2} & \scriptsize{-0.5} & \scriptsize{-0.6} & \scriptsize{0.5} & \scriptsize{-0.7} & \scriptsize{0.8} & \scriptsize{1.2} \\ 
\scriptsize{6.6--6.8} & \scriptsize{1.06} & \scriptsize{1.3} & \scriptsize{5.4} & \scriptsize{-5.4} & \scriptsize{2.0} & \scriptsize{3.0} & \scriptsize{-0.9} & \scriptsize{1.0} & \scriptsize{-0.5} & \scriptsize{0.4} & \scriptsize{-0.5} & \scriptsize{-0.4} & \scriptsize{1.0} \\ 
\scriptsize{6.8--7.0} & \scriptsize{1.08} & \scriptsize{1.3} & \scriptsize{5.4} & \scriptsize{-5.2} & \scriptsize{-0.0} & \scriptsize{3.3} & \scriptsize{-0.9} & \scriptsize{1.7} & \scriptsize{-0.1} & \scriptsize{0.1} & \scriptsize{-0.5} & \scriptsize{-1.3} & \scriptsize{0.6} \\ 
\scriptsize{7.0--7.2} & \scriptsize{1.11} & \scriptsize{1.2} & \scriptsize{4.4} & \scriptsize{-4.5} & \scriptsize{-1.9} & \scriptsize{-0.2} & \scriptsize{-1.2} & \scriptsize{1.0} & \scriptsize{-0.4} & \scriptsize{0.3} & \scriptsize{-0.1} & \scriptsize{-0.6} & \scriptsize{1.1} \\ 
\scriptsize{7.2--7.4} & \scriptsize{1.11} & \scriptsize{0.9} & \scriptsize{5.3} & \scriptsize{-5.7} & \scriptsize{-3.4} & \scriptsize{0.1} & \scriptsize{1.2} & \scriptsize{-2.5} & \scriptsize{-0.4} & \scriptsize{0.3} & \scriptsize{-0.2} & \scriptsize{-1.0} & \scriptsize{1.3} \\ 
\scriptsize{7.4--7.6} & \scriptsize{1.13} & \scriptsize{1.0} & \scriptsize{5.1} & \scriptsize{-6.1} & \scriptsize{-3.2} & \scriptsize{-0.6} & \scriptsize{-0.0} & \scriptsize{-3.3} & \scriptsize{-0.6} & \scriptsize{0.4} & \scriptsize{-0.2} & \scriptsize{0.5} & \scriptsize{1.6} \\ 
\scriptsize{7.6--7.8} & \scriptsize{1.17} & \scriptsize{1.0} & \scriptsize{5.9} & \scriptsize{-6.7} & \scriptsize{-4.0} & \scriptsize{-1.7} & \scriptsize{-0.0} & \scriptsize{-3.1} & \scriptsize{-0.3} & \scriptsize{0.2} & \scriptsize{-0.2} & \scriptsize{1.2} & \scriptsize{1.3} \\ 
\scriptsize{7.8--8.0} & \scriptsize{1.20} & \scriptsize{1.0} & \scriptsize{5.7} & \scriptsize{-5.4} & \scriptsize{-4.0} & \scriptsize{-0.5} & \scriptsize{1.0} & \scriptsize{1.8} & \scriptsize{-0.0} & \scriptsize{0.0} & \scriptsize{-0.1} & \scriptsize{-0.0} & \scriptsize{1.0} \\ 
\noalign{\smallskip}\hline \end{tabular} 
\end{table*}


\clearpage
\onecolumn
\begin{flushleft}
{\Large The ATLAS Collaboration}

\bigskip

G.~Aad$^{\rm 48}$,
B.~Abbott$^{\rm 110}$,
J.~Abdallah$^{\rm 11}$,
A.A.~Abdelalim$^{\rm 49}$,
A.~Abdesselam$^{\rm 117}$,
O.~Abdinov$^{\rm 10}$,
B.~Abi$^{\rm 111}$,
M.~Abolins$^{\rm 87}$,
O.S.~AbouZeid$^{\rm 157}$,
H.~Abramowicz$^{\rm 152}$
% ...
\bigskip

$^{1}$ University at Albany, Albany NY, United States of America\\
% ...
$^{10}$ Institute of Physics, Azerbaijan Academy of Sciences, Baku, Azerbaijan\\
$^{11}$ Institut de F\'isica d'Altes Energies and Departament de F\'isica de la Universitat Aut\`onoma  de Barcelona and ICREA, Barcelona, Spain\\
% ...
$^{48}$ Fakult\"{a}t f\"{u}r Mathematik und Physik, Albert-Ludwigs-Universit\"{a}t, Freiburg i.Br., Germany\\
$^{49}$ Section de Physique, Universit\'e de Gen\`eve, Geneva, Switzerland\\
% ...
$^{87}$ Department of Physics and Astronomy, Michigan State University, East Lansing MI, United States of America\\
% ...
$^{110}$ Homer L. Dodge Department of Physics and Astronomy, University of Oklahoma, Norman OK, United States of America\\
$^{111}$ Department of Physics, Oklahoma State University, Stillwater OK, United States of America\\
% ...
$^{117}$ Department of Physics, Oxford University, Oxford, United Kingdom\\
% ...
$^{152}$ Raymond and Beverly Sackler School of Physics and Astronomy, Tel Aviv University, Tel Aviv, Israel\\
% ...
$^{157}$ Department of Physics, University of Toronto, Toronto ON, Canada\\
% ...
$^{*}$ Deceased\end{flushleft}



\begin{thebibliography}{99}

\bibitem{:2010ir}
  ATLAS Collaboration,
  New J.\ Phys.\ \ {\bf 13} (2011) 053033
  [arXiv:1012.5104 [hep-ex]].

\bibitem{arXiv:1011.5531}
  CMS Collaboration,
  JHEP\ {\bf 1101} (2011) 079
  [arXiv:1011.5531 [hep-ex]].

\bibitem{arXiv:1004.3514}
  ALICE Collaboration,
  Eur.\ Phys.\ J.\ {\bf C68} (2010) 345
  [arXiv:1004.3514 [hep-ex]].

\bibitem{Bjorken:1992mj}
  J.~Bjorken, S.~Brodsky and H.~J.~Lu,
  Phys.\ Lett.\ {\bf B286} (1992) 153.

\bibitem{pomeron}
E.~Feinberg and I.~Pomeran\u{c}uk,
Suppl.\ Nuovo Cimento {\bf 3} (1956) 652.

\bibitem{Chew:1961ev}
  G.~Chew and S.~Frautschi,
  Phys.\ Rev.\ Lett.\ \ {\bf 7} (1961) 394.

\bibitem{Acosta:2003xi}
  CDF Collaboration,
  Phys.\ Rev.\ Lett.\ \ {\bf 93} (2004) 141601
  [hep-ex/0311023].

\bibitem{lauren}
  ATLAS Collaboration,
  Nature Comm.\ \ {\bf 2} (2011) 463.

\bibitem{Antchev:2011zz}
  TOTEM Collaboration,
  Europhys.\ Lett.\ \ {\bf 95} (2011) 41001
  [arXiv:1110.1385 [hep-ex]].

\bibitem{TOTEM:sigmatot}
  TOTEM Collaboration,
  Europhys.\ Lett.\ \ {\bf 96} (2011) 21002
  [arXiv:1110.1395 [hep-ex]].

\bibitem{Mueller:1970fa}
  A.~Mueller,
  Phys.\ Rev.\ {\bf D2} (1970) 2963.

\bibitem{Low:1975sv}
  F.~Low,
  Phys.\ Rev.\ {\bf D12} (1975) 163.

\bibitem{Nussinov:1975mw}
  S.~Nussinov,
  Phys.\ Rev.\ Lett.\ \ {\bf 34} (1975) 1286.

\bibitem{Gribov:1984tu}
  L.~Gribov, E.~Levin and M.~Ryskin,
  Phys.\ Rep.\ \ {\bf 100} (1983) 1.

\bibitem{GolecBiernat:1998js}
  K.~Golec-Biernat and M.~W\"{u}sthoff,
  Phys.\ Rev.\ {\bf D59} (1998) 014017
  [hep-ph/9807513].

\bibitem{Iancu:2000hn}
  E.~Iancu, A.~Leonidov and L.~McLerran,
  Nucl.\ Phys.\ {\bf A692} (2001) 583
  [hep-ph/0011241].

\bibitem{Jung:2009eq}
Z.~Ajaltouni {\it et al.},
{\em Proceedings of the `HERA and the LHC' workshop
series, chapter 5},
arXiv:0903.3861 [hep-ex].

\bibitem{Brower:2006ea}
  R.~Brower, J.~Polchinski, M.~Strassler and C.~Tan,
  JHEP\ {\bf 0712} (2007) 005
  [hep-th/0603115].

\bibitem{Aad:2011dr}
  ATLAS Collaboration,
  Eur.\ Phys.\ J.\ {\bf C71} (2011) 1630
  [arXiv:1101.2185 [hep-ex]].

\bibitem{Goulianos:1982vk}
  K.~Goulianos,
  Phys.\ Rep.\ \ {\bf 101} (1983) 169.

\bibitem{Alberi:1981af}
  G.~Alberi and G.~Goggi,
  Phys.\ Rep.\ \ {\bf 74} (1981) 1.

\bibitem{Zotov}
N.~Zotov and V.~Tsarev,
Sov.\ Phys.\ Usp.\ \ {\bf 31} (1988) 119.

\bibitem{Kaidalov:1979jz}
  A.~Kaidalov,
  Phys.\ Rep.\ \ {\bf 50} (1979) 157.

\bibitem{Albrow:2010yb}
  M.~Albrow, T.~Coughlin and J.~Forshaw,
  Prog.\ Part.\ Nucl.\ Phys.\ \ {\bf 65} (2010) 149
  [arXiv:1006.1289 [hep-ph]].

\bibitem{Bernard:1986yh}
  UA4 Collaboration,
  Phys.\ Lett.\ {\bf B186} (1987) 227.

\bibitem{Ansorge:1986xq}
  UA5 Collaboration,
  Z.\ Phys.\ {\bf C33} (1986) 175.

\bibitem{Amos:1992jw}
  E710 Collaboration,
  Phys.\ Lett.\ {\bf B301} (1993) 313.

\bibitem{Abe:1993wu}
  CDF Collaboration,
  Phys.\ Rev.\ {\bf D50} (1994) 5535.

\bibitem{Adloff:1997mi}
  H1 Collaboration,
  Z.\ Phys.\ {\bf C74} (1997) 221
  [hep-ex/9702003].

\bibitem{Breitweg:1997za}
  ZEUS Collaboration,
  Z.\ Phys.\ {\bf C75} (1997) 421
  [hep-ex/9704008].

\bibitem{Aktas:2006hy}
  H1 Collaboration,
  Eur.\ Phys.\ J.\ {\bf C48} (2006) 715
  [hep-ex/0606004].

\bibitem{Chekanov:2008fh}
  ZEUS Collaboration,
  Nucl.\ Phys.\ {\bf B816} (2009) 1
  [arXiv:0812.2003 [hep-ex]].

\bibitem{H1:2010kz}
  H1 Collaboration,
  Eur.\ Phys.\ J.\ {\bf C71} (2011) 1578
  [arXiv:1010.1476 [hep-ex]].

\bibitem{Affolder:2001vx}
  CDF Collaboration,
  Phys.\ Rev.\ Lett.\ \ {\bf 87} (2001) 141802
  [hep-ex/0107070].

\bibitem{Joyce:1993ru}
  D.~Joyce  {\it et al.},
  Phys.\ Rev.\ {\bf D48} (1993) 1943.

\bibitem{Brandt:2002qr}
  UA8 Collaboration,
  Eur.\ Phys.\ J.\ {\bf C25} (2002) 361
  [hep-ex/0205037].

\bibitem{Ingelman:1984ns}
  G.~Ingelman and P.~Schlein,
  Phys.\ Lett.\ {\bf B152} (1985) 256.

\bibitem{Kaidalov:1973tc}
  A.~Kaidalov, V.~Khoze, Y.~Pirogov and N.~Ter-Isaakyan,
  Phys.\ Lett.\ {\bf B45} (1973) 493.

\bibitem{Field:1974fg}
  R.~Field and G.~Fox,
  Nucl.\ Phys.\ {\bf B80} (1974) 367.

\bibitem{pythia6}
  T.~Sj\"{o}strand, S.~Mrenna and P.~Skands,
  JHEP\ {\bf 0605} (2006) 026
  [hep-ph/0603175].

\bibitem{py8}
  T.~Sj\"{o}strand, S.~Mrenna and P.~Skands,
  Comput.\ Phys.\ Commun.\ \ {\bf 178} (2008) 852
  [arXiv:0710.3820 [hep-ph]].

\bibitem{phojet}
  R.~Engel,
  Z.\ Phys.\ {\bf C66} (1995) 203.

\bibitem{Ryskin:2011qe}
  M.~Ryskin, A.~Martin and V.~Khoze,
  Eur.\ Phys.\ J.\ {\bf C71} (2011) 1617
  [arXiv:1102.2844 [hep-ph]].

\bibitem{Gotsman:2010nw}
  E.~Gotsman, E.~Levin and U.~Maor,
  Eur.\ Phys.\ J.\ {\bf C71} (2011) 1553
  [arXiv:1010.5323 [hep-ph]].

\bibitem{ATLAS_TDR}
  ATLAS Collaboration,
  JINST {\bf 3} (2008) S08003.

\bibitem{LuminosityWorkingGroup:1330358}
  ATLAS Collaboration,
{\em Updated Luminosity Determination in pp Collisions at
     $\sqrt{s} =7 \ {\rm TeV}$ using the ATLAS Detector},
ATLAS-CONF-2011-011 (2011).

\bibitem{:2010wv}
  ATLAS Collaboration,
  Eur.\ Phys.\ J.\ {\bf C71} (2011) 1512
  [arXiv:1009.5908 [hep-ex]].

\bibitem{Lampl:1099735}
  W.~Lampl {\it et al.},
{\em Calorimeter Clustering Algorithms: Description and Performance},
ATL-LARG-PUB-2008-002; ATL-COM-LARG-2008-003 (2008).

\bibitem{:2009zp}
  ATLAS Collaboration,
  Eur.\ Phys.\ J.\ C {\bf 70} (2010) 723
  [arXiv:0912.2642 [physics.ins-det]].

\bibitem{Aad:2010yt}
  ATLAS Collaboration,
  JHEP\ {\bf 1012} (2010) 060
  [arXiv:1010.2130 [hep-ex]].

\bibitem{ATLAS-CONF-2010-017}
  ATLAS Collaboration, 
{\em Response of the ATLAS Calorimeters to Single Isolated
Hadrons Produced in Proton-proton Collisions at $\sqrt{s} = 900 \ {\rm GeV}$},
ATLAS-CONF-2010-017 (2010).

\bibitem{ATLAS-CONF-2010-052}
ATLAS Collaboration, 
{\em ATLAS Calorimeter Response to Single Isolated
Hadrons and Estimation of the Calorimeter Jet Energy Scale Uncertainty},
ATLAS-CONF-2010-052 (2010).

\bibitem{ATLAS-CONF-2011-028}
ATLAS Collaboration, 
{\em ATLAS Calorimeter Response to Single Isolated Hadrons and Estimation of the Calorimeter Jet Scale Uncertainty},
ATLAS-CONF-2011-028 (2011).

\bibitem{Pinfold:2008zzb}
  J.~Pinfold {\it et al.},
  Nucl.\ Instrum.\ Meth.\ {\bf A593} (2008) 324.

\bibitem{Archambault:2008zz}
J.~Archambault {\it et al.},
JINST {\bf 3} (2008) P02002.

\bibitem{Dowler:2002zp}
  B.~Dowler {\it et al.},
  Nucl.\ Instrum.\ Meth.\ {\bf A482} (2002) 94.

\bibitem{Roy:1974wq}
  D.~Roy and R.~Roberts,
  Nucl.\ Phys.\ {\bf B77} (1974) 240.

\bibitem{Donnachie:1992ny}
  A.~Donnachie and P.~Landshoff,
  Phys.\ Lett.\ {\bf B296} (1992) 227
  [hep-ph/9209205].

\bibitem{Cudell:1996sh}
  J.~Cudell, K.~Kang and S.~Kim,
  Phys.\ Lett.\ {\bf B395} (1997) 311
  [hep-ph/9601336].

\bibitem{Good:1960ba}
  M.~Good and W.~Walker,
  Phys.\ Rev.\ \ {\bf 120} (1960) 1857.

\bibitem{Bopp:1998rc}
  F.~Bopp, R.~Engel and J.~Ranft,
{\em Rapidity gaps and the PHOJET Monte Carlo},
hep-ph/9803437.

\bibitem{py6_diff}
  G.~Schuler and T.~Sj\"{o}strand,
  Phys.\ Rev.\ {\bf D49} (1994) 2257.

\bibitem{Bruni:1993ym}
  P.~Bruni and G.~Ingelman,
  Phys.\ Lett.\ {\bf B311} (1993) 317.

\bibitem{Berger:1986iu}
  E.~Berger, J.~Collins, D.~Soper and G.~Sterman,
  Nucl.\ Phys.\ {\bf B286} (1987) 704.

\bibitem{Streng:1988hd}
K.~Streng,
{\em Hard QCD scatterings in diffractive reactions at HERA},
CERN-TH-4949 (1988).

\bibitem{Donnachie:1984xq}
  A.~Donnachie and P.~Landshoff,
  Nucl.\ Phys.\ {\bf B244} (1984) 322.

\bibitem{Jung:1993gf}
  H.~Jung,
  Comput.\ Phys.\ Commun.\ \ {\bf 86} (1995) 147.

\bibitem{Abe:1993xx}
  CDF Collaboration,
  Phys.\ Rev.\ {\bf D50} (1994) 5518.

\bibitem{Andersson:1983ia}
  B.~Andersson, G.~Gustafson, G.~Ingelman and T.~Sj\"{o}strand,
  Phys.\ Rep.\ \ {\bf 97} (1983) 31.

\bibitem{Navin:2010kk}
  S.~Navin,
{\em Diffraction in Pythia},
arXiv:1005.3894 [hep-ph].

\bibitem{MB15T}
ATLAS Collaboration,
{\em Charged particle multiplicities in $pp$ interactions at 
$\sqrt{s} = 0.9$ and $7 \ {\rm TeV}$ in a diffractive limited 
phase space measured with the ATLAS detector at the LHC and a new \py6 tune},
ATLAS-CONF-2010-031 (2010).

\bibitem{ATL-PHYS-PUB-2011-009}
ATLAS Cololaboration,
{\em ATLAS tunes of PYTHIA 6 and Pythia 8 for MC11},
ATL-PHYS-PUB-2011-009 (2009). 

\bibitem{Capella:1992yb}
  A.~Capella, U.~Sukhatme, C.~Tan and J.~Tran Thanh Van,
  Phys.\ Rep.\ \ {\bf 236} (1994) 225.

\bibitem{Abramovsky:1973fm}
  V.~Abramovsky, V.~Gribov and O.~Kancheli,
  Yad.\ Fiz.\ \ {\bf 18} (1973) 595
   [Sov.\ J.\ Nucl.\ Phys.\ \ {\bf 18} (1974) 308].

\bibitem{Capella:1994uy}
  A.~Capella, A.~Kaidalov, C.~Merino and J.~Tran Thanh Van,
  Phys.\ Lett.\ {\bf B343} (1995) 403
  [hep-ph/9407372].

\bibitem{Capella:1995pn}
  A.~Capella {\it et al.},
  Phys.\ Rev.\ {\bf D53} (1996) 2309
  [hep-ph/9506454].

\bibitem{Bahr:2008pv}
  M.~B\"{a}hr {\it et al.},
  Eur.\ Phys.\ J.\ {\bf C58} (2008) 639
  [arXiv:0803.0883 [hep-ph]].

\bibitem{Gieseke:2011xy}
  S.~Gieseke, C.~A.~R\"{o}hr and A.~Si\'{o}dmok,
{\em Multiple Partonic Interaction Developments in Herwig++},
  arXiv:1110.2675 [hep-ph].

\bibitem{HERWIG:more}
HERWIG$++$ Collaboration, 
{\em Minimum Bias and Underlying Event tunes}, URL: http://  \\
projects.hepforge.org/herwig/trac/wiki/MB\_UE\_tunes.

\bibitem{Donnachie:2004pi}
  A.~Donnachie and P.~Landshoff,
  Phys.\ Lett.\ {\bf B595} (2004) 393
  [hep-ph/0402081].

\bibitem{Gieseke:2011na}
  S.~Gieseke {\it et al.},
{\em Herwig++ 2.5 Release Note},
arXiv:1102.1672 [hep-ph].

\bibitem{:2010wqa}
  ATLAS Collaboration,
  Eur.\ Phys.\ J.\ {\bf C70} (2010) 823
  [arXiv:1005.4568 [physics.ins-det]].

\bibitem{Agostinelli:2002hh}
  GEANT4 Collaboration,
  Nucl.\ Instrum.\ Meth.\ {\bf A506} (2003) 250.

\bibitem{D'Agostini:1994zf}
  G.~D'Agostini,
  Nucl.\ Instrum.\ Meth.\ {\bf A362} (1995) 487.

\bibitem{Hocker:1995kb}
  A.~H\"{o}cker and V.~Kartvelishvili,
  Nucl.\ Instrum.\ Meth.\ {\bf A372} (1996) 469
  [hep-ph/9509307].

\bibitem{Kiryunic}
A.~Kiryunic and P.~Strizenec,
{\em Validation of the Local Hadronic Calibration Scheme
of ATLAS with Combined Beam Test Data in the Endcap and Forward Regions},
Proceedings of the 13th ICATPP Conference, Como, 
Italy, 2011.

\bibitem{springerlink:10.1140/epjc/s10052-010-1392-5}
  V.~A.~Khoze {\it et al.},
  Eur.\ Phys.\ J.\ {\bf C69} (2010) 85
  [arXiv:1005.4839 [hep-ph]].

\bibitem{Martin:2011gi}
A.~Martin, M.~Ryskin and V.~Khoze,
{\em From hard to soft high-energy $pp$ interactions},
arXiv:1110.1973 [hep-ph].

\bibitem{Ryskin:private}
M.~Ryskin, private communication.

\bibitem{hepdata}
{\em The Durham HepData Reaction Database}, URL: \\
http://durpdg.dur.ac.uk/.

\bibitem{rivet}
  B.~Waugh {\it et al.},
{\em HZTool and Rivet: Toolkit and Framework for the Comparison of Simulated Final States and Data at Colliders},
  hep-ph/0605034.



\end{thebibliography}
\end{document}